\documentclass[11pt]{book}

\usepackage{these}
\usepackage[french]{babel}
\usepackage{amsmath}
\usepackage{graphicx,epsfig}

\def\oppropto{\mathop{\propto}} 
\def\opsimeq{\mathop{\simeq}}

\def\opsim{\mathop{\sim}} 

\def\fig#1#2{\includegraphics[height=#1]{#2}}

\begin{document}

\thispagestyle{empty}

 \begin{center}
      {\LARGE 
        \textbf{Université Pierre et Marie Curie }\\[\baselineskip]
      }
       
     { Sp\'ecialit\'e : Physique Th\'eorique }
         \vskip 2cm

%      \PRESENTATION\\[\baselineskip]
      {\Large
        \textbf{ M\'emoire d'habilitation }\\[\baselineskip]
      }
      
      pr\'esent\'e par\\[\baselineskip]
      {\large 
        C\'ecile  MONTHUS }  \\[\baselineskip]
      
      \vskip 2cm
%      Titre du m\'emoire d'habilitation~:\\
     {{\Large\textsl{M\'ethodes de renormalisation dans l'espace r\'eel   \\~\\
            de type Ma-Dasgupta \\~\\
       pour divers syst\`emes d\'esordonn\'es  }}}  \\
      \vfill
      Soutenue le 1er Juin 2004 devant la commission d'examen compos\'ee de Messieurs \\[\baselineskip]
{
\begin{center}
\begin{tabular}{r@{\protect\hspace{0.5cm}}ll@{\protect\hspace{1.0cm}}l}
Pr\'esident~:& Bernard DERRIDA &ENS-Universit\'e Paris 6 \\
Rapporteurs~: & Jean-Philippe BOUCHAUD &SPEC Saclay \\
              & Hendrik-Jan HILHORST &LPT-Universit\'e Paris-Sud \\
              & Jorge KURCHAN &CNRS-ESPCI Paris \\
               Examinateurs~: &  Alan BRAY &Universit\'e de Manchester \\
               & Jean-Marc LUCK &SPhT Saclay \\
\\
\end{tabular}
\end{center}
}
    \end{center}

 \newpage\thispagestyle{empty}\addtocounter{page}{-1}

 \vspace{\stretch{1}}
  \begin{flushright}
  \emph{``Pour les esprits qui savent comprendre, \`a c\^ot\'e de la beaut\'e d'une loi
g\'en\'erale, \\
les finesses d'une analyse subtile et d\'elicate fr\^olant parfois le
paradoxe, \\
 la th\'eorie des hasards pr\'esentera un attrait et un charme tout
particuliers." }\\
  ~\\
 Louis Bachelier
  \end{flushright}
  \vspace{\stretch{2}}

 \newpage\thispagestyle{empty}\addtocounter{page}{-1}

\frontmatter

\chapter{ Avant-Propos   }
\thispagestyle{plain}
\markboth{Avant-Propos}{Avant-Propos}

%\addcontentsline{toc}{chapter}{Guide de lecture }

 Ce m\'emoire d'habilitation d\'ecrit l'application de proc\'edures de renormalisation de type Ma-Dasgupta \`a plusieurs types de syst\`emes d\'esordonn\'es, classiques ou quantiques, dynamiques ou statiques.
Afin de replacer dans une perspective commune ces diff\'erentes \'etudes,
les premiers chapitres sont consacr\'es \`a 
une pr\'esentation g\'en\'erale de la m\'ethode.

%\section*{Structure du m\'emoire}

%\addcontentsline{toc}{section}{Structure du m\'emoire}

\section*{ Les chapitres g\'en\'eraux sur la m\'ethode}

%\addcontentsline{toc}{section}{Les chapitres g\'en\'eraux sur la m\'ethode}

$\bullet$ L'introduction situe les renormalisations de type Ma-Dasgupta 
par rapport aux nombreuses approches qui existent pour les syst\`emes d\'esordonn\'es.  

 \'Etant donn\'e qu'une renormalisation de type Ma-Dasgupta est 
\`a la fois une ``fa\c{c}on de penser"
et une ``fa\c{c}on de calculer", ces deux aspects sont discut\'es
s\'epar\'ement, pour plus de clart\'e :

$\bullet$ Le Chapitre \ref{chapsignification} explique 
en d\'etails les id\'ees physiques essentielles communes
aux approches de type Ma-Dasgupta dans les diff\'erents contextes.

$\bullet$ Le Chapitre \ref{chapreglesrg} pr\'esente 
les r\`egles quantitatives de renormalisation,
qui donnent lieu \`a de jolis probl\`emes de probabilit\'es.

\section*{ Les chapitres sur des syst\`emes d\'esordonn\'es classiques}

%\addcontentsline{toc}{section}{Les chapitres sur des syst\`emes d\'esordonn\'es classiques}

La majorit\'e du m\'emoire est consacr\'ee \`a divers mod\`eles de physique statistique classique, o\`u les approches de type Ma-Dasgupta
sont utilis\'ees soit pour caract\'eriser des propri\'et\'es de dynamique 
lente, de vieillissement ou de croissance de domaines,
soit pour \'etudier les propri\'et\'es \`a l'\'equilibre thermodynamique
 de certains mod\`eles d\'esordonn\'es.
Nous discuterons en particulier les mod\`eles suivants :

$\bullet$ la dynamique d'une particule dans un milieu al\'eatoire unidimensionnel,
que ce soit dans un paysage de forces al\'eatoires (Chapitres \ref{chapsinai}
et \ref{chapsinaibiais}), ou
dans un paysage de pi\`eges al\'eatoires (Chapitres \ref{chaptrap}
et \ref{chapreponsetrap}).

%$\bullet$ les processus de r\'eaction-diffusion
%dans un potentiel Brownien unidimensionnel (Chapitre \ref{chapreadiff}).

$\bullet$ l'\'equilibre et la dynamique hors \'equilibre
de croissance de domaines 
 de cha\^{\i}nes de spins d\'esordonn\'ees classiques (Chapitre \ref{chaprfim}), 
comme la cha\^{\i}ne d'Ising en champ al\'eatoire ou
la cha\^{\i}ne verre de spin en champ magn\'etique ext\'erieur.

$\bullet$ la transition de d\'elocalisation d'un polym\`ere al\'eatoire,
compos\'e de monom\`eres hydrophobes et hydrophiles, \`a une interface entre deux solvants s\'electifs (Chapitre \ref{chappolymere}).

\section*{ Les chapitres sur des syst\`emes d\'esordonn\'es quantiques}

%\addcontentsline{toc}{section}{Les chapitres sur des syst\`emes d\'esordonn\'es %quantiques}

Apr\`es ces mod\`eles de physique statistique classique, 
nous reviendrons dans les deux derniers chapitres 
aux {\bf cha\^{\i}nes de spins quantiques d\'esordonn\'ees},
c'est \`a dire le domaine initial de la m\'ethode : 

 $\bullet$ Le Chapitre \ref{chapquantique1}
concerne la cha\^{\i}ne de spin $S=1$ antiferromagn\'etique al\'eatoire, qui pr\'esente une transition de phases en fonction du d\'esordre, 
contrairement \`a la cha\^{\i}ne $S=1/2$.

 $\bullet$ Le Chapitre \ref{chapquantique2}
est consacr\'e aux propri\'et\'es de taille finie
  de la transition de phase quantique de la
cha\^{\i}ne d'Ising avec couplages et champs transverses al\'eatoires.

 $\bullet$ Enfin, un petit aper\c{c}u des diff\'erentes
\'etudes de type Ma-Dasgupta qui existent dans la litt\'erature,
est donn\'e en Annexe.

\section*{ Publications associ\'ees }

% \addcontentsline{toc}{section}{Publications associ\'ees}

Ce m\'emoire est bas\'e sur les publications
not\'ees [P1] ... [P15], publi\'ees entre 1997 et 2004,
dont la liste se trouve \`a la fin, en page \pageref{listpubli}.
 Par rapport au m\'emoire, ces publications contiennent 
des discussions plus approfondies des mod\`eles,
des r\'esultats compl\'ementaires, 
les d\'erivations de tous les r\'esultats,
et des listes de r\'ef\'erences appropri\'ees.
Par comparaison, le but de ce m\'emoire est de donner une id\'ee g\'en\'erale de la m\'ethode, des mod\`eles qu'elle permet d'\'etudier, et des r\'esultats qu'elle permet d'obtenir, sans entrer dans une description trop d\'etaill\'ee
de chaque mod\`ele discut\'e.

\section*{ Remerciements }

%\chapter*{ Remerciements }
%\thispagestyle{plain}
%\markboth{Remerciements}{Remerciements}

%\addcontentsline{toc}{section}{Remerciements }

Je souhaite bien s\^ur 
remercier chaleureusement mes diff\'erents collaborateurs 
sur les travaux de recherche pr\'esent\'es dans ce m\'emoire.

Tout a commenc\'e `par hasard' en 1996, lors d'une rencontre
dans la salle de caf\'e du SPhT,
lorsque Thierry Jolicoeur m'a parl\'e de la m\'ethode de Ma-Dasgupta pour les cha\^{\i}nes de spins quantiques al\'eatoires. 
Alors que je ne conna\^{\i}ssais presque rien 
sur les cha\^{\i}nes de spins quantiques, j'ai tout de suite eu un coup de coeur pour cette m\'ethode qui proposait de renormaliser des distributions
de probabilit\'e de mani\`ere si simple et si jolie!
En collaboration avec Olivier Golinelli et Thierry Jolicoeur,
nous avons alors entrepris de g\'en\'eraliser la m\'ethode de Ma-Dasgupta
\`a la cha\^{\i}ne de spin $S=1$ al\'eatoire (Chapitre \ref{chapquantique1}). 
Le jour o\`u nous avons compris que la transition de phase
quantique en fonction de la largeur du d\'esordre correspondait
\`a une transition de percolation pour les amas ``VBS"
est l'un de mes meilleurs souvenirs de recherche.
Je remercie donc vivement Olivier Golinelli et Thierry Jolicoeur 
pour ce premier travail de type Ma-Dasgupta qui m'a beaucoup appris.

Je suis tout particuli\`erement reconna\^{\i}ssante \`a
 Pierre Le Doussal, qui m'a propos\'e en 1997
de travailler en collaboration avec Daniel Fisher,
sur l'application de la m\'ethode de
Ma-Dasgupta \`a la diffusion unidimensionnelle en milieu al\'eatoire. 
 Comme je conna\^{\i}ssais d\'ej\`a la m\'ethode 
par les cha\^{\i}nes de spins, et le mod\`ele de Sinai pour
y avoir travaill\'e par d'autres approches durant ma th\`ese, 
c'\'etait \'evidemment un sujet id\'eal pour moi... 
C'\'etait aussi un magnifique cadeau, car il y avait beaucoup
de probabilit\'es \`a calculer !
Je remercie donc vivement Daniel Fisher et Pierre Le Doussal
pour cette collaboration sur la marche al\'eatoire dans un potentiel
Brownien (Chapitre \ref{chapsinai}) et sur la dynamique hors \'equilibre
de cha\^{\i}nes de spins d\'esordonn\'ees classiques (Chapitre \ref{chaprfim}).
Ces travaux ont eu ensuite plusieurs d\'eveloppements en collaboration avec
Pierre Le Doussal, notamment sur
les propri\'et\'es de la localisation de Golosov
(Chapitre \ref{chapsinai}), sur la renormalisation
de potentiels unidimensionnels plus g\'en\'eraux (Chapitre \ref{chaplandscape}),
et sur d'autres sujets, qui attendent encore d'\^etre termin\'es...

Je voudrais ensuite remercier ceux qui m'ont int\'eress\'ee
\`a des mod\`eles discut\'es dans ce m\'emoire.
Ainsi, l'\'etude de la localisation du polym\`ere al\'eatoire
\`a une interface (Chapitre \ref{chappolymere}) est issue de
discussions avec Thomas Garel et Henri Orland,
alors que
l'\'etude des propri\'et\'es de vieillissement et de r\'eponse
du mod\`ele de pi\`eges
(Chapitres \ref{chaptrap} et \ref{chapreponsetrap})
a \'et\'e suscit\'ee par des questions pos\'ees par Jean-Philippe Bouchaud et \'Eric Bertin.

Enfin, je remercie tous les membres du jury pour avoir accept\'e
d'en faire partie et de
consacrer un peu de temps \`a la lecture de ce m\'emoire.

\tableofcontents

\mainmatter

\chapter*{Introduction }
\thispagestyle{plain}
\markboth{Introduction}{Introduction}

\addcontentsline{toc}{chapter}{Introduction }

  \begin{flushright}
  \emph{ O\`u l'on situe les renormalisations de type Ma-Dasgupta parmi les approches \\
 qui cherchent \`a d\'ecrire les fluctuations spatiales dans les syst\`emes d\'esordonn\'es. }\\
  ~\\
  \end{flushright}

La pr\'esence de d\'esordre dans un syst\`eme 
peut donner naissance \`a des ph\'enom\`enes physiques
tout \`a fait nouveaux,
 comme la localisation d'Anderson 
en mati\`ere condens\'ee, ou les propri\'et\'es de dynamique lente
en physique statistique. 
L'\'etude des syst\`emes d\'esordonn\'es 
a donc engendr\'e un certain nombre d'approches sp\'ecifiques
depuis une cinquantaine d'ann\'ees : le premier exemple est 
la m\'ethode de mesure invariante
 de Dyson \cite{dyson} et Schmidt \cite{schmidt} qui
permet d'obtenir des r\'esultats exacts pour les syst\`emes
unidimensionnels d\'ecrits par des produits infinis de matrices de transfert al\'eatoires \cite{luckbook,crisantibook}.
Pour comprendre l'int\'er\^et et la sp\'ecificit\'e 
 des proc\'edures de renormalisation de type
Ma-Dasgupta qui constituent le sujet de ce m\'emoire,
on peut classer 
les diff\'erentes fa\c{c}ons d'aborder les syst\`emes d\'esordonn\'ees
en deux grandes cat\'egories :

$\bullet$ il y a d'une part {\bf des approches qui commencent
 par moyenner sur le d\'esordre},
car elles se fixent pour objectif d'\'evaluer des observables
automoyennantes, comme par exemple l' \'energie libre s'il s'agit d'\'etudier
la thermodynamique du syst\`eme.  
Il existe un certain nombre de prescriptions sp\'ecifiques pour effectuer
cette moyenne sur le d\'esordre :
 la m\'ethode des r\'epliques 
\cite{replica}, la m\'ethode supersym\'etrique \cite{susy}, 
la m\'ethode dynamique
\cite{dynamicreview} 
(pour une pr\'esentation parall\`ele des trois m\'ethodes : \cite{kurchan3methods}).
 Apr\`es cette moyenne sur le d\'esordre, il n'y a plus d'h\'et\'erog\'en\'eit\'es spatiales, mais il y a en \'echange 
un syst\`eme pur avec de nouvelles interactions
effectives.

$\bullet$ il y a d'autre part 
{ \bf des approches qui cherchent \`a d\'ecrire les
h\'et\'erog\'en\'eit\'es spatiales du d\'esordre},
comme certains arguments c\'el\`ebres 
et diverses proc\'edures de renormalisation
dans l'espace r\'eel, que nous allons discuter
un peu plus en d\'etails, car ces approches  
appartiennent \`a la m\^eme `famille de pens\'ee'
que les proc\'edures de type Ma-Dasgupta.

\newpage

\section*{ Des arguments probabilistes 
sur les fluctuations spatiales du d\'esordre ... }

\addcontentsline{toc}{section}{Des arguments probabilistes 
sur les fluctuations spatiales du d\'esordre ...}

\label{argumentslocaux}

Parmi les arguments qui ont jou\'e
un grand r\^ole dans la compr\'ehension des syst\`emes d\'esordonn\'es,
on peut citer

 (a) {\bf l'argument de Lifshitz } \cite{lifshitz,lifbook}, 
qui permet de pr\'edire
les singularit\'es essentielles des densit\'es d'\'etats en bord de spectre,
en identifiant les configurations du d\'esordre qui
engendrent des \'etats dans cette zone d' \'energie, et en estimant
 les probabilit\'es de pr\'esence de ces configurations favorables.

(b) {\bf les phases de Griffiths }, dans lesquelles des r\'egions
rares localement ordonn\'ees induisent des singularit\'es essentielles  
pour l'\'equilibre \cite{griffiths}
et des lois de d\'ecroissance lente pour la dynamique
\cite{randeira,braygriffithsdyna}.

(c) {\bf le crit\`ere de Harris } \cite{harris} sur la pertinence d'un faible d\'esordre autour du point critique pur, qui estime l'influence des fluctuations
locales du d\'esordre sur les fluctuations de temp\'eratures critiques.

(d) { \bf l'argument de Imry-Ma } \cite{imryma}, qui permet de pr\'edire
l'apparition de parois de domaines dans 
les syst\`emes en champs al\'eatoires, en consid\'erant les fluctuations
spatiales d' \'energie venant des champs al\'eatoires.

(e) { \bf le th\'eor\`eme de Chayes et al. } \cite{chayes}, qui montre que les fluctuations spatiales
du d\'esordre impliquent l'in\'egalit\'e $\nu \geq 2/d$ pour l'exposant critique
$\nu$ dans un syst\`eme d\'esordonn\'e en dimension $d$.

En fait, ces diff\'erents arguments ne font intervenir que
deux propri\'et\'es statistiques diff\'erentes.
D'une part, les arguments (a) et (b), qui sont tr\`es proches \cite{theo89,luckbook},
reposent tous les deux la prise en compte d'\'ev\`enements rares :
dans toute configuration infinie de d\'esordre, 
il existe des domaines ordonn\'es arbitrairement grands 
avec des probabilit\'es exponentiellement petites.
D'autre part, les arguments (c), (d) et (e) reposent 
tous les trois sur
le comportement typique en $\sqrt N$ de la somme d'un grand nombre $N$
de variables al\'eatoires ind\'ependantes.

Ces arguments de type probabiliste sont bien fond\'es
et presque ``inattaquables". A ma conna\^{\i}ssance, 
le seul argument qui a donn\'e lieu \`a une controverse \cite{imryrf}
est l'argument de Imry-Ma qui \'etait en d\'esaccord 
avec la ``r\'eduction dimensionnelle" pr\'edite
par les approches de th\'eorie des champs,
que ce soit en perturbation \`a tous les ordres \cite{aharony,villain88}
ou dans le formalisme supersym\'etrique \cite{parisisourlas}.
Les \'etudes rigoureuses \cite{imbrie,bricmont} 
ont finalement donn\'e raison \`a ... l'argument de Imry-Ma!
Cet exemple montre que les arguments dits ``heuristiques"
m\^eme s'ils sont simples, peuvent avoir un contenu physique
non-trivial, qu'il n'est pas toujours facile de retrouver
par des m\'ethodes plus ``sophistiqu\'ees".

En conclusion, ces arguments de type probabilistes
 permettent de bien comprendre la physique,
car ils identifient les fluctuations locales du d\'esordre
qui sont responsables de tel ou tel ph\'enom\`ene.
En contrepartie, il est souvent difficile de d\'epasser
 l'aspect qualitatif et de `calculer' vraiment...
Pour aller au del\`a en restant dans le m\^eme esprit, 
l'id\'ee la plus naturelle est bien s\^ur 
 d'utiliser des proc\'edures de renormalisation dans l'espace r\'eel.

\section*{ ... aux proc\'edures de renormalisation sur le d\'esordre }

\addcontentsline{toc}{section}{... aux proc\'edures de renormalisation sur le d\'esordre}

Le choix de travailler dans l'espace r\'eel pour
d\'efinir une proc\'edure de renormalisation,
qui a d\'ej\`a un grand int\'er\^et pour les syst\`emes purs 
\cite{niemeijer,burkhardt},
devient l'unique choix possible en pr\'esence de d\'esordre 
si l'on souhaite d\'ecrire les h\'et\'erog\'en\'eit\'es spatiales.

\subsection*{ Renormalisations par blocs  }

Les renormalisations par blocs, de type d\'ecimation
ou de type Migdal-Kadanoff sont les renormalisations
les plus utilis\'ees pour les syst\`emes d\'esordonn\'es.
Les proc\'edures de type Migdal-Kadanoff \cite{migdal,kadanoff}
constituent en effet des approximations simples
pour effectuer des renormalisations par blocs
sur les r\'eseaux r\'eguliers. Elles repr\'esentent aussi
des renormalisations exactes sur certains r\'eseaux hi\'erarchiques
\cite{kaufman}.
Parmi les syst\`emes \'etudi\'es, on peut citer
par exemple le mod\`ele de Potts \cite{kinzel},
les ferromagn\'etiques dilu\'es \cite{jaya},
et surtout les verres de spin,
qui ont donn\'e lieu \`a un grand nombre de travaux :
ces \'etudes concernent les diagrammes de phases \cite{youngsgrgmk,braymoore84,brayfeng},
diverses propri\'et\'es de la phase verre de spin
\cite{thill,moore,
sasaki}, et surtout le caract\`ere chaotique
des trajectoires du flot de renormalisation
\cite{mckay,banavarbray,muriel,thill} 
qui est une grande nouveaut\'e par rapport aux syst\`emes purs.

Par ailleurs, divers syst\`emes d\'esordonn\'es
ont \'et\'e \'etudi\'es par des proc\'edures de renormalisation
sur r\'eseaux hi\'erarchiques,
en particulier le mod\`ele de Potts \cite{derridagardner},
les polym\`eres dirig\'es en milieu al\'eatoire
\cite{derridagriffiths,cookderrida,dasilveirajpb},
et les probl\`emes de mouillage al\'eatoire \cite{derridawetting,tangchate}.

Citons enfin les proc\'edures de renormalisation
par blocs pour la diffusion dans les mod\`eles de pi\`eges \cite{machta}
et pour les cha\^{\i}nes de spins quantiques al\'eatoires \cite{hirsch},
car nous discuterons dans ce m\'emoire l'application
des proc\'edures de type Ma-Dasgupta \`a ces m\^emes mod\`eles.

\subsection*{ Renormalisation fonctionnelle pour les interfaces en milieux al\'eatoires }

Pour les mod\`eles d'interfaces en milieux al\'eatoires,
il existe
une m\'ethode de renormalisation fonctionnelle \cite{danielFRG} :
c'est une m\'ethode de th\'eorie des champs qui \'etudie le flot
de la fonction de corr\'elation du d\'esordre.
Nous renvoyons \`a la revue \cite{wiese} pour la description
des divers d\'eveloppements r\'ecents, et aux r\'ef\'erences
\cite{balents,pldwiese} pour des comparaisons avec les m\'ethodes de r\'epliques.

\subsection*{ Renormalisation pour les mod\`eles XY d\'esordonn\'es en 2D }

L'introduction de d\'esordre
dans les mod\`eles bidimensionnels de type XY, dans lesquels il existe
des transitions de type Kosterlitz-Thouless dans le cas pur, 
a conduit \`a une renormalisation de type gaz coulombiens,
dans laquelle on \'etudie le flot
d'une distribution de probabilit\'e des fugacit\'es \cite{carpentier1}
afin de prendre en compte l'effet des h\'et\'erog\'en\'eit\'es spatiales
sur les d\'efauts topologiques.
Cette approche permet aussi d'\'etudier la transition vitreuse
d'une particule dans un potentiel al\'eatoire pr\'esentant
des corr\'elations logarithmiques \cite{carpentier2}.

\subsection*{ Renormalisations ph\'enom\'enologiques
pour les verres de spin  }

La ``th\'eorie des droplets" \cite{droplets}
pour les verres de spin en dimension finie,
a pour origine une renormalisation ph\'enom\'enologique
introduite par Mc Millan \cite{mcmillan}
et d\'evelopp\'ee par Bray et Moore
 \cite{braymooreheidelberg}.
Dans la formulation de Bray et Moore \cite{braymooreheidelberg},
l'id\'ee essentielle est que la distribution de probabilit\'e $P_L(J)$
des couplages effectifs $J$ \`a l'\'echelle $L$
converge vers une forme fixe, avec une largeur
$J(L)=J L^{y}$ qui d\'epend de l'\'echelle $L$. 
L'exposant $y$ et la distribution limite, qui sont calculables
exactement en $d=1$, ont \'et\'e \'etudi\'es num\'eriquement en $d=2$
($y<0$) et $d=3$ ($y>0$ couplage fort ) \cite{braymooreheidelberg}. 
Cette id\'ee d'un point fixe de couplage fort (ou de temp\'erature nulle), 
d\'ecrit par une forme d'\'echelle
pour la distribution de probabilit\'e d'une variable al\'eatoire,
correspond en fait exactement \`a la description que l'on obtient
par les proc\'edures de renormalisation de type Ma-Dasgupta
lorsqu'on peut les utiliser.

\subsection*{ Renormalisation de Ma-Dasgupta pour les cha\^{\i}nes de spins quantiques   }

La proc\'edure de renormalisation 
introduite par Ma-Dasgupta-Hu en 1979 \cite{madasgupta}  
pour \'etudier la cha\^{\i}ne de spin quantique $S=1/2$ avec interactions
antiferromagn\'etiques al\'eatoires
a pour caract\'eristique essentielle
de renormaliser de mani\`ere inhomog\`ene dans l'espace
pour mieux s'adapter aux r\'ealisations locales du d\'esordre. 
En effet, les m\'ethodes usuelles de renormalisation
traitent l'espace de mani\`ere homog\`ene, en rempla\c{c}ant
par exemple chaque bloc de spins de taille donn\'ee
par un super-spin. Si ce caract\`ere homog\`ene
 est naturel pour les syst\`emes purs,
 on peut cependant se poser la question de sa l\'egitimit\'e
pour les syst\`emes d\'esordonn\'es qui brisent
l'invariance par translation du syst\`eme.
Ma, Dasgupta et Hu ont d\'efini une proc\'edure de renormalisation sur l' \'energie, et non sur la taille d'une cellule spatiale.
La proc\'edure consiste 
\`a d\'ecimer de mani\`ere it\'erative les degr\'es de libert\'e
de plus haute  \'energie, afin d'obtenir une th\'eorie effective de basse  \'energie pour la cha\^{\i}ne de spin, qui porte aujourd'hui le nom
de ``random singlet phase".

Cette proc\'edure de renormalisation est en fait rest\'ee peu connue et peu utilis\'ee pendant de nombreuses ann\'ees ...
jusqu'aux travaux de Daniel Fisher
en 1994-1995 \cite{danielrtfic,danielantiferro}
qui lui ont donn\'e \`a la fois :

\label{introdaniel}

{\bf (i) un statut th\'eorique bien d\'efini : 
les points fixes de ``d\'esordre infini"  }
  
 Alors que la m\'ethode apparaissait jusque l\`a
comme une proc\'edure approximative peu contr\^ol\'ee,
Daniel Fisher a montr\'e que le flot de renormalisation
prenait une forme d'\'echelle qui convergeait vers un point fixe
de ``d\'esordre infini" (ce qui signifie que le d\'esordre
est de plus en plus fort \`a grande \'echelle), 
ce qui rendait la m\'ethode asymptotiquement exacte
\cite{danielrtfic,danielantiferro}. Par ailleurs, l'application de 
la m\'ethode \`a la cha\^{\i}ne de spin avec couplages et champs transverses
al\'eatoires (``Random Transverse Field Ising Chain" en anglais)
lui a permis de montrer explicitement l'exactitude
des r\'esultats obtenus gr\^ace \`a une comparaison directe
avec certaines observables calcul\'ees rigoureusement pour
le mod\`ele de McCoy et Wu \cite{mccoywu,shankar}. 
(Ce mod\`ele d\'esordonn\'e de McCoy et Wu est un mod\`ele d'Ising classique bidimensionnel
avec un d\'esordre constant par colonne, qui est \'equivalent 
\`a la cha\^{\i}ne RTFIC.)

{\bf (ii) des possibilit\'es de calculs explicites assez remarquables  }

Pour la cha\^{\i}ne RTFIC \cite{danielrtfic}, 
 la proc\'edure de Ma-Dasgupta permet  
d'obtenir beaucoup de r\'esultats nouveaux
par rapport aux m\'ethodes rigoureuses \cite{mccoywu,shankar,theohenri},
en particulier l'exposant critique exact 
$\beta=(3-\sqrt 5)/2$ pour l'aimantation spontan\'ee.
Plus surprenant, Daniel Fisher a m\^eme calcul\'e 
des observables que l'on ne conna\^{\i}t
pas pour le mod\`ele pur correspondant,
c'est \`a dire pour le mod\`ele d'Ising pur bidimensionnel !
(par exemple la fonction d'\'echelle explicite
d\'ecrivant l'aimantation en fonction du champ appliqu\'e
dans la r\'egion critique) .

Ces travaux de Daniel Fisher ont engendr\'e 
un grand int\'er\^et pour ces m\'ethodes dans le domaine des spins quantiques
(cf les Chapitres \ref{chapquantique1} et \ref{chapquantique2},
ainsi que l'Annexe).
Dans ce m\'emoire, nous allons voir que 
les proc\'edures de renormalisation de type Ma-Dasgupta 
ne se limitent pas aux 
syst\`emes de spins quantiques, mais sont aussi 
un outil id\'eal pour \'etudier une classe beaucoup plus large
de syst\`emes d\'esordonn\'es dans le domaine
de la physique statistique.

%\mainmatter

\chapter{ La signification physique
des approches de type Ma-Dasgupta} 

\label{chapsignification}

  \begin{flushright}
  \emph{ O\`u l'on explique la notion de ``point fixe de fort d\'esordre" \\
et la mani\`ere de raisonner dans une approche de type Ma-Dasgupta }\\
  ~\\
  \end{flushright}

Dans ce premier chapitre \`a caract\`ere p\'edagogique,
nous pr\'esentons en d\'etails les id\'ees physiques
essentielles communes aux diverses proc\'edures de renormalisation
utilis\'ees dans divers contextes physiques.

\section{ L'id\'ee essentielle : le d\'esordre qui domine par rapport aux fluctuations thermiques ou quantiques}

%\markboth{La signification physique
%des approches de type Ma-Dasgupta}{L'id\'ee essentielle : le d\'esordre qui %domine}

\markright{L'id\'ee essentielle : le d\'esordre qui domine}

Dans tous les mod\`eles consid\'er\'es dans ce m\'emoire,
les diff\'erentes proc\'edures de renormalisation reposent sur
la m\^eme id\'ee : \`a grande \'echelle,
le d\'esordre domine par rapport aux fluctuations thermiques ou quantiques.
En particulier, les proc\'edures de renormalisation
de type Ma-Dasgupta sont vraiment sp\'ecifiques
aux syst\`emes d\'esordonn\'es et ne peuvent m\^eme pas \^etre d\'efinies
pour les syst\`emes purs qui ne pr\'esentent pas
d'h\'et\'erog\'eit\'es spatiales.

\subsection{ Signification dans les diff\'erents contextes physiques}

D'une certaine mani\`ere, les syst\`emes purs, qui sont gouvern\'es
par les fluctuations d'origine thermique ou quantique,
 sont caract\'eris\'es
par une grande ``d\'eg\'en\'erescence", dans la mesure o\`u
tous les sites sont \'equivalents, alors que la pr\'esence de d\'esordre
l\`eve compl\`etement cette d\'eg\'en\'erescence.
Voici quelques exemples de cette id\'ee dans les diff\'erents domaines.

\subsubsection{ Exemple pour l'\'etat fondamental d'un syst\`eme quantique}

Dans une cha\^{\i}ne quantique antiferromagn\'etique
pure de spin $S=1/2$, on peut voir qualitativement l'\'etat fondamental
comme une combinaison lin\'eaire appropri\'ee
des \'etats qui correspondent
\`a toutes les fa\c{c}ons possibles d'apparier les spins
 par deux pour former des singulets. En revanche, en pr\'esence de d\'esordre,
la proc\'edure de renormalisation de Ma-Dasgupta
associe \`a chaque r\'ealisation du d\'esordre
un \'etat fondamental qui correspond \`a une seule fa\c{c}on
d'apparier les spins en singulets.
Ainsi, dans un \'echantillon d\'esordonn\'e fix\'e,
un spin donn\'e est compl\`etement corr\'el\'e \`a un seul autre spin
de la cha\^{\i}ne, qui peut \'eventuellement \^etre \`a une distance assez grande, mais n'est pratiquement pas corr\'el\'e
avec les autres spins, m\^eme ses voisins imm\'ediats sur la cha\^{\i}ne.

 \subsubsection{ Exemple pour l'\'equilibre d'une cha\^{\i}ne de spins classiques}

 Dans la cha\^{\i}ne d'Ising ferromagn\'etique pure,
 une paroi de domaine peut se trouver 
de mani\`ere \'equiprobable sur tous les liens
car ils sont tous \'equivalents.
L'argument classique  \'energie/entropie entre l' \'energie $2J$
que co\^ute une paroi de domaine et l'entropie $S \sim k \ln L$
associ\'e \`a la position arbitraire
de la paroi dans un syst\`eme fini de $L$ sites, permet de comprendre
l'absence d'ordre \`a longue port\'ee \`a temp\'erature finie
et le comportement en $L_T \sim e^{2J/T}$ de la taille typique d'un domaine.

En revanche, la pr\'esence d'un champ al\'eatoire
va lever cette \'equivalence entre tous les sites.
L'argument d'Imry-Ma \cite{imryma}, qui remplace 
l'argument  \'energie/entropie pr\'ec\'edent,
est un argument  \'energie/ \'energie :
la comparaison entre l' \'energie $2J$
que co\^ute une paroi de domaine, et l' \'energie typique $\sqrt{\sigma L}$
gagn\'ee en profitant d'une fluctuation favorable
de la somme $\sum_{i=1}^L h_i$ des champs al\'eatoires
 sur un domaine de taille $L$, conduit \`a une absence d'ordre \`a longue port\'ee, m\^eme \`a temp\'erature nulle, et \`a une longueur typique
 $L_{IM} \sim J^2/\sigma$ pour les domaines.
Nous verrons que la proc\'edure de renormalisation de type Imry-Ma permet
de d\'eterminer les positions des
parois des domaines d'Imry-Ma qui existent dans un \'echantillon donn\'e.

 \subsubsection{ Exemples pour les marches al\'eatoires en milieux al\'eatoires}

Pour une marche al\'eatoire pure, la probabilit\'e de pr\'esence
se r\'epartit de mani\`ere gaussienne autour de l'origine,
et il y a un \'etalement progressif au cours du temps.

En revanche, pour une marche al\'eatoire dans un potentiel Brownien,
il y a des r\'egions particuli\`eres \`a chaque \'echantillon
qui localisent presque toute la probabilit\'e de pr\'esence,  
et nous verrons comment les caract\'eriser
 par une proc\'edure de type Ma-Dasgupta.
En particulier, les fluctuations thermiques sont tout \`a fait sous-dominantes :
la distance entre deux particules ind\'ependantes
( c'est \`a dire deux histoires thermiques) qui diffusent dans
le m\^eme \'echantillon, reste une variable al\'eatoire finie
dans la limite de temps infini, ce qui constitue la localisation
de Golosov \cite{golosovlocali}.
 
De m\^eme, dans les mod\`eles unidimensionnels de pi\`eges
caract\'eris\'es par une distribution large des temps de pi\'egeage
$p(\tau) \sim 1/\tau^{1+\mu}$ avec $0<\mu<1$, 
le front de diffusion dans un \'echantillon
donn\'e est essentiellement localis\'e
 sur un nombre fini de pi\`eges. \`A nouveau,
nous verrons comment \'etudier
leurs propri\'et\'es statistiques par une approche
de type Ma-Dasgupta g\'en\'eralis\'ee.
Au contraire, dans la phase $\mu>1$ o\`u le temps moyen de pi\'egeage est fini,
il n'y a plus de localisation \`a temps infini, et une approche de type Ma-Dasgupta n'a plus de sens.

 \subsubsection{ Conclusion : les points fixes de fort d\'esordre }

Si l'on s'int\'eresse \`a la physique d'un syst\`eme d\'esordonn\'e
\`a grande \'echelle, 
il y a a priori trois possibilit\'es pour l'\'evolution
du d\'esordre effectif par rapport aux fluctuations thermiques.
  En effet, lorsque l'\'echelle augmente, ce d\'esordre effectif peut devenir

(i) de plus en plus petit : le syst\`eme est gouvern\'e par un point fixe pur.

(ii) de plus en plus grand : le syst\`eme est gouvern\'e par un point fixe de d\'esordre infini.

(iii) ou bien rester asymptotiquement \`a un niveau stable :
le syst\`eme est gouvern\'e par un point fixe de d\'esordre fini.

Dans certains mod\`eles, un d\'esordre initial m\^eme faible
conduit \`a un point fixe de d\'esordre infini (ii) :
c'est en particulier le cas pour la cha\^{\i}ne quantique antiferromagn\'etique
al\'eatoire $S=1/2$, pour le mod\`ele de Sinai, pour la cha\^{\i}ne d'Ising
en champs al\'eatoires...

Lorsqu'il existe un point fixe de d\'esordre fini (iii), 
il est quantifi\'e par un nombre qui varie contin\^ument, 
comme le param\`etre $\mu$ du mod\`ele de pi\`eges.
Le point fixe peut souvent \^etre qualifi\'e de fort d\'esordre dans une certaine r\'egion de param\`etres. Ainsi, dans le mod\`ele de pi\`eges
ou dans le mod\`ele de Sinai en pr\'esence d'un champ ext\'erieur,
la dynamique est gouvern\'ee par un point fixe de fort d\'esordre
dans la phase $\mu<1$ qui pr\'esente une localisation partielle
du paquet thermique : il y a une probabilit\'e finie que
deux trajectoires thermiques dans le m\^eme \'echantillon
soient \`a une distance finie \`a temps infini.

En conclusion, les m\'ethodes de renormalisation 
de type Ma-Dasgupta concernent :

$\bullet$ les points fixes de d\'esordre infini (ii).

$\bullet$ les points fixes de d\'esordre fini (iii) qui peuvent \^etre qualifi\'es de d\'esordre fort.

\subsection{ Comment savoir si le d\'esordre domine \`a grande \'echelle? }

\subsubsection{ Par des arguments th\'eoriques a priori? }

L'importance relative qu'ont les fluctuations thermiques
par rapport au d\'esordre \`a grande \'echelle 
ne se voit pas directement sur le mod\`ele microscopique
et n'est souvent pas tr\`es bien connue pour la plupart
des syst\`emes d\'esordonn\'es. 
M\^eme pour les syst\`emes en champ al\'eatoire
pour lesquels il existe un argument d'Imry-Ma  \'energie/ \'energie
discut\'e ci-dessus, il n'existe pas, \`a ma conna\^{\i}ssance, 
d'argument g\'en\'eralis\'e qui inclurait dans la discussion
les fluctuations thermiques des parois de domaines
et qui estimerait l'importance de ``l'entropie de d\'ecoupage en domaines
Imry-Ma" dans un \'echantillon.

\subsubsection{ Par des \'etudes num\'eriques? }

Dans les \'etudes num\'eriques
qui ont a priori une information directe sur diff\'erentes
configurations thermiques pour une r\'ealisation fix\'ee du d\'esordre,
il est tr\`es rare de trouver cette information sur un
 \'echantillon, car les r\'esultats publi\'es
sont en g\'en\'eral consacr\'es \`a
des moyennes de diverses observables sur les \'echantillons.
Ce qui est bien s\^ur dommage du point de vue des approches de type Ma-Dasgupta, o\`u l'information essentielle est justement
l'importance des fluctuations thermiques dans un \'echantillon fix\'e.
En effet, les moyennes sur les \'echantillons ont toujours le ``risque"
d'\^etre domin\'ees par des \'ev\`enements rares et de donner une fausse
image des comportements typiques. 
Par exemple, dans le mod\`ele de Sinai, la largeur
thermique moyenn\'ee sur les \'echantillons, qui est une observable
naturelle \`a mesurer num\'eriquement pour caract\'eriser
l'\'etalement du paquet thermique, diverge \`a grand temps.
Ce r\'esultat pourrait faire penser qu'il n'y a pas de localisation asymptotiquement,
alors qu'en fait, la distance entre deux particules ind\'ependantes
dans le m\^eme \'echantillon reste une variable al\'eatoire finie
\`a temps infini, ce qui correspond \`a une localisation tr\`es forte en loi.

\subsubsection{ Hypoth\`ese du d\'esordre fort et sa v\'erification }

En cons\'equence, la d\'emarche
g\'en\'eralement utilis\'ee dans les approches de type Ma-Dasgupta
est la suivante : {  on commence par supposer que le d\'esordre domine
\`a grande \'echelle, et on v\'erifie \`a la fin la consistance de l'hypoth\`ese.} 

Plus pr\'ecis\'ement, les proc\'edures de renormalisation
de Ma-Dasgupta contiennent leur propre test de validit\'e : 
si les distributions de probabilit\'e ont une largeur
qui cro\^{\i}t ind\'efiniment par le flot de renormalisation,
elles conduisent \`a des r\'esultats asymptotiquement exacts,
alors que si la largeur des distributions de probabilit\'e 
converge vers une valeur finie, elles donnent des r\'esultats
qui seront seulement approch\'es.

\section{ La mani\`ere de raisonner   }

Dans la renormalisation de Ma-Dasgupta
pour les cha\^{\i}nes de spins quantiques, la proc\'edure
correspond \`a une d\'ecimation de degr\'es de libert\'e sur
l'Hamiltonien : elle est donc assez proche des renormalisations usuelles,
la seule diff\'erence \'etant que la d\'ecimation se fait de mani\`ere it\'erative sur le couplage le plus fort, au lieu de se faire de mani\`ere homog\`ene sur toute la cha\^{\i}ne \`a chaque \'etape.

En revanche, dans les renormalisations de type Ma-Dasgupta
pour les syst\`emes de physique statistique qui seront
discut\'es dans ce m\'emoire, 
la mani\`ere de raisonner s'\'ecarte beaucoup plus des proc\'edures
usuelles. En effet, { le point de d\'epart 
 n'est pas} une int\'egration exacte ou approch\'ee
sur les degr\'es de libert\'e du mod\`ele microscopique, 
c'est \`a dire sur la fonction de partition
pour les probl\`emes d'\'equilibre ou sur l'\'equation ma\^{\i}tresse
pour les probl\`emes dynamiques.
Le point de d\'epart est plut\^ot un argument physique heuristique,
qui permet d'identifier a priori les degr\'es de libert\'e 
qui sont importants \`a grande \'echelle.
On d\'efinit alors directement la renormalisation sur ces degr\'es de libert\'e
jug\'es importants, et on obtient ainsi
dans les cas favorables des r\'esultats 
exacts dans la limite asymptotique o\`u la proc\'edure
de renormalisation est appliqu\'ee un grand nombre de fois.
Comme ce m\'elange entre arguments heuristiques au d\'epart
et r\'esultats exacts \`a la fin
peut para\^{\i}tre d\'econcertant au premier abord, et se heurte
m\^eme souvent 
\`a une certaine incompr\'ehension, il est utile 
d'analyser en d\'etails les diff\'erentes \'etapes du raisonnement
sur le cas du mod\`ele de Sinai (Chapitre \ref{chapsinai}),
qui est le cas le plus p\'edagogique.

\subsection{ Mod\`eles dynamiques : l'exemple du mod\`ele de Sinai  }

La physique du mod\`ele de Sinai sera d\'ecrite dans le chapitre
\ref{chapsinai} qui lui est consacr\'e. Notre but ici
est uniquement de d\'ecrire la mani\`ere de raisonner
sur cet exemple pr\'ecis.

\label{exemplesinai}

\subsubsection{ On identifie les degr\'es de libert\'e du d\'esordre
qui vont \^etre importants \`a grande \'echelle } 

 ``On part d'un argument physique qualitatif" :
il existe depuis longtemps un argument
qualitatif simple \cite{jpbreview} 
pour pr\'edire le comportement typique
du d\'eplacement en $x \sim (\ln t)^2$ dans le mod\`ele de Sinai,
au lieu du comportement habituel en $x \sim \sqrt{t}$ de la diffusion pure,
que l'on peut r\'esumer ainsi :
le temps $t(x)$ n\'ecessaire pour atteindre le point $x>0$
va \^etre domin\'e par le facteur d'Arrh\'enius $e^{\beta B_x}$
correspondant \`a la plus grande barri\`ere $B_x$
qu'il faudra franchir par activation thermique
pour passer du point de d\'epart $x=0$ au point $x$
(Cette approximation par le facteur d'Arrh\'enius
 revient \`a effectuer une m\'ethode du col sur
une expression analytique du temps du premier passage).
Dans un potentiel al\'eatoire Brownien,
le comportement typique de la barri\`ere est donn\'e par $B_x \sim \sqrt{x}$,
ce qui correspond \`a un temps d'Arrh\'enius $t \sim e^{\beta \sqrt{x} }$,
ce qui correspond bien au scaling $x \sim (\ln t)^2$ apr\`es inversion.

 ``On prend au s\'erieux cet argument pour d\'eterminer
les degr\'es de libert\'e qui vont \^etre importants \`a grande \'echelle" :
l'argument heuristique ci-dessus sugg\`ere que les degr\'es de libert\'e
qui vont \^etre importants \`a grande \'echelle sont les grandes
barri\`eres qui existent dans le potentiel al\'eatoire. 
Plus pr\'ecis\'ement, \`a un instant $t$ fix\'e, la particule
n'aura pas eu le temps de franchir par activation thermique
les barri\`eres qui sont
plus grandes qu'une \'echelle d'ordre $ (T \ln t)$.

\subsubsection{ On d\'efinit la renormalisation
directement sur les degr\'es de libert\'e importants du d\'esordre }

On d\'efinit donc une proc\'edure de renormalisation sur les barri\`eres
du potentiel al\'eatoire, dans laquelle on \'elimine de mani\`ere it\'erative la plus petite barri\`ere. Cette barri\`ere minimum 
d\'efinit l'\'echelle de renormalisation $\Gamma$ du paysage.
Le paysage renormalis\'e \`a l'\'echelle $\Gamma$ ne contient donc
que des barri\`eres plus grandes que $\Gamma$, toutes les barri\`eres plus petites ayant \'et\'e d\'ecim\'ees. 
Lorsque l'\'echelle $\Gamma$ augmente,  
la distribution $P_{\Gamma}(F)$
des barri\`eres $F$ du paysage \`a l'\'echelle $\Gamma$ 
tend vers une forme d'\'echelle $\theta(F \geq \Gamma) P^* \left( \frac{F-\Gamma}{\Gamma} \right)$, dans laquelle la distribution stationnaire $P^*$ caract\'erise le point fixe de d\'esordre infini.

\subsubsection{ On \'etablit la correspondance avec 
le mod\`ele initial }

On associe \`a chaque temps $t$ du mod\`ele initial,
le paysage renormalis\'e des barri\`eres \`a l'\'echelle $\Gamma=T \ln t$.
On d\'efinit une dynamique effective sans fluctuations thermiques,
dans laquelle la particule se trouve \`a l'instant $t$
au minimum de la vall\'ee renormalis\'ee \`a l'\'echelle $\Gamma=T \ln t$
qui contient la position initiale \`a $t=0$.
On \'etablit aussi une correspondance entre les diff\'erentes observables
physiques du mod\`ele initial et les propri\'et\'es, statiques ou dynamiques,
 du paysage renormalis\'e.

\subsubsection{ On v\'erifie la consistance
de la proc\'edure dans le r\'egime asymptotique
et on \'etudie les premi\`eres corrections }

La probabilit\'e que la particule ne soit pas
dans la vall\'ee renormalis\'ee de la dynamique effective, 
est d'ordre $1/(\ln t)$ et tend donc vers z\'ero dans la limite des temps infinis, ce qui montre la consistance de la proc\'edure de renormalisation du paysage. On \'etudie les fluctuations thermiques par rapport
\`a la dynamique effective, en consid\'erant d'une part la distribution
de probabilit\'e \`a l'int\'erieur de la vall\'ee renormalis\'ee,
et d'autre part les \'ev\`enements rares d'ordre $1/(\ln t)$
o\`u la particule n'est pas
dans la vall\'ee renormalis\'ee de la dynamique effective.

\subsubsection{ Discussion }

Cet exemple montre bien comment cette fa\c{c}on de raisonner
permet d'obtenir une description tr\`es compl\`ete
de la dynamique asymptotique \`a grand temps.
Il montre aussi tout l'int\'er\^et   
de renormaliser de mani\`ere inhomog\`ene dans l'espace
pour mieux s'adapter aux extrema locaux du d\'esordre
aux diff\'erentes \'echelles,
par rapport aux renormalisations usuelles
qui traitent l'espace de mani\`ere homog\`ene
avec des cellules de taille fix\'ee \`a chaque \'etape.

D'une certaine mani\`ere, la d\'emarche que l'on vient de d\'ecrire
 utilise au maximum
les id\'ees qualitatives contenues dans la notion de renormalisation,
comme la non-pertinence des d\'etails du mod\`ele microscopique
sur les comportements \`a grande \'echelle,
et la convergence vers des th\'eories plus simples repr\'esentant
des classes d'universalit\'e,
{ \it avant } de d\'efinir une proc\'edure quantitative.
Il ne faut donc pas reprocher aux
approches de type Ma-Dasgupta de prendre comme point
de d\'epart des arguments physiques qualitatifs,
car c'est de l\`a justement que vient toute leur efficacit\'e !
En effet, pour tous les mod\`eles de physique statistique
que nous allons consid\'erer,
alors qu'une renormalisation usuelle sur le mod\`ele microscopique
d\'esordonn\'e n'aurait aucun espoir d'\^etre ferm\'ee et 
d'aboutir \`a la fin \`a des r\'esultats exacts,
l'approche de Ma-Dasgupta permet d\'ecrire une renormalisation ferm\'ee
asymptotiquement exacte, directement sur les degr\'es de libert\'e 
qui sont vraiment importants \`a grande \'echelle.

\subsection{ Mod\`eles \`a l'\'equilibre : de l'argument Imry-Ma \`a 
la renormalisation Ma-Dasgupta ! }

Pour les syst\`emes d\'esordonn\'es \`a l'\'equilibre
qui seront discut\'es dans ce m\'emoire, 
comme la cha\^{\i}ne d'Ising classique en champ al\'eatoire
(ou la cha\^{\i}ne verre de spin en champ magn\'etique ext\'erieur
qui lui est \'equivalente par transformation de jauge),
ou l'h\'et\'eropolym\`ere \`a une interface, la renormalisation
de type Ma-Dasgupta est un moyen de construire 
les domaines Imry-Ma dans chaque \'echantillon. 
Les degr\'es de libert\'e importants
sont les sommes des  \'energies al\'eatoires par domaines, et la renormalisation permet d'\'eliminer,
de mani\`ere it\'erative, les
domaines qui sont instables vis \`a vis d'un retournement global,
et d'obtenir ainsi, \`a la fin, la structure en domaines stables.

 \section{ Conclusion }

Le message essentiel de ce chapitre est donc que
  les approches de type Ma-Dasgupta
ont un sens chaque fois que les h\'et\'erog\'en\'eit\'es spatiales
du d\'esordre d\'eterminent l'\'etat dominant du syst\`eme, 
alors que les fluctuations
thermiques ou quantiques ne fournissent que des corrections sous-dominantes.
Leur but est alors de construire l'\'etat dominant du syst\`eme
pour chaque r\'ealisation du d\'esordre, g\'en\'eralement en fonction d'un param\`etre, qui
est le temps pour les mod\`eles dynamiques,
ou la temp\'erature pour les probl\`emes d'\'equilibre
thermodynamique.

Pour mettre en oeuvre ce programme, il faut maintenant pr\'eciser
les m\'ethodes de calculs utilis\'ees pour 
effectuer quantitativement la renormalisation
 sur le d\'esordre :
c'est l'objet du Chapitre qui suit.

\chapter{ Les r\`egles de renormalisation 
de type Ma-Dasgupta  } 

\label{chapreglesrg}

  \begin{flushright}
  \emph{ 
O\`u l'on pr\'esente les r\`egles quantitatives de renormalisation de type Ma-Dasgupta, \\
en insistant sur la classe d'universalit\'e des extrema Browniens. }\\
  ~\\
  \end{flushright}

Alors que les renormalisations pour les syst\`emes purs
portent sur un nombre fini de constantes de couplages,
les renormalisations pour les syst\`emes d\'esordonn\'es
font intervenir des distributions de probabilit\'e,
c'est \`a dire des fonctions qui appartiennent \`a un espace de dimension infinie, ce qui complique \'evidemment beaucoup l'analyse des flots de renormalisation
et la recherche de points fixes.
Cette difficult\'e conduit donc le plus souvent \`a des \'etudes
compl\`etement num\'eriques,
ou alors, sur le plan analytique, \`a des approximations
suppl\'ementaires qui consistent \`a projeter sur des espaces de dimension
finie, c'est \`a dire \`a choisir une certaine forme analytique
de distributions contenant quelques param\`etres dont on \'etudie
les \'evolutions.
Dans ce chapitre, nous allons voir que les r\`egles de renormalisation
de type Ma-Dasgupta engendrent, dans un certain nombre de cas
favorables, des flots de renormalisation assez simples
pour permettre un calcul exact de distributions de points fixes,
qui ont souvent une interpr\'etation probabiliste int\'eressante.

\section{ Qu'est-ce qu'une r\`egle de renormalisation de type Ma-Dasgupta ?  }

Si l'on souhaite  
englober les diff\'erentes renormalisations de type Ma-Dasgupta
qui ont \'et\'e d\'evelopp\'es dans les diff\'erents contextes physiques,
 on peut prendre comme d\'efinition g\'en\'erale 
la structure suivante :
dans un syst\`eme d\'ecrit par une collection de variables al\'eatoires
$\{z_i\}$ qui ont des relations donn\'ees de voisinage dans l'espace physique,
la renormalisation consiste \`a d\'ecimer 
de mani\`ere it\'erative la plus petite valeur $min \{z_i\}$,
avec \'eventuellement ses variables al\'eatoires voisines,
afin de reconstruire une nouvelle collection de variables al\'eatoires
$\{z_i'\}$ reli\'ees par de nouvelles relations de voisinage dans l'espace physique.

Au del\`a de cette d\'efinition g\'en\'erale un peu trop formelle,
les deux caract\'eristiques essentielles sont les suivantes :

$\bullet$ la renormalisation s'effectue sur la valeur extr\^eme 
d'une variable al\'eatoire.  Cette valeur extr\^eme qui \'evolue
par renormalisation constitue l'\'echelle de renormalisation :
c'est le ``cut-off" de la distribution renormalis\'ee.

$\bullet$ la renormalisation est locale dans l'espace : 
\`a chaque \'etape, il n'y a que le voisinage
physique imm\'ediat de la variable al\'eatoire extr\^eme qui est concern\'e
par la renormalisation.

Nous allons maintenant d\'ecrire plus pr\'ecis\'ement les r\`egles
qui seront utiles dans les diff\'erents syst\`emes d\'esordonn\'es
unidimensionnels discut\'es dans ce m\'emoire.
(Quelques exemples en dimension sup\'erieure seront \'evoqu\'es 
dans le Chapitre \ref{chaptrap} et dans l'Annexe.)

\section{ La classe d'universalit\'e des extrema Browniens  }

Dans les syst\`emes unidimensionnels avec d\'esordre gel\'e local,
les approches de type Ma-Dasgupta conduisent souvent \`a une m\^eme classe
d'universalit\'e qui a une interpr\'etation probabiliste simple
en termes d'extrema d'un potentiel Brownien \`a grande \'echelle.

\subsection{ Les deux exemples quantiques fondamentaux   }

\subsubsection{ La cha\^{\i}ne antiferromagn\'etique al\'eatoire }

La r\`egle de d\'ecimation propos\'ee par Ma-Dasgupta-Hu \cite{madasgupta}
pour une cha\^{\i}ne de spin $S=1/2$ caract\'eris\'ee par 
une suite $\{J_i\}$ de couplages al\'eatoires positifs ind\'ependants
est la suivante :
on cherche le couplage maximum $\Omega$ dans toute la cha\^{\i}ne,
supposons pour fixer les notations que ce soit le couplage $J_2=\Omega$.
On \'elimine ce couplage maximum et ses deux voisins,
c'est \`a dire les trois couplages cons\'ecutifs $(J_1,J_2,J_3)$,
 et on les remplace par un nouveau couplage effectif
\begin{eqnarray}
J'=\frac{J_1 J_3}{2 \Omega} 
\label{rules1sur2}
\end{eqnarray}
La signification quantique de cette r\`egle sera discut\'ee dans
le Chapitre \ref{chapquantique1}, car ici notre but est de discuter
uniquement l'aspect probabiliste de cette r\`egle.
Comme le nouveau couplage $J'$ introduit est statistiquement ind\'ependant
de tous les autres couplages qui restent dans la cha\^{\i}ne,
cette r\`egle de d\'ecimation engendre un 
{ flot de renormalisation ferm\'e 
pour la distribution } $P_{\Omega}(J)$ des couplages $J$
qui sont pr\'esents \`a l'\'echelle $\Omega$ de renormalisation,
o\`u $\Omega$ repr\'esente le couplage maximum qui \'evolue au cours
de la renormalisation
\begin{eqnarray} 
-{{\partial P(J,\Omega)} \over {\partial
\Omega}}=P(\Omega,\Omega) \int_0^{\Omega} dJ_a \int_0^{\Omega} dJ_b \
P(J_a,\Omega) \ P(J_b,\Omega) \ \delta\left(J-{{J_a J_b} \over
{2\Omega}} \right) 
\end{eqnarray}

\subsubsection{ La cha\^{\i}ne d'Ising avec couplages et champs transverses al\'eatoires 
(RTFIC en anglais)  }

Dans le mod\`ele RTFIC, il y une alternance le long de la cha\^{\i}ne
de couplages al\'eatoires $J_i$
et de champs al\'eatoires $h_i$, qui sont des variables al\'eatoires positives
ind\'ependantes, tir\'ees a priori avec deux distributions diff\'erentes.
La physique du mod\`ele sera discut\'ee dans le chapitre \ref{chapquantique2}.
Ici, notre but est simplement de d\'ecrire les r\`egles de d\'ecimation
\cite{danielrtfic} :
on choisit \`a chaque \'etape le maximum des couplages et des champs
$ \Omega=max \{ J_i,h_j \}$.
Si c'est le champ $h_2=\Omega$, on l'\'elimine avec ses deux couplages voisins $(J_2,J_3)$
pour former un nouveau couplage effectif  
\begin{eqnarray}
J'=\frac{J_2 J_3}{ \Omega}
\label{rulertfic1} 
\end{eqnarray}
Si c'est le couplage $J_2=\Omega$, on l'\'elimine avec ses deux champs voisins $(h_1,h_2)$
pour former un nouveau champ effectif  
\begin{eqnarray}
h'=\frac{h_1 h_2}{ \Omega} 
\label{rulertfic2} 
\end{eqnarray}
Ici encore, les nouveaux couplages introduits sont statistiquement 
ind\'ependants de tous les autres couplages qui restent dans la cha\^{\i}ne,
ce qui permet d'\'ecrire un syst\`eme ferm\'e de 
deux flots de renormalisation  
pour les distributions $P_{\Omega}(J)$ et $P_{\Omega}(h)$
des couplages et champs effectifs \`a l'\'echelle $\Omega$.

Nous allons maintenant expliquer comment ces r\`egles qui apparaissent 
dans les cha\^{\i}nes de spins quantiques peuvent s'interpr\'eter
en termes de statistique des extrema d'une marche al\'eatoire
\`a grande \'echelle.

\subsection{ Extrema des marches al\'eatoires 
\`a grande \'echelle  }

Dans plusieurs mod\`eles de physique statistique que nous allons consid\'erer, 
il existe un seul type de variables al\'eatoires $(f_i)$ sur une ligne, mais
ces variables al\'eatoires peuvent \^etre positives ou n\'egatives.
Ces variables al\'eatoires repr\'esentent par exemple
les forces locales dans le mod\`ele de Sinai (Chapitres \ref{chapsinai}
et \ref{chapsinaibiais}), les champs al\'eatoires dans la
cha\^{\i}ne d'Ising classique en champ al\'eatoire (Chapitre \ref{chaprfim}), les charges
dans le mod\`ele du polym\`ere hydrophile/hydrophobe (Chapitre \ref{chappolymere}).
Dans les approches de type Ma-Dasgupta
pour ces diff\'erents mod\`eles, les degr\'es de libert\'e
importants du d\'esordre sont les extrema du potentiel associ\'e
$U(i)=\sum_{j=0}^i f_j$ \`a grande \'echelle.

\begin{figure}

\centerline{\includegraphics[height=8cm]{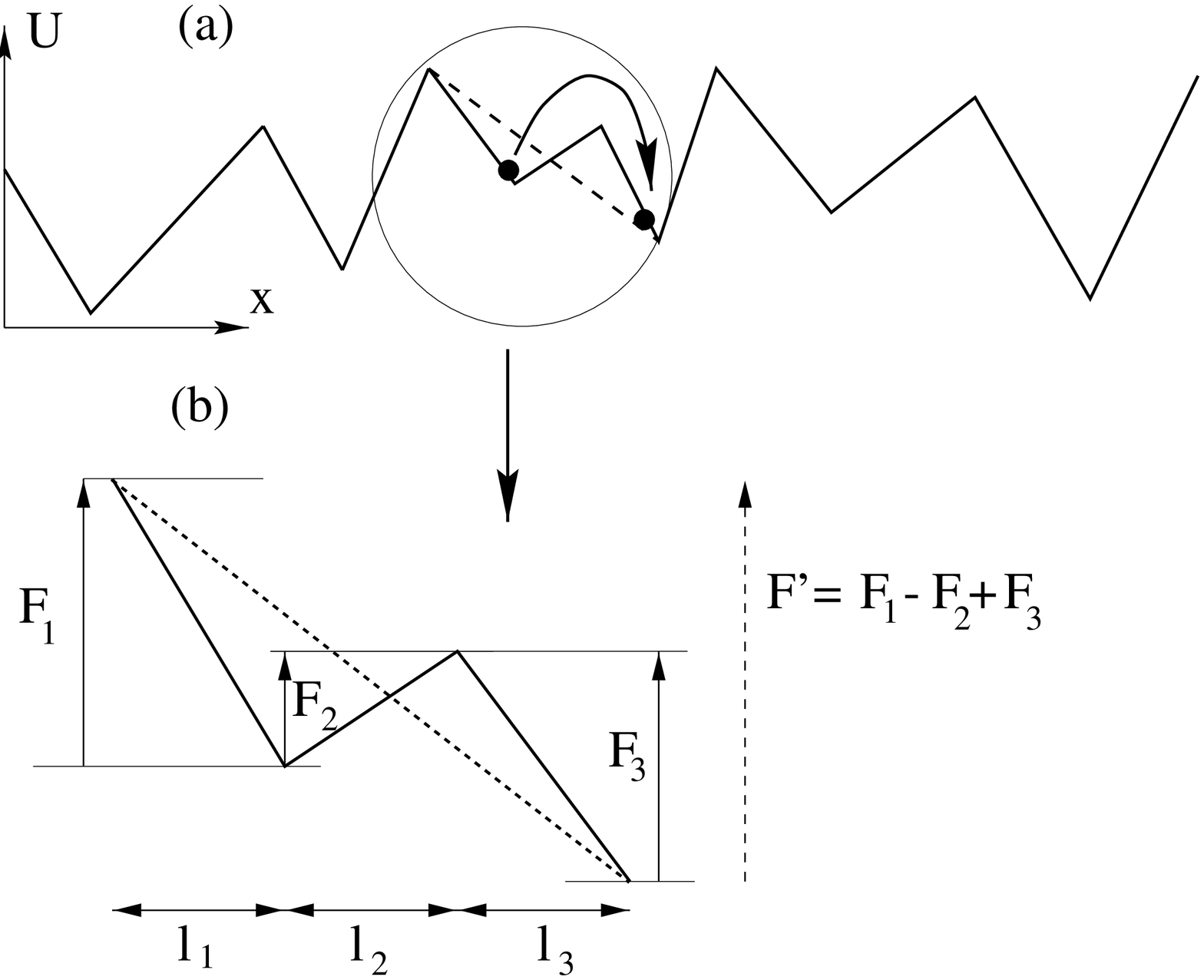}} 
\caption{\it Illustration des r\`egles de renormalisation pour un potentiel al\'eatoire unidimensionnel. } 
\label{figrulerg}
\end{figure}

Pour d\'efinir les extrema du potentiel initial, il faut donc
commencer par grouper ensemble toutes les forces $f_i$
 cons\'ecutives de m\^eme signe :
on obtient alors un paysage constitu\'e par une alternance
de liens descendants $(F_i^+,l_i^+)$ et de liens montants $(F_i^-,l_i^-)$.
Les barri\`eres $F$ et les longueurs $l$
sont maintenant des variables al\'eatoires positives
qui repr\'esentent respectivement les diff\'erences de potentiel 
et les distances entre deux extrema locaux cons\'ecutifs du mod\`ele initial.
La r\`egle de renormalisation du paysage est alors la suivante 
(Publications [P3,P4]):
on choisit la plus petite barri\`ere $\Gamma=min \{ F_i^+,F_i^- \}$.
Si la plus petite barri\`ere est un lien descendant $F_2^+=\Gamma$,
on l'\'elimine avec ces deux liens montants voisins $F_1^-$ et $F_2^-$
pour former un nouveau lien montant
\begin{eqnarray}
(F^-)' =F_1^- + F_2^- -\Gamma
\end{eqnarray}
Si la plus petite barri\`ere est un lien montant $F_1^-=\Gamma$,
on l'\'elimine avec ces deux liens descendants voisins $F_1^+$ et $F_2^+$
pour former un nouveau lien descendant
\begin{eqnarray}
(F^+)'=F_1^+ + F_2^+ -\Gamma
\end{eqnarray}
Ces r\`egles sont donc compl\`etement 
\'equivalentes aux r\`egles de d\'ecimations 
(\ref{rulertfic1} \ref{rulertfic2}) du mod\`ele RTFIC
par une simple transformation logarithmique sur les variables.
Si l'on souhaite garder une information sur
les distances initiales dans l'espace physique, il suffit
d'\'ecrire la r\`egle d'\'evolution de la longueur 
lors d'une d\'ecimation : la longueur du nouveau lien renormalis\'e
est simplement \'egale \`a la somme  
des trois liens \'elimin\'es
\begin{eqnarray}
l'=l_1+l_2+l_3
\label{rulelongueur}
\end{eqnarray}
Cette nouvelle longueur a encore la propri\'et\'e
d'\^etre ind\'ependante statistiquement des autres longueurs
 qui restent dans la syst\`eme. 
En revanche, la longueur va \^etre corr\'el\'ee avec
la barri\`ere $F$ qui existe sur le m\^eme lien.
Ainsi, on peur \'ecrire un syst\`me ferm\'e
de deux \'equations de flots
pour les deux lois jointes $P^{\pm}(F,l)$.

\subsection{ Distribution des extrema du potentiel Brownien pur  }

Le cas sym\'etrique $\overline{f_i}=0$ des mod\`eles statistiques
correspond au point critique quantique du mod\`ele
RTFIC $ \overline{\ln J}=\overline{ \ln h}$
o\`u l'aimantation spontan\'ee s'annule.
A grande \'echelle, les liens descendants et les liens montants 
deviennent statistiquement \'equivalents, et le seul param\`etre physique
pertinent est la variance $\overline{f_i^2}=2 \sigma$ du mod\`ele initial,
comme dans le th\'eor\`eme de la Limite Centrale.
Il existe alors un point fixe de ``d\'esordre
infini" caract\'eris\'e 
par une distribution jointe $P^*(\eta,\lambda)$
des variables d'\'echelle $\eta= \frac{F-\Gamma}{\Gamma}$ pour les barri\`eres 
et $\lambda=\frac{ \sigma l}{\Gamma^2}$ pour les longueurs.
Cette loi jointe qui caract\'erise le point fixe,
 est d\'efinie par sa transform\'ee de Laplace 
par rapport \`a la longueur $\lambda$ \cite{danielrtfic}
\begin{eqnarray}
\int_0^{+\infty} d\lambda e^{-p \lambda}
  {\cal P}^*(\eta,\lambda) = \theta( \eta>0) 
\frac{ {\sqrt p} }{ \sinh {\sqrt p} } e^{- \eta {\sqrt p} \coth {\sqrt p}}
\label{ptfixeetalambda}
\end{eqnarray}
En particulier, la distribution des barri\`eres seules est 
une simple exponentielle
\begin{eqnarray}
{\cal P}^*(\eta) = \theta( \eta>0) e^{- \eta}
\label{ptfixeeta}
\end{eqnarray}
alors que la distribution des longueurs seules
a la forme d'une s\'erie infinie d'exponentielles
\begin{eqnarray}
{\cal P}^*(\lambda) = LT^{-1}_{p \to \lambda} 
\left( \frac{ 1 }{ \cosh {\sqrt p} } \right)
&& = \sum_{n=-\infty}^{+\infty} \pi (-1)^n \left( n+\frac{1}{2} \right)
 e^{- \pi^2 \left( n+\frac{1}{2} \right)^2 \lambda}  \\
&& = \frac{1}{{\sqrt \pi} \lambda^{3/2}}
\sum_{m=-\infty}^{+\infty}  (-1)^m \left( m+\frac{1}{2} \right)
 e^{-  \left( m+\frac{1}{2} \right)^2 \frac{1}{\lambda} } \nonumber
\label{ptfixelambda}
\end{eqnarray}
(les deux s\'eries se correspondent par une formule d'inversion de Poisson).

La convergence vers la solution de point fixe (\ref{ptfixeetalambda})
est d'ordre $1/\Gamma$ pour les marches al\'eatoires \cite{danielrtfic}.
En revanche, pour le mouvement Brownien unidimensionnel
qui repr\'esente d\'ej\`a la limite continue universelle des marches al\'eatoires, le point fixe (\ref{ptfixeetalambda}) est un r\'esultat exact
\`a toute \'echelle $\Gamma$, comme le montre un calcul direct
par des int\'egrales de chemin contraintes (Publication [P9]).

\subsection{ Distribution des extrema
du potentiel Brownien biais\'e  }

Dans le cas dissym\'etrique o\`u il existe un biais $\overline{f_i}=f_0>0$,
les \'equations de renormalisation conduisent
\`a une famille de solutions \`a un param\`etre not\'e $ \delta$ 
\cite{danielrtfic} : les deux distributions jointes
$P_{\Gamma}^{\pm}(F,l)$ des liens descendants et montants
ont pour transform\'ees de Laplace
\begin{eqnarray}
\int_0^{+\infty} dl e^{-p l}
 P_{\Gamma}^{*\pm}(F,l)  = \theta( F >\Gamma) 
\frac{ {\sqrt {p+\delta^2} } e^{\mp \delta \Gamma } }
{ \sinh {\sqrt {p+\delta^2}} } e^{- (F-\Gamma)
\left[ {\sqrt {p+\delta^2}} \coth {\sqrt {p+\delta^2}} \mp \delta  \right] }
\label{ptfixeetalambdadelta}
\end{eqnarray}
En particulier, les distributions de barri\`eres seules ont les formes 
exponentielles suivantes
\begin{eqnarray}
 P_{\Gamma}^{*+}(F) &&  = \theta( F >\Gamma) 
\frac{ 2 \delta } { e^{2 \delta \Gamma } -1} 
 e^{- (F-\Gamma) \frac{ 2 \delta } { e^{2 \delta \Gamma } -1}  }
\opsimeq_{\Gamma \to \infty} 
\frac{ 2 \delta } { e^{2 \delta \Gamma } } 
 e^{- (F-\Gamma) \frac{ 2 \delta } { e^{2 \delta \Gamma }}  } \\
 P_{\Gamma}^{*-}(F) &&  = \theta( F >\Gamma) 
\frac{ 2 \delta } { 1-e^{-2 \delta \Gamma }} 
 e^{- (F-\Gamma) \frac{ 2 \delta } { 1-e^{-2 \delta \Gamma }}  }
\opsimeq_{\Gamma \to \infty} \theta( F >\Gamma) 
 2 \delta  
 e^{- (F-\Gamma)  2 \delta  }
\label{soluppdelta}
\end{eqnarray}
Du point de vue du paysage renormalis\'e, la signification
du param\`etre $2 \delta$ est donc claire :
la distribution $P_{\Gamma}^{*-}(F)$
des grandes barri\`eres oppos\'ees au biais $f_0$
reste stable \`a grande \'echelle (en dehors de la pr\'esence du cut-off
$\Gamma$)
et $(2 \delta)$ est le coefficient de la d\'ecroissance asymptotique
exponentielle de cette distribution.
Il faut maintenant pr\'eciser la signification du param\`etre
$\delta$ par rapport au mod\`ele microscopique de d\'epart. 

Pr\`es du point critique correspondant \`a $ \delta=0$,
le param\`etre $\delta$ peut \^etre d\'evelopp\'e au premier ordre
dans le biais \cite{danielrtfic}
\begin{eqnarray}
\delta = \frac{ f_0}{\sigma} +O(f_0^2)
\label{smalldelta}
\end{eqnarray}

Si l'on s'\'ecarte du voisinage imm\'ediat du point critique, la bonne d\'efinition non-perturbative du param\`etre $ \delta$
en terme de la distribution initiale $Q(f)$ des variables $(f_i)$
du mod\`ele microscopique de d\'epart
est que $\delta$ repr\'esente la solution de l'\'equation 
(Publication [P4])
\begin{eqnarray}
\overline{ e^{- 2 \delta f_i} } 
\equiv \int_{-\infty}^{+\infty} df Q(f) e^{- 2 \delta f} =1
\label{bonnedefdelta}
\end{eqnarray}
Cette d\'efinition correspond bien s\^ur exactement,
au changement de notation pr\`es $2 \delta=\mu/T$, \`a la d\'efinition
du param\`etre sans dimension $\mu$ qui repr\'esente l'exposant de la phase
de diffusion anormale $x \sim t^{\mu}$ des mod\`eles de Sinai avec biais 
\cite{kestenetal,derridapomeau,jpbreview}, dont nous reparlerons dans
le chapitre \ref{chapsinaibiais}.
Le d\'eveloppement dans les deux premiers cumulants redonne bien
l'expression simple (\ref{smalldelta}), qui n'est exacte \`a tous les ordres
que pour une distribution gaussienne. 
 
A nouveau, si l'on consid\`ere directement 
le mouvement Brownien unidimensionnel biais\'e
comme limite continue des marches al\'eatoires biais\'ees, 
les solutions (\ref{ptfixeetalambdadelta}) sont des r\'esultats exacts
\`a toute \'echelle $\Gamma$, comme le montre un calcul direct
par des int\'egrales de chemin contraintes (Publication [P9]).

Alors que la solution (\ref{ptfixeetalambda}) du cas sym\'etrique $\delta=0$ 
est un point fixe de ``d\'esordre infini",
la solution (\ref{soluppdelta}) du cas dissym\'etrique  $\delta >0$
est un point fixe de ``d\'esordre fini'', car la distribution
$P_{\Gamma}^{*-}(F)$ des grandes barri\`eres oppos\'ees au biais 
a asymptotiquement une largeur finie $\frac{1}{2 \delta}$. 
La renormalisation usuelle de type Ma-Dasgupta ne donne donc des r\'esultats
exacts que dans la limite $\delta \to 0$. Nous discuterons en d\'etails
dans le Chapitre \ref{chapsinaibiais} 
comment g\'en\'eraliser la m\'ethode usuelle lorsque le param\`etre $\delta$ est petit.

\section{ Renormalisation 
d'un potentiel al\'eatoire unidimensionnel quelconque }

Comme la renormalisation de type Ma-Dasgupta consiste \`a \'etudier
les extrema d'un potentiel unidimensionnel
lorsqu'on ne garde que les barri\`eres sup\'erieures
 \`a une certaine \'echelle $\Gamma$,
il est possible de d\'efinir la proc\'edure pour 
un paysage al\'eatoire quelconque.
En particulier, la proc\'edure de renormalisation peut \^etre
impl\'ement\'ee num\'eriquement pour des potentiels corr\'el\'es,
ce qui a \'et\'e fait pour le cas de corr\'elations logarithmiques
\cite{castillo}.
La Publication [P9] \'etudie 
ce que l'on peut dire de mani\`ere analytique
pour le cas des potentiels Markoviens. Nous d\'ecrirons en particulier
la solution explicite pour le cas d'un potentiel Brownien en pr\'esence
d'un potentiel d\'eterministe harmonique
dans le Chapitre \ref{chaplandscape}.

\section{ R\`egles de d\'ecimation plus g\'en\'erales }

Dans certains syst\`emes unidimensionnels, en particulier
pour la cha\^{\i}ne quantique de spin $S=1$ discut\'ee
dans le Chapitre \ref{chapquantique1}, on est amen\'e \`a d\'efinir
des proc\'edures de renormalisation plus compliqu\'ees
que la renormalisation d'un potentiel unidimensionnel, ce qui peut donner
lieu \`a des ph\'enom\`enes plus riches : en particulier,
pour la cha\^{\i}ne quantique de spin $S=1$, 
cela permet de d\'ecrire une transition de phase de type percolation
\`a temp\'erature nulle.

\section{ Variables auxiliaires et exposants critiques }

Dans les renormalisations de type Ma-Dasgupta,
on appelle `variables auxiliaires' les variables associ\'ees aux liens
qui vont \'evoluer selon des r\`egles de d\'ecimation parall\`eles 
lorsque leurs variables principales associ\'ees sont renormalis\'ees.
Le premier exemple important est la longueur $l$ 
que nous avons d\'ej\`a rencontr\'ee (\ref{rulelongueur}).
C'est une variable auxiliaire car ce n'est pas elle qui d\'etermine la renormalisation, mais la barri\`ere associ\'ee $F$ :
en effet \`a chaque \'etape, on ne choisit pas la longueur minimale,
on choisit la barri\`ere $F$ minimale.

Une autre variable auxiliaire importante pour le RTFIC
est l'aimantation $m$ des amas de spins :
cette variable n'existe qu'en association avec les champs al\'eatoires $h_i$
 et \'evolue en association avec la r\`egle (\ref{rulertfic2}) selon
\cite{danielrtfic}
\begin{eqnarray}
m'=m_1+m_2 
\label{rulertficm} 
\end{eqnarray}
Plus g\'en\'eralement, dans les diff\'erents mod\`eles discut\'es dans ce m\'emoire, que ce soit pour la diffusion de Sinai 
(Chapitre \ref{chapsinai}),
pour la cha\^{\i}ne d'Ising classique en champ al\'eatoire 
(Chapitre \ref{chaprfim}),
et pour les probl\`emes de r\'eaction-diffusion (Publication [P5]),
 on est 
conduit \`a \'etudier diverses variables auxiliaires 
qui \'evoluent selon la r\`egle g\'en\'erale
\begin{eqnarray}
m'=a m_1+b m_2+c m_3
\label{rulegeneauxi}
\end{eqnarray}
o\`u $(a,b,c)$ sont des constantes.

Nous avons d\'ej\`a vu comment 
l'\'etude jointe des barri\`eres et des longueurs 
caract\'erisent la statique du
 paysage renormalis\'e \`a une \'echelle donn\'ee  (\ref{ptfixeetalambda}) :
en particulier, dans le paysage renormalis\'e \`a l'\'echelle $\Gamma$,
dans lequel il ne reste que des barri\`eres $F>\Gamma$, les longueurs
ont pour scaling $l \sim \Gamma^2$, ce qui est le scaling Brownien
usuel comme il se doit. 
En revanche, en dehors de ce cas tr\`es particulier $a=b=c=1$,
les variables auxiliaires (\ref{rulegeneauxi})
conduisent \`a des exposants non-triviaux $m \sim \Gamma^{\Phi}$ 
qui refl\`etent des propri\'et\'es `dynamiques' de la renormalisation
sur toutes les \'echelles pr\'ec\'edentes $\Gamma'<\Gamma$, et qui
contiennent plus d'informations que le paysage \`a l'\'echelle $\Gamma$
seulement. Par exemple pour le RTFIC, la r\`egle 
tr\`es simple (\ref{rulertficm})
correspondant \`a $a=c=1$ et $b=0$
conduit pour l'aimantation $m \sim \Gamma^{\phi}$ \`a l'exposant
irrationnel \'egal au nombre d'or \cite{danielrtfic}
\begin{eqnarray}
\phi(a=c=1,b=0)=\frac{1+\sqrt 5}{2}
\end{eqnarray}
Cet exemple montre qu'une simple marche al\'eatoire unidimensionnelle
contient des exposants non-triviaux si l'on s'int\'eresse \`a des propri\'et\'es
un peu `subtiles' qui concernent la dynamique du paysage des extrema
en fonction de l'\'echelle.

Plus g\'en\'eralement, pour une variable auxiliaire de type (\ref{rulegeneauxi}), 
 on obtient que, lorsque la condition $a+c=2$ est satisfaite 
(ce qui arrive assez souvent en pratique pour les observables physiques
les plus naturelles), l'exposant associ\'e 
 satisfait une \'equation du second degr\'e et vaut
\begin{eqnarray}
\phi(a+c=2,b)=\frac{1+\sqrt {5+4 b} }{2}
\end{eqnarray}
En revanche, lorsque $a+c \neq 2$, l'exposant $\phi(a,b,c)$ satisfait une
\'equation plus compliqu\'ee faisant intervenir la fonction hyperg\'eom\'etrique
confluente $U(A,B,z)$ (Publications [P4] et [P5]).

\section{ Comparaison avec certains mod\`eles de croissance }

Les d\'ecimations de type Ma-Dasgupta ont une grande parent\'e
avec certains mod\`eles de croissance, qui ont \'et\'e introduits de
mani\`ere compl\`etement ind\'ependante.
Ces mod\`eles g\'eom\'etriques de croissance 
 consid\`erent une suite d'intervalles sur une ligne
qui \'evolue par une transformation it\'erative sur
le plus petit segment qui reste, avec les diverses 
r\`egles suivantes : 

(i) dans le ``cut-in-two model" \cite{yekutieli},
l'intervalle le plus petit est \'elimin\'e et donne
une moiti\'e de sa longueur \`a chacun de ces deux voisins.
Ce mod\`ele introduit donc des corr\'elations entre les intervalles voisins
et a \'et\'e \'etudi\'e num\'eriquement \cite{yekutieli}.

(ii) dans le ``paste-all model" \cite{yekutieli},
l'intervalle le plus petit est \'elimin\'e et donne
toute sa longueur \`a l'un de ces deux voisins 
tir\'e au hasard de mani\`ere \'equiprobable.
Ce mod\`ele n'introduit pas de corr\'elations entre les intervalles voisins,
et la distribution invariante des longueurs a \'et\'e calcul\'ee \cite{yekutieli}.

(iii) dans le ``instantaneous collapse model" \cite{rutenberg},
 l'intervalle le plus petit est \'elimin\'e avec ses
deux voisins pour former un seul nouveau domaine $l'=l_1+l_2+l_3$.
Ce mod\`ele d\'ecrit en fait la dynamique effective \`a grand temps
d'un champs scalaire unidimensionnel qui \'evolue
selon une \'equation de Ginzburg-Landau \`a temp\'erature nulle
\cite{nagai,rutenberg}, et   
a donc suscit\'e un grand int\'er\^et
en tant que mod\`ele soluble de croissance.
Les r\'esultats exacts concernent 
la distribution invariante des longueurs \cite{rutenberg}),
l'exposant de persistance \cite{phi4persist}
qui caract\'erise la variable auxiliaire $d'=d_1+d_3$,
l'exposant d'autocorr\'elation \cite{phi4autocorre}
qui caract\'erise la variable auxiliaire $q'=q_1-q_2+q_3$
et enfin l'exposant de persistance g\'en\'eralis\'e \cite{majbray}
qui caract\'erise la variable auxiliaire $m'=m_1+p m_2+ m_3$
avec un param\`etre $p$.

La seule diff\'erence `technique' avec les r\`egles de d\'ecimation
de type Ma-Dasgupta est que, dans les mod\`eles de croissance,
c'est la longueur qui est la variable principale qui 
d\'etermine la renormalisation,
alors que dans les mod\`eles d\'esordonn\'es,
la longueur n'est qu'une variable auxiliaire, la variable principale
qui d\'efinit la dynamique \'etant une variable de d\'esordre.
Cette diff\'erence explique les diff\'erences analytiques
entre les solutions de points fixes et d'exposants 
dans les deux types de mod\`eles.

D'un point de vue physique, cet exemple montre qu'une dynamique 
sans d\'esordre intrins\`eque, d\'efinie \`a partir d'une condition initiale al\'eatoire,
peut \^etre gouvern\'ee, d'une certaine mani\`ere, par un `point fixe de d\'esordre infini'.

\section{ Conclusion }

La sp\'ecificit\'e des r\`egles de renormalisation de type Ma-Dasgupta 
est de d\'ecimer localement une variable extr\^eme de d\'esordre
de mani\`ere it\'erative. Cette structure tr\`es particuli\`ere
permet souvent d'obtenir en dimension $d=1$ des solutions explicites 
pour les distributions de probabilit\'e du paysage et
pour les exposants critiques associ\'ees aux variables auxiliaires
repr\'esentant des observables physiques int\'eressantes.

Ce Chapitre 2 termine la pr\'esentation g\'en\'erale des m\'ethodes
de type Ma-Dasgupta. Le reste du m\'emoire est consacr\'e 
\`a l'application de ces m\'ethodes \`a divers mod\`eles d\'esordonn\'es
de physique statistique ou de mati\`ere condens\'ee.

%%%%%%%%%%%%%%%%%%%%%%%%%%%%%%%%%%%%%%%%%%%%%%%%%%%%%%%%%%%%%%%%%

\chapter
{Marche al\'eatoire dans un potentiel Brownien}

\label{chapsinai}

 % \begin{flushright}
%  \emph{  
% }\\  ~\\
 % \end{flushright}

\section{ Pr\'esentation du mod\`ele }

Le mod\`ele d'une marche al\'eatoire dans un potentiel al\'eatoire
Brownien,
qui porte aujourd'hui le nom de ``mod\`ele de Sinai",
a beaucoup int\'eress\'e les math\'ematiciens probabilistes
depuis les travaux de Solomon \cite{solomon}
Kesten {\it et al.} \cite{kestenetal} et Sinai \cite{sinai},
ainsi que les physiciens des syst\`emes d\'esordonn\'es
depuis les travaux initiaux de 
Alexander {\it et al.}\cite{alexander}
 et Derrida-Pomeau \cite{derridapomeau}.
Depuis, il y a eu de nombreux d\'eveloppements dans les deux communaut\'es :
la revue r\'ecente \cite{zhanshi} contient une pr\'esentation
des r\'ef\'erences importantes pour les math\'ematiciens probabilistes,
alors que les diff\'erentes approches des physiciens
sont pr\'esent\'ees dans les revues \cite{haus,havlin,jpbreview}.

Pour les physiciens, l'int\'er\^et du mod\`ele de Sinai est double :
d'une part, la marche de Sinai
 repr\'esente un mod\`ele dynamique simple en pr\'esence de d\'esordre,
sur lequel on peut tester un certain nombre de concepts 
g\'en\'eraux,
et d'autre part, elle appara\^{\i}t naturellement dans
divers contextes, par exemple dans la dynamique d'une paroi de domaine
dans la cha\^{\i}ne d'Ising en champs al\'eatoires (cf Chapitre \ref{chaprfim})
ou dans
la dynamique d'ouverture de la double h\'elice d'ADN en pr\'esence
d'une force ext\'erieure \cite{lubenskinelson}.

\subsection{ Version continue :
 Diffusion dans un potentiel Brownien }

La version continue du mod\`ele de Sinai correspond
\`a l'\'equation de Langevin \cite{jpbreview}
\begin{eqnarray}
\frac{dx}{dt} =    -  U'(x(t))  + \eta(t) 
\label{langevin}
\end{eqnarray}
dans laquelle $\eta(t)$ repr\'esente le bruit thermique usuel
\begin{eqnarray}
<\eta(t) \eta(t') >  = 2 T \delta(t-t') 
\end{eqnarray}
et $U(x)$ est un potentiel al\'eatoire Brownien
\begin{eqnarray}
\overline {(U(x)-U(y))^2 } = 2 \sigma \vert x-y \vert 
\label{defsigma}
\end{eqnarray}
Plus g\'en\'eralement, dans tout le m\'emoire, la moyenne thermique
d'une observable $f$
sera not\'ee $<f>$, et la moyenne sur le d\'esordre 
d'une observable $f$ sera not\'ee $\overline{f}$.

\subsection{ Version discr\`ete :
 Marche al\'eatoire sur r\'eseau 1D }

\begin{figure}[ht]
\centerline{\includegraphics[height=2cm]{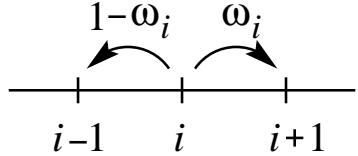}} 
\caption{\it Version discr\`ete du mod\`ele de Sinai   } 
\label{figdefsinai}
\end{figure}

Dans la version discr\`ete sur r\'eseau (Figure \ref{figdefsinai}), 
la particule qui se trouve sur le site $i$ a une probabilit\'e
$\omega_i$ de sauter vers la droite et une probabilit\'e $(1-\omega_i)$
de sauter vers la gauche. Les $\omega_i$
sont des variables al\'eatoires ind\'ependantes dans $]0,1[$.
La marche al\'eatoire est r\'ecurrente seulement si $\overline{ \ln \omega_i}
= \overline{ \ln (1-\omega_i)}$, ce qui constitue le cas de Sinai,
et qui correspond \`a l'absence de biais du potentiel $U(x)$ (\ref{defsigma})
de la version continue.

\section{ Principe de l'approche de renormalisation}

Nous avons d\'ej\`a expliqu\'e dans 
le premier chapitre (Page \pageref{exemplesinai})
la mani\`ere de raisonner sur l'exemple du mod\`ele
de Sinai.

L'id\'ee essentielle de notre approche est de d\'ecomposer
le processus $x_{U,\eta}(t)$, repr\'esentant la position de la marche
al\'eatoire engendr\'ee par le bruit thermique $\eta(t)$
dans le potentiel al\'eatoire Brownien $U(x)$, selon
une somme de deux termes
\begin{eqnarray} 
x_{\{U,\eta\}}(t) = m_{\{U\}}(t) + y_{\{U,\eta\}}(t)
\end{eqnarray}
 
$ \bullet $ { \bf Le processus $m_{\{U\}}(t)$ porte le nom de 
  ``dynamique effective" } dans les publications
 et repr\'esente la position la plus probable
de la particule \`a l'instant $t$ :
elle correspond au meilleur minimum local du potentiel al\'eatoire $U(x)$
 que la particule a eu typiquement le temps d'atteindre depuis sa condition
initiale pendant l'intervalle de temps $t$.
Comme le franchissement d'une barri\`ere de potentiel $F$ n\'ecessite 
un temps d'Arrh\'enius d'ordre $t_F=\tau_0 e^{\beta F}$,
on peut \'etudier en d\'etail cette dynamique effective 
en utilisant une proc\'edure de renormalisation de type Ma-Dasgupta,
qui consiste \`a d\'ecimer de mani\`ere it\'erative les plus petites barri\`eres qui existent dans le syst\`eme. On associe alors au temps $t$
le paysage renormalis\'e dans lequel il ne reste plus que des barri\`eres
plus grandes que l'\'echelle de renormalisation
$\Gamma=T \ln t$, ce qui correspond \`a une \'echelle $\Gamma^2=(T \ln t)^2$
pour les longueurs. La position $m_{\{U\}}(t)$ correspond
alors au minimum de la vall\'ee renormalis\'ee \`a l'\'echelle
$\Gamma=T \ln t$ qui contient le point initial.

$ \bullet $ Le processus $y_{\{U,\eta\}}(t)$ repr\'esente l'\'ecart par rapport
\`a la dynamique effective. Dans la limite de temps infini, 
c'est un variable al\'eatoire qui reste finie,
ce qui constitue le ph\'enom\`ene de { \bf localisation de Golosov} :
toutes les particules qui diffusent dans le m\^eme \'echantillon
\`a partir du m\^eme point de d\'epart avec des bruits thermiques $\eta$
diff\'erents sont asymptotiquement concentr\'ees dans la m\^eme vall\'ee
de minimum $m_{\{U\}}(t)$. Plus pr\'ecis\'ement, si
on consid\`ere les premi\`eres corrections \`a grand temps, 
la probabilit\'e qu'une
particule ne soit pas dans la vall\'ee correspondant \`a
la dynamique effective $m_{\{U\}}(t)$ est d'ordre $1/ (\ln t)$,
auquel cas la particule se trouve \`a une distance d'ordre $(\ln t)^2$
de $m_{U}(t)$.
Ces \'ev\`enements sont donc rares (leur probabilit\'e tend
vers z\'ero \`a grand temps) mais ils dominent cependant certaines observables,
comme par exemple 
la largeur thermique $\Delta x^2(t) \sim \overline {< y^2(t) >}
\sim (\ln t)^3$ qui diverge.

\begin{figure}

\centerline{\includegraphics[height=8cm]{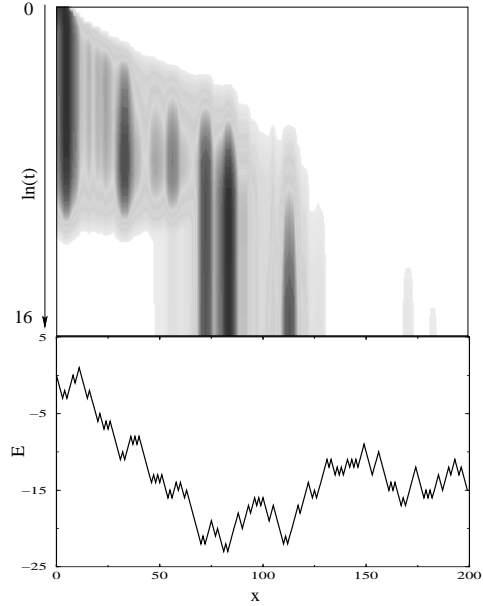}} 
\caption{\it Figure tir\'ee de la r\'ef\'erence \cite{chave} de J. Chave and E. Guitter, avec sa l\'egende ``Evolution with time (in logarithmic scale) of the distribution $P(x,t)$ in a given energy landscape (drawn below). 
The evolution runs over $10^7$ iterations. The intinsity in the grey scale is proportional to $(-\ln P(x,t))$,
i.e. darker regions correspond to higher values of $P(x,t)$."   } 
\label{figchave}
\end{figure}

\section{ \'Etude de la dynamique effective}

\subsection{Front de diffusion }  

A cause du ph\'enom\`ene de localisation de Golosov, la distribution 
de la variable d'\'echelle $X=\frac{x_{\{U,\eta\}}}{ (T \ln t)^2}$,
par rapport au bruit $\eta$ dans un \'echantillon fix\'e
est asymptotiquement une distribution delta de Dirac $\delta(X-M)$ 
o\`u $M=\frac{m_{\{U\}}(t)}{ (T \ln t)^2}$ est la variable d'\'echelle
de la dynamique effective. Si l'on souhaite maintenant
calculer la moyenne du front de diffusion moyen sur l'ensemble des \'echantillons
(ou l'ensemble des conditions initiales, ce qui revient ici au m\^eme),
il suffit d'\'etudier la distribution de $M$, et on 
retrouve ainsi tr\`es simplement la loi de Kesten 
\begin{eqnarray} 
P(X)= LT^{-1}_{p \to \vert X \vert } \left[ \frac{1}{p}
\left( 1-\frac{1}{\cosh \sqrt p} \right) \right] = \frac{4}{\pi} \sum_{n=0}^{+\infty} \frac{(-1)^n}{2n+1} e^{- \frac{\pi^2}{4} (2n+1) \vert X \vert }
\end{eqnarray}
qui est un r\'esultat exact
des math\'ematiciens probabilistes \cite{kestenlaw,golosovhitting}.
Cet exemple montre explicitement comment la proc\'edure de
renormalisation de type Ma-Dasgupta
permet d'obtenir des r\'esultats asymptotiques exacts,
et donne confiance dans les r\'esultats nouveaux qu'elle donne
pour des propri\'et\'es plus fines pour lesquelles il n'existe
pas de ``th\'eor\`eme" correspondant.

\begin{figure}

\centerline{\includegraphics[height=8cm]{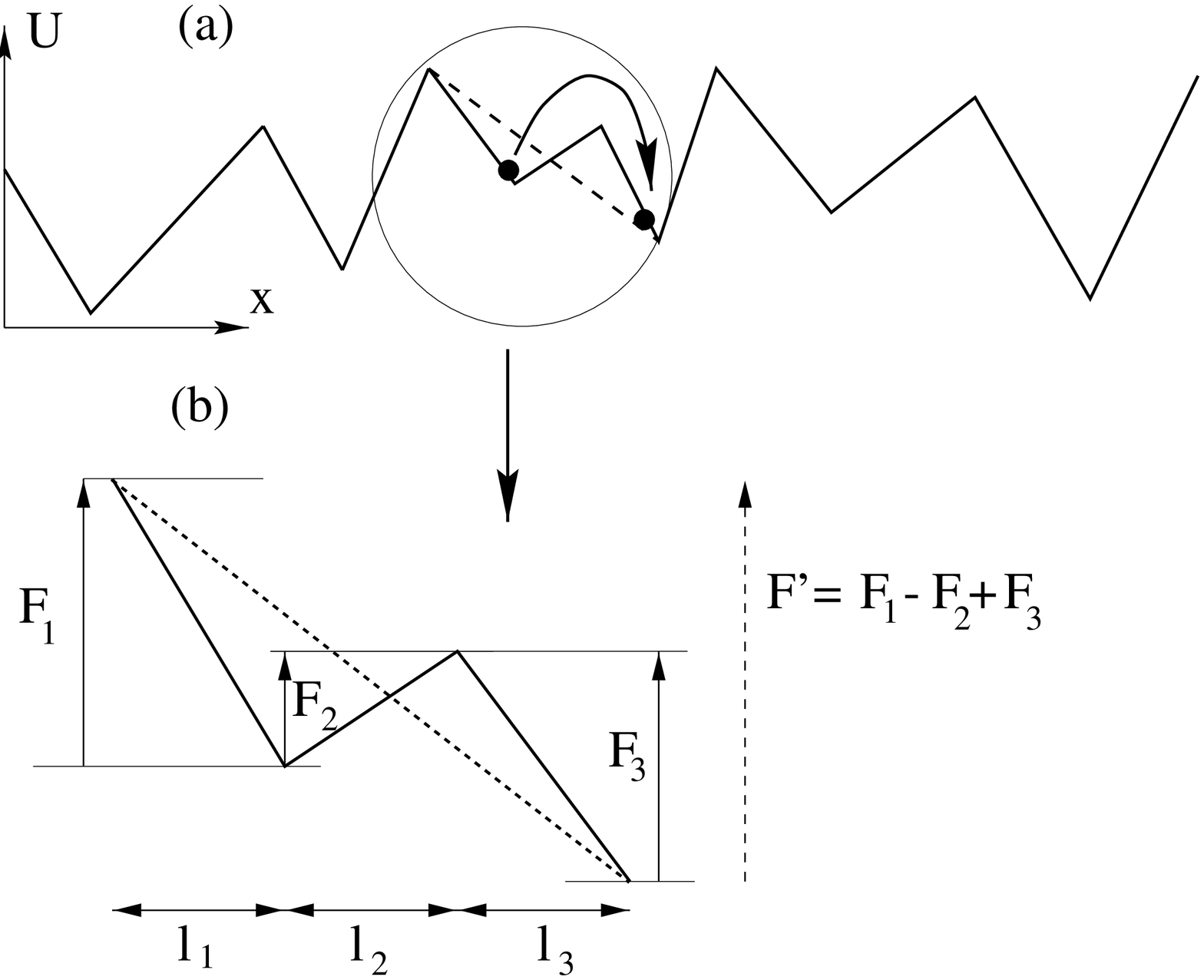}} 
\caption{\it Principe de la renormalisation pour \'etudier la
dynamique effective $m(t)$ } 
\label{figsinairg}
\end{figure}

\subsection{Distribution de l' \'energie }

De m\^eme, la variable d'\'echelle  pour l' \'energie 
\begin{eqnarray} 
w=\frac{U(x(0))-U(x(t))}{(T \ln t)} \simeq \frac{U(m(0))-U(m(t))}{(T \ln t)}
\nonumber
\end{eqnarray}
est enti\`erement d\'etermin\'ee \`a grand temps par la seule dynamique
effective. Cette variable r\'eduite 
a pour loi limite quand $t \to \infty$ :
\begin{eqnarray} 
 {\cal D}(w)= \theta(w < 1)  \left(4- 2 w-4 e^{-w} \right)
 +  \theta(w \geq 1)  \left( 2 e - 4 \right) e^{-w}
\nonumber
\end{eqnarray}

\begin{figure}

\centerline{\includegraphics[height=8cm]{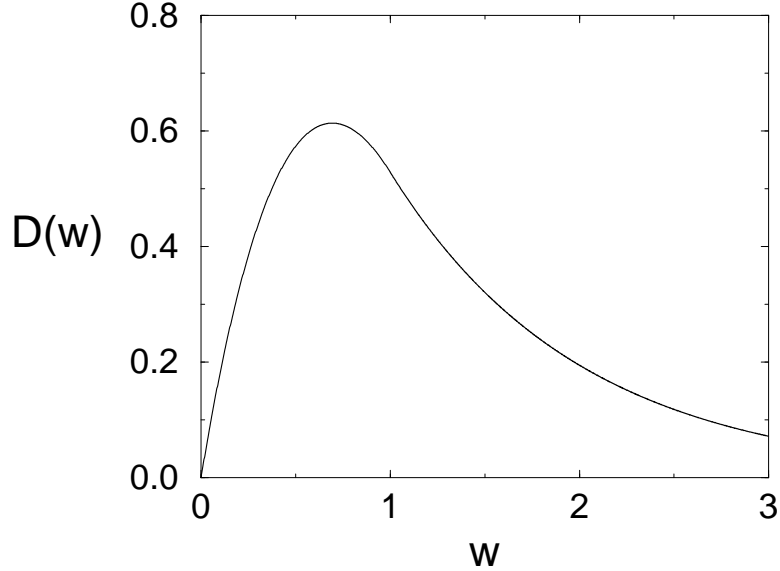}} 
\caption{\it Distribution limite de la variable d'\'echelle pour l' \'energie. } 
\label{figenergy}
\end{figure}

La loi est continue, ainsi que sa d\'eriv\'ee en $w=1$, mais la d\'eriv\'ee
seconde est discontinue en $w=1$, ce qui peut sembler surprenant!
En effet, pour tout temps fini,
la distribution de l' \'energie est analytique,
et ce n'est que dans la limite de temps infini que la loi limite
de la variable d'\'echelle pr\'esente une discontinuit\'e de la d\'eriv\'ee seconde.
Il est int\'eressant de noter que dans le travail r\'ecent
d'un math\'ematicien probabiliste \cite{hu}
sur le temps de retour \`a l'origine apr\`es l'instant $t$,
il appara\^{\i}t aussi une distribution asymptotique non-analytique 
pour une variable d'\'echelle qui repr\'esente aussi
une variable de type  \'energie et qui correspond
aussi au point $w=1$ dans nos notations.

Nous avons aussi calcul\'e  
la loi limite jointe de la position
$X= \frac{x(t)-x(0)}{(T \ln t)^2}$ 
et de l' \'energie $ w=\frac{U(x(0))-U(x(t))}{(T \ln t)}$,
qui est d\'efinie par les deux expressions suivantes 
en transformation de Laplace
\begin{eqnarray} 
 && \int_0^{+\infty} dX e^{-sX} {\cal P}(X,w>1)   =  \frac{ \sinh {\sqrt s} }{ \sqrt s}
\left( e^{ {\sqrt s} \coth {\sqrt s} }-  2 \cosh {\sqrt s}   
\right)  e^{- w {\sqrt s} \coth {\sqrt s} } \nonumber \\
&& \int_0^{+\infty} dX e^{-sX} {\cal P}(X,w<1) 
 =   \frac{ \sinh {\sqrt s} (2-w)}{ {\sqrt s} }
- \frac{ \sinh 2 {\sqrt s} }{ {\sqrt s} } e^{- w {\sqrt s} \coth {\sqrt s}}
 \end{eqnarray}

\subsection{Propri\'et\'es de vieillissement} 

Le front de diffusion \`a deux temps
$\overline{P(x,t ; x', t' \vert 0,0)}$ pr\'esente
 un r\'egime de vieillissement en $({{\ln t}/{\ln t'}})$.
Dans les variables d'\'echelle $X={(x / \ln^2t)}$ et $X'={(x' / \ln^2t)}$,
le front de diffusion est d\'etermin\'e par la dynamique effective.
La proc\'edure de renormalisation permet
de calculer la loi jointe des positions $\{m(t),m(t_w)\}$
\`a deux temps successifs $t \geq t_w$ (Publication [P4] ).
En particulier, ce front de diffusion \`a deux temps
pr\'esente une fonction $\delta(X-X')$ de Dirac,
qui traduit
 le fait qu'une particule peut se trouver pi\'eg\'ee
 dans une vall\'ee dont elle n'arrive pas \`a sortir entre $t'$ et $t$.
Le poids $D(t,t_w)$ de cette fonction $\delta(X-X')$
qui repr\'esente la probabilit\'e d'avoir $m(t)=m(t_w)$ 
a pour expression
\begin{eqnarray} 
D(t,t_w)= \frac{1}{3} \left( \frac{\ln t_w}{\ln t}\right)^2 \left( 5-2 
e^{ 1- \left( \frac{\ln t}{\ln t_w} \right) }  \right)
\label{dttw}
\end{eqnarray}
Ce r\'esultat caract\'erise bien ce qu'on appelle ``vieillissement" :
plus $t_w$ est grand, plus la particule a trouv\'e une ``bonne" vall\'ee
et plus il faudra de temps pour en sortir.

\subsection{Statistique des retours \`a l'origine de $m(t)$ }

La proc\'edure de renormalisation permet aussi de montrer
que la distribution de la s\'equence $\Gamma_1=T \ln t_1$, $\Gamma_2=T \ln t_2$
... des temps de retour \`a l'origine de la dynamique effective
$m(t)$ a une structure simple : 
c'est un processus Markovien multiplicatif
d\'efini par la r\'ecurrence $\Gamma_{k+1}=\alpha_k \Gamma_k$,
dans laquelle les coefficients $\{\alpha_i\}$ 
sont des variables al\'eatoires ind\'ependantes, de loi
\begin{eqnarray} 
\rho(\alpha)= \frac{1}{\sqrt{5} } \left( \frac{1}{\alpha^{1+\lambda_-}}-
 \frac{1}{\alpha^{1+\lambda_+} }\right)  \ \hbox{avec} \  
\lambda_{\pm} = \frac{3 \pm \sqrt 5}{2}
\end{eqnarray}

Deux cons\'equences importantes : 

$\bullet$ Le nombre total $R(t)$ de retours \`a l'origine pendant $[0,t]$
se comporte en 
\begin{eqnarray} 
R(t) \sim \frac{1}{3}  \ln (T \ln t) 
\end{eqnarray}
alors que le nombre total $S(t)$ de sauts pendant $[0,t]$ 
se comporte en 
\begin{eqnarray} 
S(t) \sim \frac{4}{3}  \ln (T \ln t) 
\end{eqnarray}
(mais ici il y a des corr\'elations entre les temps de sauts.)

$\bullet$  La probabilit\'e que $m(\tau)>0$ pour $\tau \in ]0,t]$
met en jeu un exposant de persistance irrationnel
\begin{eqnarray} 
\Pi(t) \sim \left[\frac{1}{(T \ln t)^2}\right]^{\overline{ \theta}}
\ \ \hbox{avec} \ \ 
{\overline{\theta}=\frac{3-\sqrt{5}}{4}=0.19...}
\nonumber
\end{eqnarray}
alors que la probabilit\'e qu'un marcheur donn\'e
$x(t)$ ne repasse pas par l'origine $x(o)$ pendant $]0,t]$
a un exposant de persistance simple
\begin{eqnarray} 
\Pi_1(t) \sim \left[\frac{1}{(T \ln t)^2}\right]^{ \theta}
\ \ \hbox{avec} \ \  {\theta=\frac{1}{2} }
\nonumber
\end{eqnarray}

\section{ \'Etude des propri\'et\'es de localisation } 

\subsection{Distribution asymptotique du paquet thermique}

La distribution de la position relative $y= x(t)-m(t)$ par rapport
\`a la position $m(t)$ de la dynamique effective
a pour limite \`a temps infini une distribution de Boltzmann 
dans une vall\'ee Brownienne infinie
\begin{eqnarray} 
 P(y) = 
    \left<  \frac{e^{- \beta U_1(y \vert)) } }
{ \int_0^{\infty} dx e^{- \beta U_1(x)} 
+ \int_0^{\infty} dx e^{-\beta U_2(x)} }  \right>_{\{U_1,U_2\}}
\end{eqnarray}
o\`u l'on moyenne sur deux trajectoires Browniennes ${\{U_1,U_2\}}$
qui forment une vall\'ee infinie.
Cette formulation est bien \'equivalente 
au th\'eor\`eme de Golosov \cite{golosovlocali}.
La loi $P(y)$ peut \^etre calcul\'ee en transform\'ee
de Laplace en termes de fonctions de Bessel (Publication [P8])
En particulier, on obtient le comportement 
asymptotique alg\'ebrique  
\begin{eqnarray} 
P(y ) \opsim_{y \to \infty} \frac{1}{y^{3/2}}
\label{loi3/2}
\end{eqnarray}
qui peut \^etre interpr\'et\'e ainsi :
alors que les configurations $U(y)$ typiques donnent une
 d\'ecroissance en $e^{- \beta \sqrt{\sigma y} }$
pour le facteur de Boltzmann, il existe 
des configurations rares qui reviennent pr\`es de $U \sim 0$
\`a une grande distance $y$ avec une probabilit\'e en $1/(y^{3/2})$.
Nous avons aussi calcul\'e 
la fonction de corr\'elation \`a deux particules
\begin{eqnarray} 
C(l) = \lim_{t \to \infty} 2 \int_{-\infty}^{+\infty} dx 
 \overline{ \left[  P(x,t|x_0,0)  P(x+l,t|x_0,0) \right] }
\end{eqnarray}
qui d\'ecro\^{\i}t aussi alg\'ebriquement en $1/l^{3/2}$.

\subsection{Param\`etres de localisation}

Les param\`etres de localisation, qui mesurent les probabilit\'es moyennes de trouver $k$ particules au m\^eme point \`a temps infini
\begin{eqnarray} 
Y_k = \lim_{t \to \infty}
 \int_{-\infty}^{+\infty} dx  \overline{ \left[ P(x,t|x_0,0) \right]^k }
= \frac{\Gamma^3(k)}{\Gamma(2 k)}
(  \sigma \beta^2)^{k-1}
\end{eqnarray}
ont un comportement \`a grand $k$ qui est domin\'e par les vall\'ees
tr\`es \'etroites qui ont une petite fonction de partition
(Publication [P8]).

\subsection{ Comparaison avec les fonctions d'\'equilibre }

Ces diff\'erentes observables qui caract\'erisent
la statistique du paquet thermique dans la diffusion de Sinai
co\"{\i}ncident en fait avec leurs analogues statiques d\'efinies par
la limite thermodynamique
de la distribution de Boltzmann dans un potentiel Brownien
sur un intervalle (Publication [P8]).
Cette convergence vers l'\'equilibre du paquet thermique
(alors que la dynamique effective reste ind\'efiniment hors \'equilibre)
n'est plus vraie
 d\`es qu'on ajoute une force constante (cf Chapitre \ref{chapsinaibiais})

\subsection{ Largeur thermique et \'ev\`enements rares }

\begin{figure}[h]

\centerline{\includegraphics[height=3cm]{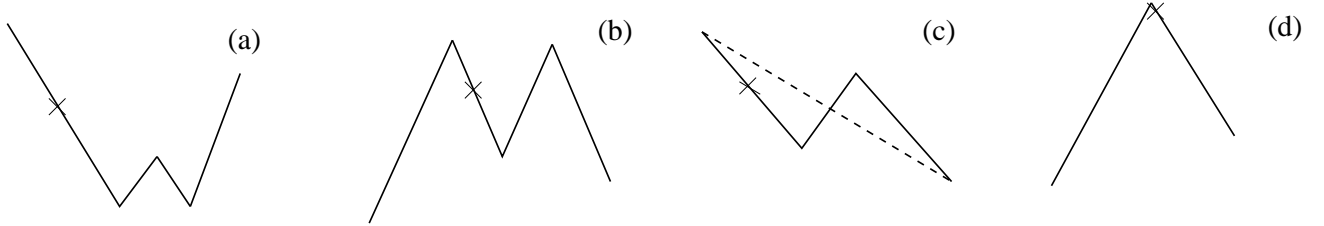}} 
\caption{\it Repr\'esentation des trois types (a) (b) (c) 
d'\'ev\`enements rares d'ordre $1/\Gamma$
qui dominent la largeur du paquet thermique \`a grand temps
( la croix repr\'esente la condition intiale).
Le cas (d) est un exemple d'\'ev\`enement rare d'ordre $1/\Gamma^2$
qui donne une contribution sous-dominante.} 
\label{figrare}
\end{figure}

La d\'ecroissance alg\'ebrique (\ref{loi3/2}) de la distribution
asymptotique de la position relative implique que le second moment
$\overline{<y^2>}$ diverge \`a temps infini.
Pour obtenir son comportement \`a grand temps, il faut prendre en compte
les \'ev\`enements rares suivants (Publication [P6]) :

(a) une vall\'ee renormalis\'ee peut avoir deux minima \'eloign\'es
dans l'espace qui sont presque d\'eg\'en\'er\'es en  \'energie.

(b) deux barri\`eres voisines peuvent \^etre presque d\'eg\'en\'er\'ees.

(c) une barri\`ere peut \^etre en train d'\^etre d\'ecim\'ee
$(\Gamma+\epsilon)$

Ces \'ev\`enements rares, repr\'esent\'es sur la Figure \ref{figrare},
 ont tous les trois une probabilit\'e
faible d'ordre $1/\Gamma$, mais ils donnent lieu \`a division du paquet
thermique en deux sous-paquets, s\'epar\'es par une grande distance d'ordre $\Gamma^2$. En cons\'equence, ce sont ces \'ev\`enements
 qui dominent la largeur thermique
\begin{eqnarray} 
\overline{<x^2(t)>-<x(t)>^2} \oppropto_{t \to \infty} \frac{T}{\Gamma}
(\Gamma^2)^2 = T (T \ln T)^3
\end{eqnarray}
et plus g\'en\'eralement tous les moments divergents d'ordre $k>1/2$
\begin{eqnarray} 
\overline{ \vert x(t)-<x(t)> \vert^k} \opsimeq_{t \to \infty}
c_k T (T \ln T)^{2k-1}
\end{eqnarray}
Les constantes $c_k$ en pr\'efacteur peuvent \^etre calcul\'ees
\`a partir des propri\'et\'es statistiques des \'ev\`enements rares 
(a,b,c) d\'ecrits ci-dessus (Publication [P4]).

\newpage

\section{ Relation avec la th\'eorie g\'en\'erale
des dynamiques lentes et des \'etats m\'etastables}

\subsection{ \'Etats m\'etastables }

Dans la description qualitative usuelle des dynamiques lentes,
que ce soit pour les verres, les milieux granulaires ou les syst\`emes
d\'esordonn\'es, la notion d'\'etat m\'etastable joue un grand r\^ole.
Comme les \'etats m\'etastables au sens strict n'existent
que dans les approximations de champ moyen ou dans la limite
de temp\'erature nulle, si on veut utiliser ce concept
pour des syst\`emes en dimension finie \`a temp\'erature finie,
il faut consid\'erer des \'etats m\'etastables de temps de vie finie
\cite{biroli}, en s\'eparant la dynamique en deux 
\'echelles de temps :
il y a d'une part des degr\'es de libert\'e ``rapides", 
qui atteignent vite un quasi-\'equilibre local, ce qui correspond
aux ``\'etats m\'etastables", et il y a d'autre part
une dynamique lente hors \'equilibre, qui correspond
\`a l' \'evolution des \'etats m\'etastables.

Dans ce langage, notre description de la marche al\'eatoire
de Sinai peut \^etre reformul\'ee ainsi :

$ \bullet $ Les \'etats m\'etastables \`a l'instant $t$ sont 
les vall\'ees
du paysage renormalis\'e \`a l'\'echelle $\Gamma= T \ln t$ :
en effet, les marcheurs qui sont partis \`a $t=0$ d'un point de cette vall\'ee
n'ont pas eu le temps d'en sortir \`a l'instant $t$.

$ \bullet $ Dans chaque vall\'ee renormalis\'ee, 
il y a un quasi-\'equilibre d\'ecrit par une
distribution de Boltzmann \`a l'int\'erieur de la vall\'ee.

$\bullet$ La dynamique lente correspond 
\`a l'\'evolution du paysage renormalis\'e avec l'\'echelle $\Gamma= T \ln t$  : certains \'etats m\'etastables disparaissent
et sont absorb\'es par un voisin.

\subsection{ Observables \`a un temps et Conjecture d'Edwards    }

Comme la `Conjecture d'Edwards', qui propose de calculer les quantit\'es dynamiques
par une moyenne plate sur les \'etats m\'etastables, a donn\'e lieu
\`a beaucoup de travaux r\'ecents \cite{barratetc,desmedt},
il est int\'eressant de reconsid\'erer de ce point de vue notre proc\'edure.

 Dans notre approche, toutes les observables \`a un temps se calculent 
effectivement par une moyenne sur les vall\'ees renormalis\'ees
(qui sont les \'etats m\'etastables), 
 mais avec une mesure qui d\'epend de l'observable :

$\bullet$ 
Pour une condition initiale uniforme, la 
taille du bassin d'attraction d'une vall\'ee est donn\'ee par sa longueur,
et donc on utilise une mesure pond\'er\'ee par la longueur $\int dl l P(l)$ 
pour calculer la distribution de la position, de l' \'energie, etc...

$\bullet$  Pour les observables du paquet thermique,
nous avons utilis\'e une mesure plate sur les vall\'ees 
Browniennes infinies, car il y a
une ind\'ependance entre la taille de la vall\'ee et
 la statistique du fond de la vall\'ee.

\subsection{ D\'ecomposition du front de diffusion sur les \'etats m\'etastables}

Si l'on veut
d\'ecrire \`a la fois la dynamique effective des vall\'ees
et l'\'equilibre de Boltzmann dans chaque vall\'ee renormalis\'ee,
on peut \'ecrire le front de diffusion dans un \'echantillon sous la forme
\begin{eqnarray}
P(x t|x_0 0) \simeq \sum_{V_\Gamma} \frac{1}{Z_{V_\Gamma}}
e^{- \beta U(x)} \theta_{V_\Gamma}(x)
\theta_{V_\Gamma}(x_0)
\label{valleysum}
\end{eqnarray}
La somme porte sur toutes les vall\'ees renormalis\'ees $V_\Gamma$
\`a l'\'echelle $\Gamma=T \ln t$. La notation 
$\theta_{V}(x)$ d\'esigne la fonction caract\'eristique
de la vall\'ee $V$, c'est \`a dire $\theta_{V}(x)=1$ si $x$ appartient \`a la vall\'ee et $\theta_{V}(x)=0$ sinon. Enfin  
$Z_{V} = \int_{V} dx e^{- \beta U(x)}$ repr\'esente la fonction de partition 
de la vall\'ee $V$.

Cette expression du front
de diffusion (\ref{valleysum}) correspond
 tout \`a fait \`a la construction g\'en\'erale en pr\'esence
d'\'etats m\'etastables (cf \cite{biroli,kurchanfermion} et les r\'ef\'erences incluses), dans laquelle 
l'op\'erateur d'\'evolution $e^{-t H_{FP} }$ est remplac\'e par un projecteur
sur les \'etats $(i)$ d'\'energie $E_i  < 1/t $
\begin{eqnarray}
e^{-t H_{FP} } \sim \sum_{i} \vert P_i > < Q_i \vert 
\end{eqnarray}
L'interpr\'etation est claire : ``everything fast has happened and everything slow has not taken place" \cite{kurchanfermion}.
Pour le mod\`ele de Sinai, les expressions explicites sont les suivantes : les \'etats \`a droite
\begin{eqnarray}
P_i(x)= \frac{ e^{-\beta U(x)} } { \int_{V_{\Gamma}^{(i)}}  dy   e^{-\beta U(x)
} } \theta ( x \in V_{\Gamma}^{(i)} )
\label{pi}
\end{eqnarray}
sont bien positifs, normalis\'es, avec des supports 
qui ne se recouvrent pas, alors que les \'etats \`a gauche
\begin{eqnarray}
Q_i(x)=  \theta ( x \in V_{\Gamma}^{(i)} )
\end{eqnarray}
sont simplement \'egaux \`a 1 sur le support de leur vecteur propre \`a droite associ\'e, et nuls ailleurs. 

 Dans le mod\`ele de Sinai, on peut en fait aller au del\`a
de cette description \`a un temps en terme de projection sur les \'etats m\'etastables, en consid\'erant la dynamique de ces \'etats m\'etastables
pour obtenir des informations sur les propri\'et\'es spectrales 
de l'op\'erateur de Fokker-Planck.

\section{ \'Etude des fonctions propres de l'op\'erateur de Fokker-Planck }

\subsection{ Construction des fonctions propres de l'op\'erateur de Fokker-Planck associ\'e \`a la dynamique effective }

Il est int\'eressant de voir comment change 
l'expression (\ref{valleysum}) lors de la d\'ecimation d'une vall\'ee renormalis\'ee, c'est \`a dire lors de la disparition d'un \'etat m\'etastable.
La d\'ecomposition de l'op\'erateur d'\'evolution sur les
valeurs propres $E_n \geq 0 $ et les fonctions propres 
\`a droite $\Phi^R_n$ et \`a gauche $\Phi^L_n$ de l'op\'erateur de Fokker-Planck $H_{FP}$ s'\'ecrit
\begin{eqnarray}
{ P}(x t|x_0 0) = <x \vert e^{-t H_{FP}} \vert x_0> = \sum_{n} 
e^{- { E_n} t}  { \Phi}^R_n(x) 
{ \Phi}^L_n(x_0)  
\label{relationschro}
\end{eqnarray}
En dehors de l'\'etat fondamental $n=0$ d'\'energie nulle $E_0=0$
qui correspond \`a l'\'equilibre de Boltzmann sur l'\'echantillon total
\begin{eqnarray}
\Phi_0^L(x) && = 1/\sqrt{Z_{tot}} \\
 \Phi_0^R(x) && = e^{- U(x)/T}/\sqrt{Z_{tot}}
\end{eqnarray}
la comparaison avec l'\'equation (\ref{valleysum}) conduit aux
identifications suivantes pour les \'etats excit\'es $n \geq 1$ :
les \'energies ${ E_n}$ sont d\'etermin\'ees 
par les \'echelles de renormalisation $\Gamma_n=T \ln t_n
= - T \ln { E_n}$ qui correspondent aux d\'ecimations de barri\`eres.
Lors d'une d\'ecimation, deux vall\'ees $V_1$ et $V_2$ 
se joignent en une seule nouvelle vall\'ee renormalis\'ee
 $V'$, et les fonctions propres associ\'ees s'\'ecrivent
\begin{eqnarray}
&& { \Phi}_n^L(x) = \sqrt{\frac{Z_{V_1} Z_{V_2}}{Z_{V_1} + Z_{V_2}}} 
(\frac{1}{Z_{V_1}} \theta_{V_1}(x) - \frac{1}{Z_{V_2}} \theta_{V_2}(x) ) \\
&& { \Phi}_n^R(x) = e^{- U(x)/T} { \Phi}_n^L(x) 
\label{eigen}
\end{eqnarray}
On peut v\'erifier que ces fonctions propres ont toutes les propri\'et\'es requises d'orthonormalisation
\begin{eqnarray}
\int dx  { \Phi}_n^L(x) { \Phi}_m^R(x)  = \delta_{n,m}
\end{eqnarray}
et de normalisation de la probabilit\'e totale du paquet thermique
\begin{eqnarray}
\int dx  { \Phi}_n^R(x)   = 0
\end{eqnarray}
 
 Au del\`a du mod\`ele de Sinai, la structure (\ref{eigen}) en termes de fonctions de partition partielles semble donc d\'ecrire 
plus g\'en\'eralement les fonctions propres de l'op\'erateur de Fokker-Planck
pour les dynamiques lentes dans lesquelles les \'etats m\'etastables disparaissent de mani\`ere hi\'erarchique et emboit\'ee.

\subsection{ Structure spatiale des fonctions propres : 2 pics et 3 \'echelles de longueur ! }

Un \'etat propre (\ref{eigen}) pr\'esente deux pics qui correspondent aux 
minima des vall\'ees $V_1$ et $V_2$. Chacun des deux pics a une largeur finie,
qui repr\'esente la longueur caract\'eristique associ\'e poids de Boltzmann \`a temp\'erature $T$ autour du minimum d'une vall\'ee.
La distance entre les deux pics est d'ordre $l(E) \sim \Gamma^2 
\sim (\ln E)^2$. Loin des minima, mais \`a l'int\'erieur de la vall\'ee
renormalis\'ee $ r \leq \Gamma^2$, la fonction propre $\Phi_n^R(x)$
a une d\'ecroissance gouvern\'ee par le poids de
Bolzmann $e^{-\beta U(r)}$ de comportement typique $e^{- c \sqrt{r} }$.
En particulier au bord de la vall\'ee $r \sim \Gamma^2$,
l'amplitude typique est d'ordre $e^{- c' \Gamma }$.
Au del\`a des deux vall\'ees en jeu, l'approximation simple
(\ref{eigen}) avec des fonctions theta devient insuffisante.
Pour estimer la d\'ecroissance d'un \'etat propre sur une distance 
  $r \geq \Gamma^2$, il faut consid\'erer \cite{eigenhuse} 
que les deux points sont s\'epar\'es par un nombre de vall\'ees renormalis\'ees
d'ordre $\frac{r}{\Gamma^2}$
et que le recouvrement entre deux vall\'ees voisines n'est pas z\'ero,
mais d'ordre $e^{- c'' \Gamma }$. Une th\'eorie de perturbation conduit
alors \`a une d\'ecroissance asymptotique exponentielle
en $e^{- c'' \frac{r}{\Gamma} }$, ce qui correspond
en effet \`a la longueur de localisation 
$\lambda(E) \sim  \Gamma \sim (-T \ln E)$ 
calcul\'ee exactement par une m\'ethode de type Dyson-Schmidt
pour le probl\`eme de Schr\"odinger associ\'e \cite{jpbreview}.

En conclusion, les propri\'et\'es des fonctions propres font donc intervenir
trois \'echelles de longueur qui coexistent :

$ \bullet$ une \'echelle finie $l \sim 1$ qui caract\'erise la largeur d'un pic,
et qui est reli\'ee \`a la localisation de Golosov du paquet thermique. 

$ \bullet$ une \'echelle $ l(E)  
\sim (\ln E)^2 $ qui repr\'esente la distance entre les deux pics
et qui est reli\'ee \`a la distance totale parcourue au temps $t \sim 1/E$. 

$ \bullet$ une \'echelle $ \lambda(E)  
\sim (-\ln E) $ qui caract\'erise la d\'ecroissance exponentielle asymptotique
de la fonction d'onde, et qui correspond \`a la longueur de localisation  
du probl\`eme de Schr\"odinger associ\'e. 

\label{twolengtheigen}

\section{ Conclusion }

Le mod\`ele de Sinai est 
un exemple parfait de `point fixe de d\'esordre infini'.
La proc\'edure de renormalisation donne
une image tr\`es compl\`ete de la dynamique asymptotique.
Elle permet d'obtenir des r\'esultats exacts explicites 
sur la dynamique effective d'une particule, sur les propri\'et\'es
de vieillissement, sur les propri\'et\'es
de localisation du paquet thermique, et sur les \'ev\`enements rares
qui gouvernent la largeur thermique.
Par ailleurs, la description du mod\`ele de Sinai en termes
de vall\'ees renormalis\'ees est un exemple explicite de la th\'eorie
g\'en\'erale des \'etats m\'etastables de temps de vie fini, 
et permet d'obtenir une image
tr\`es claire de la structure des fonctions propres de l'op\'erateur de Fokker-Planck.

Mentionnons pour terminer que certains r\'esultats obtenus
 par notre approche de renormalisation
de type Ma-Dasgupta ont \'et\'e depuis confirm\'es par des \'etudes
de math\'ematiciens probabilistes, que ce soit sur le poids
de la partie singuli\`ere du front de diffusion \`a deux temps \cite{dembo}
ou sur la statistique des retours \`a l'origine de la dynamique effective
 \cite{cheliotis}.

\subsection*{  Publications associ\'ees }

$\bullet$ Principe de la renormalisation et premiers r\'esultats : Publication [P3] 

$\bullet$ Dynamique effective, \'ev\`enements rares et vieillissement : Publication  [P4]

$\bullet$ Sur la dynamique de l'\'energie : Publication [P10]

$\bullet$  Sur la localisation de Golosov et l'op\'erateur de Fokker-Planck : Publication [P8]

%%%%%%%%%%%%%%%%%%%%%%%%%%%%%%%%%%%%%%%%%%%%%%%%%%%%%%%%%%%%%%%%%

\chapter
{Marche al\'eatoire dans un potentiel Brownien biais\'e}

\label{chapsinaibiais}

%  \begin{flushright}
%  \emph{  
% }\\  ~\\
%  \end{flushright}

\section{ Pr\'esentation du mod\`ele }

\subsection{ La phase de diffusion anormale }

 L'introduction d'une force constante $F_0$ dans l'\'equation
de Langevin (\ref{langevin}) du mod\`ele de Sinai
discut\'e au Chapitre pr\'ec\'edent est \'evidemment 
tr\`es naturelle. Ce mod\`ele avec force
a m\^eme encore plus
int\'eress\'e les math\'ematiciens et physiciens
depuis longtemps, 
car il 
pr\'esente une s\'erie de transitions de phase dynamiques
\cite{kestenetal,derridapomeau,jpbreview}
en fonction du param\`etre sans dimension $\mu=F_0 T/\sigma$.
En particulier, il existe une phase de diffusion anormale 
 $0<\mu<1$ qui est caract\'eris\'ee par le comportement
asymptotique sous-lin\'eaire
\begin{eqnarray} 
\overline{ <x(t)>} \opsimeq_{t \to \infty} t^{\mu}
\end{eqnarray}
alors que pour $\mu>1$, la vitesse devient finie :
$\overline{ <x(t)>} \sim V(\mu) t$.

\subsection{ Le mod\`ele dirig\'e de pi\`eges }

Il a \'et\'e propos\'e depuis longtemps un argument heuristique \cite{feigelman,jpbreview}  selon lequel le mod\`ele de Sinai biais\'e
\'etait asymptotiquement \'equivalent 
\`a un mod\`ele dirig\'e de pi\`eges d\'efini par l' \'equation ma\^{\i}tresse 
\begin{eqnarray}
\frac{d  P_t(n)}{dt}  =  - \frac{P_t(n)}{\tau_n}+\frac{P_t(n-1)}{\tau_{n-1}}
\label{master}
\end{eqnarray}
dans laquelle les $\tau_n$ sont des variables al\'eatoires
ind\'ependantes distribu\'ees avec une loi alg\'ebrique
\begin{eqnarray} 
q(\tau) \opsimeq_{\tau \to \infty} \frac{\mu}{ \tau^{1+\mu}} 
\label{lawtrap}
\end{eqnarray}
La phase de diffusion anormale $0<\mu<1$ 
correspond alors \`a la phase dans laquelle la moyenne
des temps de pi\'egeage est infini.

\section{ Principe de la renormalisation }

\begin{figure}[ht]
\centerline{\includegraphics[height=6cm]{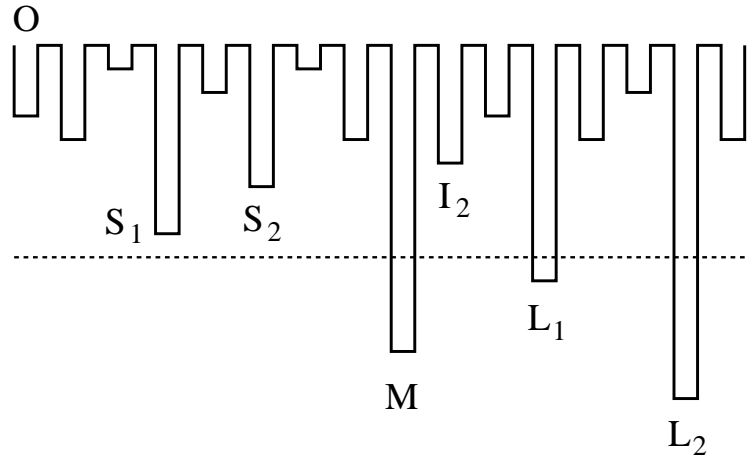}} 
\caption{\it Structure des pi\`eges importants pour une particule partie de
$x=0$ \`a $t=0$ :
La ligne pointill\'ee s\'epare les ``petits" pi\`eges 
(qui ont un temps de pi\'egeage $\tau_i<t$) et les ``grand" pi\`eges
 (qui ont un temps de pi\'egeage $\tau_i>t$). 
Le premier grand pi\`ege not\'e $M$ est le pi\`ege principal,
qui est occup\'e avec un poids d'ordre $O(1)$. 
Le grand pi\`ege suivant $L_1$ et le plus grand pi\`ege $S_1$ 
parmi les petits pi\`eges situ\'es avant $M$ 
sont les pi\`eges secondaires occup\'es avec un poids d'ordre $O(\mu)$.
Le troisi\`eme grand pi\`ege $L_2$, le plus grand pi\`ege $I_2$ 
parmi les petits pi\`eges situ\'es entre $M$ et $L_1$,
et le second plus grand pi\`ege $S_2$ parmi les petits
pi\`eges situ\'es avant $M$ sont les pi\`eges tertiaires
occup\'es avec un poids d'ordre $O(\mu^2)$.  } 
\label{figpieges}
\end{figure}

La renormalisation pr\'esent\'ee dans le Chapitre \ref{chapsinai}
pour la marche de Sinai peut \^etre appliqu\'ee
en pr\'esence d'un biais, mais
elle ne donne des r\'esultats exacts
dans la limite de temps infini $t \to \infty$ 
que si le biais tend vers z\'ero $\mu \to 0$ :
par exemple, la renormalisation donne comme
front de diffusion une loi exponentielle 
pour la variable d'\'echelle $X= \frac{x(t)}{t^{\mu}}$
ce qui ne co\"{\i}ncide avec le r\'esultat exact
mettant en jeu une loi de L\'evy 
\cite{kestenetal,jpbreview} que dans la limite $\mu \to 0$.
Le fait que la dynamique effective ne soit plus exacte
lorsque $\mu$ est fini a pour origine le fait
que la distribution des barri\`eres $F_-$ oppos\'ees
au biais (\ref{soluppdelta}) ne devient pas une distribution
de largeur infinie par renormalisation, mais qu'elle 
converge vers une distribution exponentielle de largeur finie
proportionnelle \`a $1/(2\delta) = T/\mu$.
Ceci montre que la propri\'et\'e de concentration
du paquet thermique tout entier
dans la m\^eme vall\'ee renormalis\'ee \`a grand temps,
qui est valide dans la limite $\mu \to 0$,
et qui constitue la localisation de Golosov,
n'est plus exacte pour $\mu$ fini.
Nous avons donc propos\'e
de g\'en\'eraliser la m\'ethode de renormalisation  
en incluant un certain \'etalement du paquet thermique
sur plusieurs vall\'ees renormalis\'ees.
Nous allons d'abord discuter cette proc\'edure g\'en\'eralis\'ee
pour le mod\`ele dirig\'e de pi\`eges :
dans un \'echantillon donn\'e, la distribution de probabilit\'e
d'un marcheur est une somme de distributions $\delta$ de Dirac,
avec la structure hi\'erarchique 
d\'ecrite dans la l\'egende de la figure (\ref{figpieges}).

\section{ R\'esultats pour le mod\`ele dirig\'e de pi\`eges }

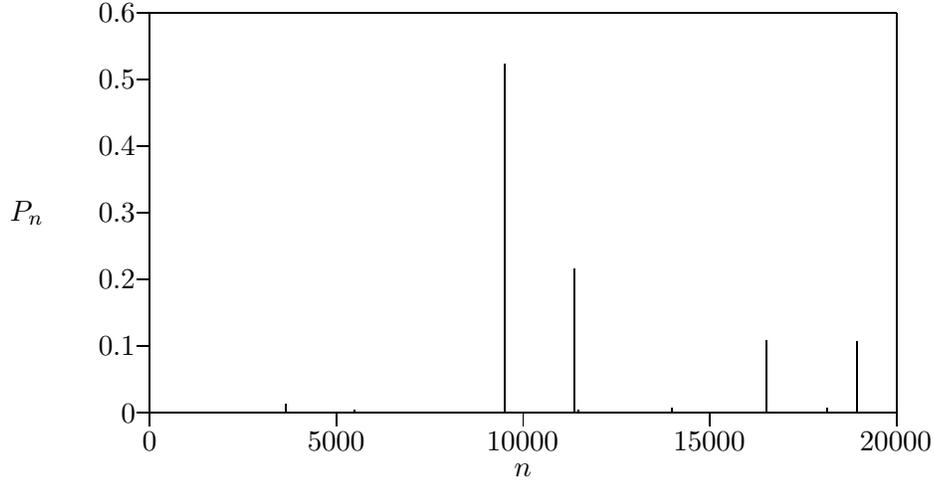
\begin{figure}[thb]
%[htbp]
% GNUPLOT: LaTeX picture
\setlength{\unitlength}{0.240900pt}
\ifx\plotpoint\undefined\newsavebox{\plotpoint}\fi
\sbox{\plotpoint}{\rule[-0.175pt]{0.350pt}{0.350pt}}%
\begin{picture}(1500,900)(0,0)
\sbox{\plotpoint}{\rule[-0.175pt]{0.350pt}{0.350pt}}%
\put(264,158){\rule[-0.175pt]{282.335pt}{0.350pt}}
\put(264,158){\rule[-0.175pt]{0.350pt}{151.526pt}}
\put(242,158){\makebox(0,0)[r]{0}}
\put(244,158){\rule[-0.175pt]{4.818pt}{0.350pt}}
\put(242,263){\makebox(0,0)[r]{0.1}}
\put(244,263){\rule[-0.175pt]{4.818pt}{0.350pt}}
\put(242,368){\makebox(0,0)[r]{0.2}}
\put(244,368){\rule[-0.175pt]{4.818pt}{0.350pt}}
\put(242,473){\makebox(0,0)[r]{0.3}}
\put(244,473){\rule[-0.175pt]{4.818pt}{0.350pt}}
\put(242,577){\makebox(0,0)[r]{0.4}}
\put(244,577){\rule[-0.175pt]{4.818pt}{0.350pt}}
\put(242,682){\makebox(0,0)[r]{0.5}}
\put(244,682){\rule[-0.175pt]{4.818pt}{0.350pt}}
\put(242,787){\makebox(0,0)[r]{0.6}}
\put(244,787){\rule[-0.175pt]{4.818pt}{0.350pt}}
\put(264,113){\makebox(0,0){0}}
\put(264,138){\rule[-0.175pt]{0.350pt}{4.818pt}}
\put(557,113){\makebox(0,0){5000}}
\put(557,138){\rule[-0.175pt]{0.350pt}{4.818pt}}
\put(850,113){\makebox(0,0){10000}}
\put(850,138){\rule[-0.175pt]{0.350pt}{4.818pt}}
\put(1143,113){\makebox(0,0){15000}}
\put(1143,138){\rule[-0.175pt]{0.350pt}{4.818pt}}
\put(1436,113){\makebox(0,0){20000}}
\put(1436,138){\rule[-0.175pt]{0.350pt}{4.818pt}}
\put(264,158){\rule[-0.175pt]{282.335pt}{0.350pt}}
\put(1436,158){\rule[-0.175pt]{0.350pt}{151.526pt}}
\put(264,787){\rule[-0.175pt]{282.335pt}{0.350pt}}
\put(45,472){\makebox(0,0)[l]{\shortstack{$P_n$}}}
\put(850,68){\makebox(0,0){$n$}}
\put(264,158){\rule[-0.175pt]{0.350pt}{151.526pt}}
\put(264,158){\usebox{\plotpoint}}
\put(264,158){\rule[-0.175pt]{51.312pt}{0.350pt}}
\put(477,158){\rule[-0.175pt]{0.350pt}{3.132pt}}
\put(477,158){\rule[-0.175pt]{0.350pt}{3.132pt}}
\put(477,158){\rule[-0.175pt]{15.418pt}{0.350pt}}
\put(541,158){\usebox{\plotpoint}}
\put(541,158){\usebox{\plotpoint}}
\put(541,158){\rule[-0.175pt]{4.577pt}{0.350pt}}
\put(560,158){\usebox{\plotpoint}}
\put(560,159){\usebox{\plotpoint}}
\put(561,158){\rule[-0.175pt]{5.782pt}{0.350pt}}
\put(585,158){\rule[-0.175pt]{0.350pt}{0.964pt}}
\put(585,158){\rule[-0.175pt]{0.350pt}{0.964pt}}
\put(585,158){\rule[-0.175pt]{56.852pt}{0.350pt}}
\put(821,158){\rule[-0.175pt]{0.350pt}{132.013pt}}
\put(821,158){\rule[-0.175pt]{0.350pt}{132.013pt}}
\put(821,158){\rule[-0.175pt]{3.373pt}{0.350pt}}
\put(835,158){\usebox{\plotpoint}}
\put(835,158){\usebox{\plotpoint}}
\put(835,158){\rule[-0.175pt]{23.126pt}{0.350pt}}
\put(931,158){\rule[-0.175pt]{0.350pt}{54.684pt}}
\put(931,158){\rule[-0.175pt]{0.350pt}{54.684pt}}
\put(931,158){\rule[-0.175pt]{1.204pt}{0.350pt}}
\put(936,158){\rule[-0.175pt]{0.350pt}{0.964pt}}
\put(936,158){\rule[-0.175pt]{0.350pt}{0.964pt}}
\put(936,158){\rule[-0.175pt]{35.653pt}{0.350pt}}
\put(1084,158){\rule[-0.175pt]{0.350pt}{1.927pt}}
\put(1084,158){\rule[-0.175pt]{0.350pt}{1.927pt}}
\put(1084,158){\rule[-0.175pt]{7.227pt}{0.350pt}}
\put(1114,158){\usebox{\plotpoint}}
\put(1114,158){\usebox{\plotpoint}}
\put(1114,158){\rule[-0.175pt]{18.790pt}{0.350pt}}
\put(1192,158){\usebox{\plotpoint}}
\put(1192,158){\usebox{\plotpoint}}
\put(1192,158){\rule[-0.175pt]{9.636pt}{0.350pt}}
\put(1232,158){\rule[-0.175pt]{0.350pt}{27.703pt}}
\put(1232,158){\rule[-0.175pt]{0.350pt}{27.703pt}}
\put(1232,158){\rule[-0.175pt]{1.686pt}{0.350pt}}
\put(1239,158){\usebox{\plotpoint}}
\put(1239,158){\rule[-0.175pt]{16.381pt}{0.350pt}}
\put(1307,158){\usebox{\plotpoint}}
\put(1307,158){\usebox{\plotpoint}}
\put(1307,158){\rule[-0.175pt]{4.818pt}{0.350pt}}
\put(1327,158){\rule[-0.175pt]{0.350pt}{1.927pt}}
\put(1327,158){\rule[-0.175pt]{0.350pt}{1.927pt}}
\put(1327,158){\rule[-0.175pt]{11.081pt}{0.350pt}}
\put(1373,158){\rule[-0.175pt]{0.350pt}{26.981pt}}
\put(1373,158){\rule[-0.175pt]{0.350pt}{26.981pt}}
\put(1373,158){\rule[-0.175pt]{15.177pt}{0.350pt}}
\end{picture}
\caption{\it Reproduction d'une figure de la r\'ef\'erence 
\cite{compte} de A. Compte et J.P. Bouchaud avec sa description :
``Distribution of probability after a time $t=7 \times 10^{10}$ for a particular sample of disorder in our 1D
directed random walk model with $\mu = 0.4$. The simulation was done
with $1.000$ particles in a lattice of $20.000$ sites.
This probability distribution is made of several sharp
peaks that gather a finite fraction of the particles. However, the
position of these peaks is scattered on a region of space of width
$t^\mu$. As time progresses, the position and relative weights
of these peaks of course change, but at any given (large) time only a
finite number of peaks, corresponding to very large trapping times,
contain most of the particles."}
\label{profile}
\end{figure}

A partir de cette description du front de diffusion dans chaque \'echantillon, il est possible de calculer 
exactement des s\'eries perturbatives en $\mu$ pour toutes les observables.
En particulier, les calculs explicites jusqu'\`a l'ordre $\mu^2$
(Publication [P11])
du front de diffusion de la variable d'\'echelle $X=\frac{x}{t^{\mu}}$,
de la largeur thermique
\begin{eqnarray}
 \lim_{t \to \infty} \frac{\overline { < \Delta n^2 (t) >} }{ t^{2 \mu} } 
=  \mu (2 \ln 2) + \mu^2  [ - \frac{\pi^2}{6}
 + 2 \ln 2 (\ln2 -2+2 \gamma_E)  ] 
+O(\mu^3)
\label{widthexact}
\end{eqnarray}
et du param\`etre de localisation
\begin{eqnarray}
Y_2 (\mu) = 1 - \mu (2 \ln 2) + \mu^2 ( 4 \ln 2- \frac{\pi^2}{6}) +O(\mu^3)
\label{y2exactexpansion}
\end{eqnarray}
co\"{\i}ncident avec les d\'eveloppements des r\'esultats exacts
obtenus par d'autres m\'ethodes 
pour le front de diffusion \cite{kestenetal,jpbreview},
pour la largeur thermique \cite{aslangul}
et pour le param\`etre de localisation \cite{compte}.
Ces comparaisons avec des r\'esultats exacts obtenus par d'autres m\'ethodes
montre que la proc\'edure de renormalisation g\'en\'eralis\'ee
est exacte ordre par ordre en $\mu$.
Plus g\'en\'eralement, pour calculer les observables \`a l'ordre $\mu^n$,
il suffit de consid\'erer que le front de diffusion
s'\'etale sur au plus $(1+n)$ pi\`eges
 et de moyenner sur les \'echantillons avec la mesure appropri\'ee.
Et comme nous avons une description du front de diffusion \'echantillon par \'echantillon, cette m\'ethode permet de calculer perturbativement
toutes les observables que l'on souhaite (Publication [P11]).

Cette approche permet par ailleurs de bien comprendre
comment, dans la phase de diffusion anormale $0<\mu<1$,
il y a \`a la fois une largeur thermique
qui croit en $t^{2\mu}$ (\ref{widthexact})
et une probabilit\'e finie $Y_2(\mu)$ (\ref{y2exactexpansion})
de trouver deux particules au m\^eme point \`a temps infini.
Notre description de la r\'epartition du paquet thermique
dans un \'echantillon donn\'e est tout \`a fait en accord
 avec les simulations num\'eriques de A. Compte et J.P. Bouchaud \cite{compte},
dont la figure reproduite ici (\ref{profile}) montre que m\^eme pour $\mu=0.4$,
il n'y a typiquement que quatre pi\`eges qui jouent un r\^ole important
dans un \'echantillon donn\'e \`a un instant donn\'e.

\section{ \'Equivalence quantitative entre le mod\`ele de Sinai biais\'e
et le mod\`ele dirig\'e de pi\`eges}

\begin{figure}[ht]

\centerline{\fig{6cm}{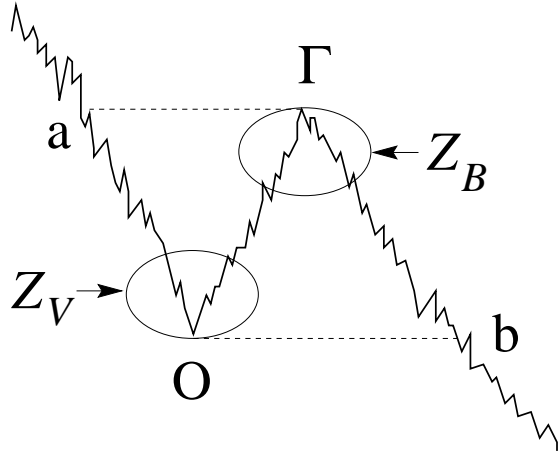}} 
\caption{\it Calcul du temps de sortie d'une vall\'ee renormalis\'ee
de barri\`ere $\Gamma$ : l'expression exacte du temps de premier passage
en $b$ pour une particule qui part de $0$
fait intervenir une int\'egrale double
qui est domin\'ee par le facteur d'Arrh\'enius $e^{\beta \Gamma}$,
avec un pr\'efacteur qui fait intervenir deux fonctions de partition :
$Z_V$ repr\'esente la fonction de partition de la vall\'ee et $Z_B$ 
repr\'esente la fonction de partition du potentiel $(-V)$ 
autour du sommet de la barri\`ere $\Gamma$.}
 \label{figescapeti}  

\end{figure}

Une analyse par la m\'ethode du col du temps de sortie d'une vall\'ee renormalis\'ee donn\'ee
de barri\`ere $\Gamma$ 
montre qu'il est distribu\'e selon une loi exponentielle,
comme dans le mod\`ele de pi\`ege, 
avec un temps de pi\`ege 
\begin{eqnarray}
\theta  \opsimeq_{\Gamma \to \infty} \beta Z_B Z_V
 e^{\beta \Gamma}
\label{theta1V}
\end{eqnarray}
qui d\'epend surtout de la barri\`ere $\Gamma$ \`a travers
de facteur d'Arrh\'enius usuel $e^{\beta \Gamma}$, 
mais qui d\'epend aussi des d\'etails de la vall\'ee \`a travers le pr\'efacteur
qui contient deux fonctions de partition de vall\'ees Browniennes ind\'ependantes
$Z_V$ et $Z_B$ (Figure \ref{figescapeti}).

La distribution du temps de pi\`ege $\tau_V$ 
sur l'ensemble des vall\'ees du paysage renormalis\'e
\`a l'\'echelle $\Gamma$ peut \^etre calcul\'ee
\`a partir de la distribution des barri\`eres et du pr\'efacteur :
on retrouve alors la m\^eme expression 
que dans le paysage renormalis\'e du mod\`ele dirig\'e de pi\`ege
\begin{eqnarray}
q_t(\tau)= \theta(t<\tau) \frac{\mu}{\tau} 
\left( \frac{t}{\tau} \right)^{\mu}
\label{qttau}
\end{eqnarray}
\`a condition de choisir l'\'echelle de renormalisation $\Gamma$
du paysage de Sinai en fonction du temps selon
\begin{eqnarray}
\Gamma(t) = T \ln \left[ t \sigma^2 \beta^3 
\left( \Gamma^{2} (1+\mu) \right)^{\frac{1}{\mu}}  \right]
\label{gammat}
\end{eqnarray}
L'\'echelle de longueur associ\'ee
\begin{eqnarray}
b(t)
= \frac{ \Gamma^2 (\mu) }{\sigma \beta^2 } 
\left[ t \sigma^2 \beta^3   \right]^{\mu}
\label{bt}
\end{eqnarray}
correspond alors exactement  
\`a la constante du front de diffusion 
qui a \'et\'e calcul\'ee r\'ecemment par les math\'ematiciens probabilistes
\cite{hushiyor}.

Finalement, notre approche \'etablit que le mod\`ele de Sinai biais\'e 
et le mod\`ele dirig\'e de pi\`eges sont asymptotiquement \'equivalents
du point de vue de leurs descriptions renormalis\'ees \`a grande \'echelle,
\`a une \'echelle de longueur pr\`es que nous avons d\'etermin\'ee.

\section{ R\'esultats pour le mod\`ele de Sinai biais\'e }

\begin{figure}[ht]

\centerline{\fig{6cm}{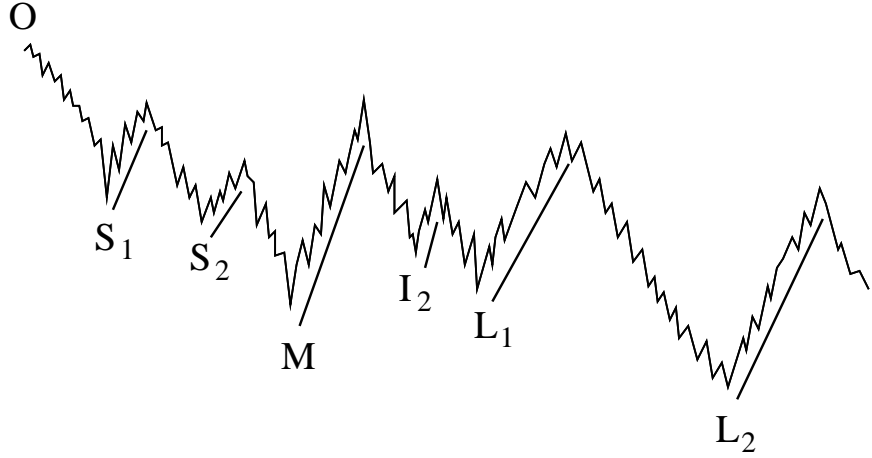}} 
\caption{\it Structure hi\'erarchique des vall\'ees importantes pour une particule partant de l'origine.
Les barri\`eres oppos\'ees au biais, qui sont soulign\'ees par des lignes droites, correspondent aux profondeurs du mod\`ele de pi\`eges de la Figure \ref{figpieges}.  Le fond $M$ de la vall\'ee renormalis\'ee
qui contient l'origine \`a l'\'echelle $\Gamma$ est 
avec une probabilit\'e d'ordre $O(1)$. Le fond $L_1$ de la vall\'ee
renormalis\'ee suivante et le fond $S_1$ de la plus grande sous-vall\'ee 
avant $M$ sont occup\'ees avec des poids d'ordre $O(\mu)$.
Le fond $L_2$ de la deuxi\`eme vall\'ee renormalis\'ee apr\`es $M$, 
la plus grande sous-vall\'ee $I_2$ entre $M$ et $L_1$,
et la seconde plus grande sous-vall\'ee $S_2$ avant $M$ sont occup\'ees
avec des probabilit\'es d'ordre $O(\mu^2)$. }
\label{figsinaib}  

\end{figure}

Tous les r\'esultats du mod\`ele dirig\'e de pi\`eges
peuvent donc \^etre traduits
pour le mod\`ele de Sinai biais\'e, en rempla\c{c}ant les pi\`eges ponctuels
par des vall\'ees renormalis\'ees. Le front de diffusion
dans un \'echantillon fix\'e a une structure hi\'erarchique analogue, 
repr\'esent\'ee par la Figure \ref{figsinaib}
qui est l'\'equivalent de la figure \ref{figpieges}.
Dans les r\'esultats quantitatifs, il suffit de remplacer
la variable d'\'echelle $X=\frac{n}{t^{\mu}}$ 
par la variable d'\'echelle $X=\frac{x(t)}{b(t)}$ adapt\'ee
au mod\`ele de Sinai avec biais (\ref{bt}).
En particulier, la largeur thermique admet le d\'eveloppement suivant
\begin{eqnarray}
\frac{ \overline { < \Delta x^2(t) >} }{t^{2 \mu}}
 = \frac {  \left(  \sigma^2 \beta^3   \right)^{2\mu}}
{  \sigma^2 \beta^4 } \left[  
 \frac{(2 \ln 2)}{\mu^3} +   [ - \frac{\pi^2}{6}
 + 2 \ln 2 (\ln2 -2-2 \gamma_E)  ] \frac{1}{\mu^2}
+O(\frac{1}{\mu}) \right]
\label{widthsinaidv}
\end{eqnarray}
Et si on traduit le r\'esultat exact obtenu pour le mod\`ele dirig\'e de pi\`eges \cite{aslangul}, on obtient m\^eme le r\'esultat suivant pour toute
la phase $0<\mu<1$ de diffusion anormale
\begin{eqnarray}
\frac{ \overline { < \Delta x^2(t) >} }{t^{2 \mu}}
  = \frac {  \left(  \sigma^2 \beta^3   \right)^{2\mu}}
{  \sigma^2 \beta^4 }
\frac{\Gamma^4 (\mu)}{\Gamma (2 \mu) } \left( \frac{ \sin \pi \mu}{\pi \mu} \right)^3 I(\mu) 
\label{widthsinai}
\end{eqnarray}
en termes de l'int\'egrale \cite{aslangul}
\begin{eqnarray}
I(\mu) 
 =  \int_0^{1} dz \frac{ (1+z) z^{\mu} (1-z)^{2 \mu} }
 {  z^{2 \mu +2} + 2 \cos \pi \mu z^{\mu+1} +1 }
\label{integralmu}
\end{eqnarray}

\section{ Conclusion }

La phase de diffusion anormale
$x \sim t^{\mu}$ avec $0 <\mu<1$ 
dans le mod\`ele de Sinai biais\'e 
est caract\'eris\'ee par une localisation
sur plusieurs vall\'ees renormalis\'ees,
dont les positions et les poids peuvent
\^etre caract\'eris\'es dans chaque \'echantillon.
La proc\'edure de renormalisation g\'en\'eralis\'ee
permet de calculer toutes les observables souhait\'ees par
un d\'eveloppement perturbatif exact en $\mu$.

Du point de vue des m\'ethodes de renormalisation de type Ma-Dasgupta,
cette \'etude montre que la proc\'edure usuelle
qui est exacte lorsque le d\'esordre \'evolue vers
un point fixe de d\'esordre infini, est une approximation qui
garde un grand int\'er\^et lorsque le point fixe
est caract\'eris\'e par un d\'esordre fini assez grand :
la proc\'edure usuelle constitue alors l'ordre dominant
d'un d\'eveloppement perturbatif syst\'ematique.

\subsection*{  Publication associ\'ee  [P11] }

%%%%%%%%%%%%%%%%%%%%%%%%%%%%%%%%%%%%%%%%%%

\chapter
{Cha\^{\i}nes de spin classiques d\'esordonn\'ees }

\label{chaprfim}

%  \begin{flushright}
%  \emph{ 
%  }\\  ~\\
%  \end{flushright}

\section{ Pr\'esentation des mod\`eles }

L'\'equilibre thermodynamique
des cha\^{\i}nes de spins classiques d\'esordonn\'ees
peut \^etre formul\'e comme un produit de matrices de transfert $2 \times 2$
al\'eatoires. En particulier,
l' \'energie libre par spin correspond \`a l'exposant de Lyapunov
et peut donc \^etre \'etudi\'ee par la m\'ethode de Dyson-Schmidt : le
livre de J.M. Luck \cite{luckbook} expose les nombreux
r\'esultats obtenus dans ce cadre.
Dans ce chapitre, nous allons \'etudier l'\'equilibre et la
dynamique de ces mod\`eles, en termes des domaines Imry-Ma.

\subsection{ La cha\^{\i}ne d'Ising en champ al\'eatoire } 

La cha\^{\i}ne d'Ising en champ al\'eatoire a pour Hamiltonien 
\begin{eqnarray}
{\cal H } = - J \sum_{i=1}^{n=N-1}  S_i S_{i+1} - \sum_{i=1}^{i=N} h_i S_i
\label{hamilrfim}
\end{eqnarray}
Les champs $\{h_i\}$ sont des variables al\'eatoires ind\'ependantes,
de moyenne nulle $ \overline{h_i}=0$ 
(cf la publication [P7] pour le cas d'une moyenne non-nulle )
et de variance 
\begin{eqnarray}
g \equiv\overline{h_i^2}
\end{eqnarray}
D'apr\`es l'argument de Imry-Ma \cite{imryma}, l'\'etat fondamental \`a temp\'erature nulle est d\'esordonn\'e : il y a une alternance
de domaines de spins $(+)$ et de domaines de spins $(-)$, 
la taille typique d'un domaine \'etant la longueur de Imry-Ma
\begin{eqnarray}
L_{IM} \approx \frac{4 J^2}{g}
\end{eqnarray}
L'argument de Imry-Ma est le suivant \cite{imryma} :
la cr\'eation d'un domaine de taille $L$
co\^ute une \'energie $4J$ (cr\'eation de deux parois), qui est ind\'ependante de $L$, mais elle permet de gagner une \'energie typique 
$| 2 \sum_{i=x}^{x+L} h_i |_{\text{typ}} \sim 2 \sqrt{  g L}$ . 
En cons\'equence, pour $L>L_{IM}$, il devient \'energ\'etiquement favorable
de cr\'eer un domaine pour profiter d'une fluctuation favorable des champs
al\'eatoires.

\subsection{ La cha\^{\i}ne verre de spin en champ magn\'etique ext\'erieur }

L'Hamiltonien du verre de spin en champ magn\'etique s'\'ecrit
\begin{eqnarray}
{ \cal H }= -\sum_{i=1}^{i=N-1} J_i \sigma_i \sigma_{i+1} - \sum_{i=1}^N h \sigma_i
\end{eqnarray}
o\`u les couplages $\{J_i\}$ sont des variables al\'eatoires ind\'ependantes.
Dans le cas particulier $J_i=\pm J$ avec probabilit\'e $(1/2,1/2)$,
il existe une \'equivalence avec la cha\^{\i}ne en champ al\'eatoire 
(\ref{hamilrfim}) avec $h_i=\pm h$ :

\begin{itemize} 
\item On pose $J_i = J \epsilon_i$ avec $\epsilon_i = \pm 1$
\item  On effectue une transformation de jauge des spins 
$\sigma_i = \epsilon_{1}..\epsilon_{i-1}  S_i$
\item  On pose  $h_i = h \epsilon_{1}..\epsilon_{i-1}$ 
\end{itemize}

L'interpr\'etation physique de cette correspondance entre les deux mod\`eles est la suivante :

$\bullet$ En champ nul $h=0$, les deux \'etats fondamentaux du verre de spin
 $\pm \sigma_i^{(0)}$, avec $ \sigma_i^{(0)} = \pm \epsilon_{1}..\epsilon_{i-1}$,
correspondent aux \'etats ferromagn\'etiques de la cha\^{\i}ne d'Ising pure  $S^{(0)}_i=+1$ et $S^{(0)}_i=-1$.

$\bullet$  En pr\'esence de $h>0$, l'\'etat fondamental est une succession 
des deux fondamentaux de champ nul de taille typique $L_{IM}=\frac{4 J^2}{h^2}$.

$\bullet$  Les parois de domaines de la cha\^{\i}ne en champ al\'eatoire repr\'esentent les liens frustr\'es
$J_i \sigma_i \sigma_{i+1} =J S_i S_{i+1} <0$ du verres de spin.

Dans la suite, nous ne d\'ecrirons donc les r\'esultats que dans le langage
de cha\^{\i}ne en champ al\'eatoire, car il est imm\'ediat de les traduire
pour la cha\^{\i}ne verre de spin en champ ext\'erieur

\section{ Principe de la renormalisation }

\subsection{ Hamiltonien pour les parois de domaines }

L'Hamiltonien (\ref{hamilrfim}) pour les spins correspond
\`a l'Hamiltonien suivant pour les parois de domaines
$A_{\alpha}(+\vert-)$
 et $B_{\alpha}(-\vert+)$ dans l'ordre $A_1 B_1 A_2 B_2 ...$ :
\begin{eqnarray}
{\cal H } =  2 J (N_A + N_B)  + \sum_{\alpha=1}^{N_A} V(a_\alpha) 
- \sum_{\alpha=1}^{N_B} V(b_\alpha) 
\label{hparois}
\end{eqnarray}
en terme du potentiel de Sinai
\begin{eqnarray}
 V(x) = - 2 \sum_{i=1}^{x} h_i
\label{potsinai}
\end{eqnarray} 
En cons\'equence :   
\begin{itemize}
\item Chaque paroi (A ou B) co\^ute une  \'energie $2J$
\item Les parois $A_{\alpha}(+\vert-)$
voient le potentiel $V(x)$ 
\item Les parois $B_{\alpha}(-\vert+)$ voient le potentiel oppos\'e $(-V(x))$ 
\end{itemize}

\subsection{ Dynamique de croissances de domaines }

On s'int\'eresse \`a la
dynamique de Glauber \`a partir d'une configuration initiale al\'eatoire : 
le taux de transition est
$W(S_j \to -S_j) = \frac{ e^{-\beta \Delta E } }
{ e^{\beta \Delta E } + e^{-\beta \Delta E }}$ avec
%  $ \Delta E=2 J S_j (S_{j-1} + S_{j+1}) + 2 h_j S_j$ :
\begin{eqnarray}
&&  \Delta E \{ \hbox{ cr\'eation de 2 parois} \}= 4 J  \pm  2 h_j 
\nonumber \\
&&  \Delta E \{ \hbox{ diffusion d'une paroi} \}=  \pm  2 h_j \nonumber
\\
&&  \Delta E \{ \hbox{ annihilation de 2 parois} \}= -4 J \pm  2 h_j
\end{eqnarray} 

Dans le r\'egime $\{h_i\} \sim T \ll J$ qui va nous int\'eresser dans toute la suite de ce chapitre, la { croissance de domaines }
 depuis $l \sim 1$ jusqu'\`a $L_{IM} \gg 1$ est d\'ecrite par le mod\`ele suivant de R\'eaction-diffusion dans le potentiel de Sinai

\begin{figure}[ht]

\centerline{\includegraphics[height=5cm]{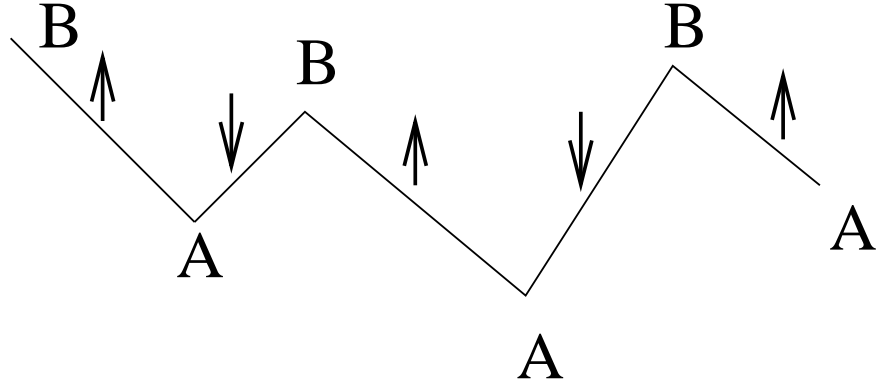}} 
\caption{\it La ligne en zig-zag repr\'esente la renormalisation du potentiel de Sinai (\ref{potsinai}) vu par les parois (\ref{hparois}). 
Les parois de domaines 
de type $A(+\vert-)$ occupent les minima, alors que 
les parois de domaines 
de type $B(-\vert+)$ occupent les maxima. Les descentes contiennent donc des spins $(+)$, alors que
les mont\'ees contiennent des spins $(-)$.   } 
\label{figparois}
\end{figure}

\begin{itemize} 
\item Les parois A diffusent vers les minima
\item  Les parois B diffusent vers les maxima
\item  Annihilation lors des rencontre $A+B \to \emptyset$
\end{itemize}
On peut de plus montrer (Publication [P7]) qu'\`a grand temps, tous les maxima
et minima sont occup\'es par des parois, comme sur la Figure (\ref{figparois}),
ce qui simplifie l'analyse du processus de r\'eaction-diffusion
par la renormalisation du paysage.

\subsection{ \'Equilibre }

 Le processus de r\'eaction-diffusion d\'ecrivant la croissance de domaines s'arr\^ete \`a l'\'echelle de renormalisation 
\begin{eqnarray}
\Gamma_{eq}=T \ln t_{eq} = 4J
\end{eqnarray} 
 lorsqu'on atteint l'
\'equilibre avec la cr\'eation de paires de domaines.

\section{ R\'esultats }

\subsection{Densit\'e de parois}

La densit\'e de parois d\'ecro\^{\i}t selon 
 $\displaystyle n(t) =  \frac{4 g}{ (T \ln t)^2} $
jusqu'\`a l'\'equilibre $n_{eq} = 1/L_{IM} $.

\subsection{Distribution des longueurs des domaines} 

Ce sont des variables ind\'ependantes (ce qui n'est pas vrai pour le cas pur
\cite{zeitak}), distribu\'ees selon la loi 
\begin{eqnarray}  
 P^*(\lambda)  
=  \pi \sum_{n = -\infty}^{\infty} \left(n+\frac{1}{2}\right)
 (-1)^n e^{- \pi^2 \lambda \left(n+\frac{1}{2}\right)^2}
\end{eqnarray}
avec la variable d'\'echelle $\lambda =\frac{2 g l}{(T \ln t)^2}$
au cours de la croissance de domaines,
et $\lambda=\frac{l}{2 L_{IM}}$ \`a l'\'equilibre.

\subsection{Fonction de corr\'elation spatiale}

\begin{eqnarray} 
\overline{\langle S_0(t)S_{x}(t) \rangle}= \sum_{n=-\infty }^{\infty } 
\frac{48 + 64 (2n+1)^2\pi^2 g \frac{\vert x \vert}{\Gamma^2 }}{(2n+1)^4 \pi^4} 
e^{-(2n+1)^2\pi^2 2 g \frac{\vert x \vert}{\Gamma^2}}  
\end{eqnarray}
avec $\Gamma =(T \ln t)$
au cours de la croissance de domaines,
et $\Gamma_{eq}=4J$ \`a l'\'equilibre.

\subsection{Fonction de corr\'elation temporelle} 

La fonction d'autocorr\'elation d'un spin d\'ecro\^{\i}t selon
\begin{eqnarray}
\overline{ \langle S_i(t) S_i(t') \rangle }
=\frac{4}{3}  \left(\frac {\ln t'}{\ln t} \right)
-\frac{1}{3} \left( \frac {\ln t'} {\ln t} \right)^2
\end{eqnarray}
ce qui correspond \`a un exposant d'autocorr\'elation \`a grand temps $\lambda=1/2$.

\subsection{Exposants de persistance} 

$\bullet$ La probabilit\'e qu'un spin ne se retourne pas dans $[0,t]$ 
a une d\'ecroissance 

$  \left( \frac{1}{(T \ln t)^2 }\right)^{\theta} $ avec 
{$\theta=1$}.

$\bullet$ La probabilit\'e que la valeur moyenne thermique
$<S_i(t)>$ ne change pas de signe 
dans $[0,t]$ d\'ecro\^{\i}t selon :

$  \left( \frac{1}{(T \ln t)^2} \right)^{\overline \theta} $ avec 
{$\overline \theta= \frac{3-\sqrt 5}{4}$}.

$\bullet$ La probabilit\'e qu'un domaine initial n'ait pas disparu 
\`a l'instant $t$ d\'ecro\^{\i}t selon :

$  \left( \frac{1}{(T \ln t)^2 }\right)^{\psi} $ avec 
{$\psi= \frac{3-\sqrt 5}{4}$}. (De mani\`ere g\'en\'erale, on a 
l'in\'egalit\'e $\psi \leq \overline \theta$).

\subsection{Violation du th\'eor\`eme de fluctuation-dissipation} 

\subsubsection{ Rapport $X$ pour mesurer le caract\`ere ``hors \'equilibre" }

Le param\`etre $X$ de violation du th\'eor\`eme de Fluctuation-Dissipation
est d\'efini par \cite{cuku}
\begin{eqnarray}
T \ R(t,t_w)= X(t,t_w) \ \partial_{t_w} C(t,t_w)
\end{eqnarray}
o\`u $C(t,t_w)$ repr\'esente la fonction de corr\'elation thermique tronqu\'ee 
\begin{eqnarray}
C(t,t_w)= \sum_x \overline{< S_0(t) S_x(t_w) > -< S_0(t)> <S_x(t_w)> }
\end{eqnarray}
et o\`u $R(t,t_w)$ repr\'esente la fonction de r\'eponse lin\'eaire
lorsqu'on applique un champ uniforme $H$ \`a partir de $t_w$
\begin{eqnarray}
 \overline{< S_0(t)>} = H \int_{t_w}^t du R(t,u)
\end{eqnarray}

Pour la cha\^{\i}ne en champ al\'eatoire, on trouve trois r\'egimes

(i) un quasi-\'equilibre des parois dans les vall\'ees
\begin{eqnarray}
X(t,t_w)= 1 
 \qquad  \text{pour} ~~  0< 
\frac{\ln (t-t_w)}{\ln t_w} <1 
\end{eqnarray}

(ii) un rapport X non trivial lorsque la dynamique effective des vall\'ees repart 
\begin{eqnarray}
X(t,t_w)=  \frac{t + t_w}{t} 
\qquad  \text{pour} ~~  
\frac{t-t_w}{t_w} ~~~ \text{fix\'e}   
\end{eqnarray}

(iii) un r\'egime final de vieillissement
\begin{eqnarray}
X(t,t_w)= \frac{t}{t_w \ln t_w} ( 1 + \frac{24}{7} \frac{\ln^2 t_w}{\ln^2 t} )
  ~ \text{pour} ~ 
\frac{\ln t}{\ln t_w} > 1 
\end{eqnarray}
En particulier, {$X$
croit vers $+\infty$}, car les corr\'elations thermiques tronqu\'ees
sont tr\`es faibles par rapport \`a la r\'eponse \`a un champ.

\subsubsection{Comparaison avec les mod\`eles de champ moyen}

Ici, le rapport $X$ n'est pas une fonction de la corr\'elation $C(t,t_w)$ :
c'est \`a cause des deux \'echelles  $(t,t_w)$ et $(\ln t, \ln t_w)$).

D'autre part, en champ moyen le rapport $X$ appartient \`a l'intervalle $ [0,1]$ et s'interpr\`ete comme l'inverse
d'une temp\'erature effective  $X = 1/T_{eff}$ \cite{cukupeliti}.

Ici, on trouve $T_{eff} \to 0$, ce que l'on peut interpr\'eter
comme un point fixe de temp\'erature nulle. 

\subsubsection{Comparaison avec la croissance dans les ferromagn\'etiques purs}

Il est int\'eressant de comparer avec ce qui est
connu sur le r\'egime de croissance dans les ferromagn\'etiques purs :

$\bullet$ Au point critique $T=T_c$, il existe un rapport
$X(\frac{t}{t_w})$ non-trivial 
qui correspond \`a un rapport d'amplitude \cite{golutc}
Par exemple, pour la cha\^{\i}ne d'Ising pure unidimensionnelle
\`a temp\'erature nulle $T_c=0$, le rapport d'amplitude \cite{golutc} vaut $X(t,t_w)=\frac{t+t_w}{2t}$ : il
d\'ecro\^{\i}t de $X(t=t_w,t_w)=1$ \`a $X(t \to \infty,t_w)=\frac{1}{2}$.

$\bullet$ Pour $T<T_c$, le rapport $X$ est nul $X=0$
\cite{barratgrowing,berthiergrowing}, ce qui est interpr\'et\'e
de la mani\`ere suivante : les parois r\'epondent par un facteur $O(1)$
mais elles n'occupent qu'une fraction $1/L(t_w)$ du volume.

Par comparaison, on voit que pour la cha\^{\i}ne en champ al\'eatoire, 
seule une  
petite fraction $1/\Gamma_w$ des parois r\'epondent, 
mais avec une r\'eponse tr\`es grande, qui correspond au renversement de tout un domaine d'ordre $ \Gamma_w^2$.

\section{ Conclusion }

La formulation des cha\^{\i}nes de spins classiques d\'esordonn\'ees
en termes de parois de domaines qui voient un potentiel de Sinai,
permet d'\'etudier \`a la fois les propri\'et\'es d'\'equilibre thermodynamique
et la dynamique de croissances de domaines \`a partir d'une condition initiale
al\'eatoire. La possibilit\'e d'obtenir des r\'esultats exacts vient
de la tr\`es forte localisation des parois de domaines par le d\'esordre.

\subsection*{ Publications associ\'ees : }

$\bullet$ Principe de la renormalisation et premiers r\'esultats : Publication [P3] 

$ \bullet$ \'Etude d\'etaill\'ee : Publication [P7] 

$\bullet$ \'Etude de processus de r\'eaction-diffusion plus g\'en\'eraux dans un potentiel de Sinai : Publication [P5].

%%%%%%%%%%%%%%%%%%%%%%%%%%%%%%%%%%%%%%%%%%%%%%%%%%%%%

\chapter
{Localisation d'un polym\`ere al\'eatoire \`a une interface}

\label{chappolymere}

%  \begin{flushright}
%  \emph{ 
% }\\  ~\\
%  \end{flushright}

\section{ Pr\'esentation du mod\`ele }

Ce chapitre est consacr\'e \`a un mod\`ele 
introduit par Garel, Huse, Leibler et 
Orland \cite{garelpoly} : un h\'et\'eropolym\`ere 
constitu\'e de monom\`eres de charges $q_i$ al\'eatoires,
est en pr\'esence d'une interface situ\'ee en $z=0$ qui s\'epare
un milieu $z>0$ favorable aux charges positives $q>0$
et un milieu $z<0$ favorable aux charges n\'egatives $q<0$.
Plus pr\'ecis\'ement, la version continue de ce mod\`ele est 
d\'efinie par la fonction de partition suivante
sur les trajectoires Browniennes $\{z(s)\}$ \cite{garelpoly}
\begin{eqnarray} \label{functional}
Z_L(\beta;\{q(s)\}) = \int {\cal D} z(s)
\exp \left( - \frac{1}{2 D} \int_0^L ds \left( \frac{dz}{ds}\right)^2 
+ \beta \int_0^L ds q(s) {\rm sgn }(z(s)) \right)
\qquad .
\end{eqnarray}
A haute temp\'erature, des arguments de type Imry-Ma ont \'et\'e propos\'es
\cite{garelpoly} pour le cas sym\'etrique $\overline{q_i}=0$
et pour le cas biais\'e $\overline{q_i}>0$ : ce sont des arguments  \'energie/entropie, \`a la diff\'erence des arguments usuels
d'Imry-Ma qui sont de type  \'energie/ \'energie.

\subsection{ Argument de type Imry-Ma pour le cas sym\'etrique  } 

L'argument de type Imry-Ma pour le cas sym\'etrique \cite{garelpoly}
est le suivant. On suppose que la cha\^{\i}ne est localis\'ee
autour de l'interface, avec des boucles de longueur typique $l$
dans chaque solvant :

$\bullet$  L' \'energie typique gagn\'ee par boucle est d'ordre
 $ \displaystyle \sum_i^{i+l} q_i \sim \sqrt{\sigma l}$

$\bullet$ La perte d'entropie pour une boucle est d'ordre $(\ln l)$

L'optimisation de l' \'energie libre par monom\`ere
\begin{eqnarray} 
f(l) \sim - \sqrt{ \frac{  \sigma }{ l }}  +  T \frac{\ln l}{l}
\end{eqnarray}
par rapport \`a la longueur $l$ conduit \`a
\begin{eqnarray} 
\frac{l}{(\ln l)^2} \sim \frac{ T^2 } { \sigma } 
\qquad . 
\end{eqnarray}
Cet argument pr\'edit donc une localisation du polym\`ere sym\'etrique \`a toute temp\'erature, avec les comportements d'\'echelle suivants pour la longueur
typique d'une boucle et l' \'energie libre
\begin{eqnarray} 
&& l(T) \sim \frac{ T^2 (\ln T)^2} { \sigma }  \\
&& f(T) \sim  -\frac{\sigma}{T \ln T} \qquad 
\label{imrymasym}
\end{eqnarray}

\subsection{ Argument de type Imry-Ma pour le cas dissym\'etrique  }

L'argument de type Imry-Ma pour le cas dissym\'etrique \cite{garelpoly}
est en fait beaucoup plus subtil que dans le cas sym\'etrique, 
parce que la description correcte des boucles dans le solvant
 $(-)$ n\'ecessite la consid\'eration des ``\'ev\`enements
rares" o\`u la somme de $l_-$ variables al\'eatoires
$q_i$, de moyenne positive $\overline{q_i}=q_0>0$,
se trouve \^etre assez n\'egative pour rendre favorable
une excursion dans le solvant $(-)$. Plus pr\'ecis\'ement,
l'argument propos\'e \cite{garelpoly} est le suivant : 
on s'attend \`a ce que le polym\`ere soit dans le solvant pr\'ef\'er\'e $(+)$,
sauf lorsqu'une boucle de longueur $l^-$ dans le solvant $(-)$ 
devient favorable \'energ\'etiquement, avec une charge 
$Q_-=-\sum_{i=j}^{j+l^-} q_i>0$ 
suffisante. Comme la probabilit\'e d'avoir $\sum_{i=j}^{j+l^-} q_i=-Q^-$
\begin{eqnarray} 
{\rm Prob}(Q^-)= \frac{1}{\sqrt{4 \pi \sigma l^-}} 
e^{-\frac{(Q^-+q_0 l^-)^2}{4 \sigma l^-} }
\end{eqnarray}
est faible, la distance typique $l^+$
entre deux tels \'ev\`enements se comporte
 comme l'inverse de cette probabilit\'e
\begin{eqnarray} 
l^+ \sim e^{
\frac{(Q^-+ q_0 l^-)^2}{4 \sigma l^-} }
\qquad .
\end{eqnarray}
Cet argument probabiliste de type `\'ev\`enement rare',
qui est inhabituel dans les arguments de type Imry-Ma qui concernent
les \'ev\`enements typiques, donne que l' \'energie gagn\'ee $Q^-$ 
dans une boucle du solvant $(-)$
se comporte en 
\begin{eqnarray} 
Q^- \sim \sqrt{4 \sigma l^- \ln l^+ }-q_0 l^-
\qquad , 
\end{eqnarray}
alors que la perte d'entropie associ\'ee est d'ordre 
$ (\ln l^+)$.
La diff\'erence d' \'energie libre par monom\`ere entre cet \'etat localis\'e
avec des boucles $(l_+,l_-)$ par rapport \`a l'\'etat d\'elocalis\'e
dans le solvant pr\'ef\'er\'e $(+)$
est d'ordre 
\begin{eqnarray} 
f(T,l^+,l^-)-f_{deloc}(T) 
\sim \frac{1}{l^+}  \left(- Q^- + T \ln l^+ \right)
\sim \frac{1}{l^+}  \left( q_0 l^- - \sqrt{4 \sigma l^- \ln l^+ }
+ T \ln l^+ \right)
\qquad .
\end{eqnarray}
L'optimisation par rapport \`a la longueur $l^-$ donne
\begin{eqnarray} 
\label{relationlmlp}
l^- \sim \frac{\sigma}{q_0^2} \ln l^+
\end{eqnarray}
et donc finalement
\begin{eqnarray} \label{qmlog}
Q^- \sim \frac{\sigma}{q_0} \ln l^+
\qquad 
\end{eqnarray}
Ainsi, cet argument pr\'edit qu'\`a la fois l' \'energie gagn\'ee $Q^-$ 
et le co\^ut en entropie ont la m\^eme d\'ependance en $(\ln l^+)$  :
la diff\'erence d' \'energie libre se factorise selon
\begin{eqnarray}
\label{relationflp} 
f(T,l^+)-f_{deloc}(T) 
\sim (T- T_c) \frac{\ln l_+}{l_+}
\end{eqnarray}
Cet argument conduit donc finalement \`a une transition
\`a la temp\'erature critique 
 \begin{eqnarray} 
T_c \sim \frac{\sigma}{q_0}
\end{eqnarray}
entre la phase localis\'ee $T<T_c$
et la phase d\'elocalis\'ee $T>T_c$.
Contrairement au cas sym\'etrique, 
les comportements de l' \'energie libre et de la longueur $l_+$
par rapport \`a la temp\'erature 
ne sont pas d\'etermin\'es par l'argument.
L'objet de la renormalisation de type Ma-Dasgupta va \^etre justement
de trouver ces comportements en construisant explicitement les boucles
d'Imry-Ma du polym\`ere autour de l'interface en fonction de la r\'ealisation du d\'esordre.

\section{ Principe de la renormalisation }

\label{RG}

\subsection{ Construction de la structure optimale en boucles }

\begin{figure}[thb] 
\null
\vglue 1.0cm
\centerline{\epsfxsize=10.0 true cm \epsfbox{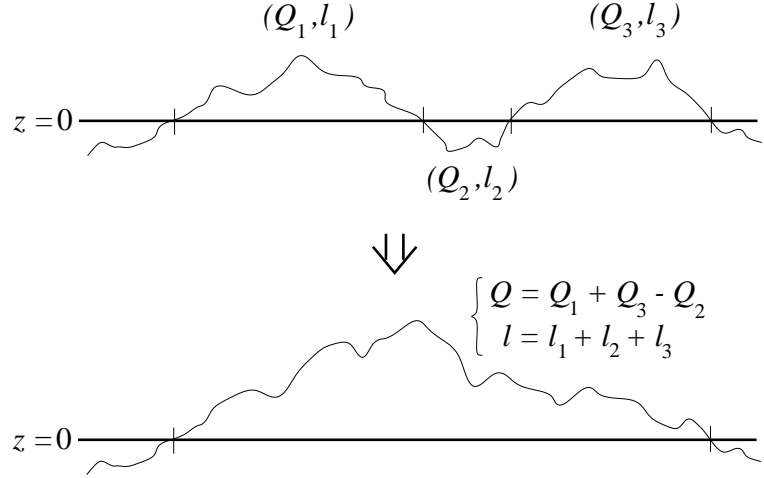}}
\vskip 1.0cm
\caption{\it
{\label{figRG} Illustration de la proc\'edure de renormalisation dans l'espace r\'eel
: le transfert d'une boucle $(Q_2=\Gamma,l_2)$ entour\'ee par les boucles voisines $(Q_1,l_1)$ et $(Q_3,l_3)$
donne naissance \`a une nouvelle boucle $(Q,l)$ avec les r\`egles de renormalisation (\ref{RGrules}). }} 
\end{figure}

A temp\'erature nulle $T=0$, chaque monom\`ere veut \^etre dans son solvant pr\'ef\'er\'e ${\rm sgn}(z_i)={\rm sgn}(q_i)$ : le polym\`ere est d\'ecompos\'e
en boucles $\alpha$ contenant $l_{\alpha}$ monom\`eres cons\'ecutifs de m\^eme signe, et portant une certaine charge absolue $Q_{\alpha}$.
Lorsque la temp\'erature $T$ croit, on consid\`ere les configurations 
de la cha\^{\i}ne qui peuvent \^etre obtenues \`a partir de la structure de l'\'etat fondamental en transf\'erant dans le solvant oppos\'e
les boucles de plus petite 
charge absolue $Q_{min} \equiv \Gamma$ de mani\`ere it\'erative.
Quand on transf\`ere une boucle $(Q_2=\Gamma,l_2)$ 
qui est entour\'ee de deux boucles $(Q_1,l_1)$ et $(Q_3,l_3)$, 
on obtient une nouvelle boucle de charge absolue $Q$
et de longueur $l$ donn\'ees par les r\`egles (cf Figure \ref{figRG})
\begin{eqnarray}
&& Q=Q_1+Q_3-Q_2  \\
&& l=l_1+l_2+l_3 \nonumber
\label{RGrules}
\end{eqnarray}

Ces r\`egles correspondent aux r\`egles de renormalisation
des extrema Browniens d\'ecrites dans le Chapitre \ref{chapreglesrg} : 
ici la marche al\'eatoire correspondante est la somme des charges
$\sum_{0}^i q_j$ en fonction du monom\`ere final $i$,
et la proc\'edure de renormalisation construit la meilleure
structure en boucles avec la contrainte qu'on n'autorise
que les boucles de charge absolue plus grande que $\Gamma$.

Pour \'etablir la correspondance entre l'\'echelle de renormalisation $\Gamma$ et la temp\'erature $T$,
il reste \`a \'etudier dans quelles conditions le transfert d'une boucle
 $(Q_2,l_2)$ dans le solvant oppos\'e est effectivement favorable pour l' \'energie libre :
le co\^ut en  \'energie est
\begin{eqnarray}
\Delta E^{flip} = 2 Q_2 
\end{eqnarray}
alors que le gain en entropie est
\begin{eqnarray} \label{dsflip}
\Delta S^{flip} = \ln ({\cal M} (l_1+l_2+l_3)) 
-  \ln [{\cal M} (l_1) {\cal M} (l_2) {\cal M} (l_3) ]
\end{eqnarray}
o\`u ${\cal M} (l)=c^l/l^{3/2}$ repr\'esente le nombre de marches al\'eatoires
de $l$ pas qui vont de $z=0$ \`a $z=0$ en pr\'esence d'une fronti\`ere absorbante en $z=0^-$, ce qui conduit au bilan d' \'energie libre
\begin{eqnarray}
\Delta F^{flip} = \Delta E^{flip} - T \Delta S^{flip}
= 2 Q_2 -  T \ln \left(\frac{ {\cal M} (l_1+l_2+l_3) }
{{\cal M} (l_1) {\cal M} (l_2) {\cal M} (l_3)} \right)
\label{flipcondition}
\end{eqnarray}
Pour obtenir la structure optimale \`a temp\'erature $T$, il faut donc
poursuivre la renormalisation tant qu'elle permet d'abaisser 
l' \'energie libre ($\Delta F^{flip} <0$), et il faut s'arr\^eter
\`a la premi\`ere it\'eration qui conduirait \`a une augmentation
($\Delta F^{flip} >0$).
On d\'efinit $\Gamma_{eq}(T)$ comme l'\'echelle de renormalisation $\Gamma$ 
o\`u il faut arr\^eter la renormalisation.

\subsection{ Validit\'e de la m\'ethode }

La m\'ethode de renormalisation est justifi\'ee si l'\'echelle
de renormalisation $\Gamma_{eq}(T)$ est grande, 
ce qui correspond \`a la r\'egion des hautes temp\'eratures.

\section{R\'esultats pour le cas sym\'etrique}

\label{summary}

\subsection {\'Energie libre }

L' \'energie libre trouv\'ee
\begin{eqnarray}
 f(T) \sim - \frac{ \sigma}{  T \ln T} 
\end{eqnarray}
est en accord avec l'argument de Imry-Ma (\ref{imrymasym}).

\subsection {Distribution des longueurs de boucles }

Les longueurs de boucles sont des variables al\'eatoires ind\'ependantes, et la variable d'\'echelle
\begin{eqnarray}
 \lambda=\frac{ \sigma l}{ 9 T^2 (\ln T)^2}
\end{eqnarray}
est distribu\'ee avec la loi 
\begin{eqnarray}  
  P(\lambda)  
&& = \sum_{n = -\infty}^{\infty} \left(n+\frac{1}{2}\right)
\pi (-1)^n e^{- \pi^2 \lambda \left(n+\frac{1}{2}\right)^2} 
\opsimeq_{\lambda \to \infty} 
\pi  e^{- \frac{\pi^2}{4} \lambda } \\
&& = \frac{1}{\sqrt \pi \lambda^{3/2}}
\sum_{m = -\infty}^{\infty} (-1)^m (m+\frac{1}{2})
 e^{-  \frac{1}{\lambda} (m+\frac{1}{2})^2}
 \opsimeq_{\lambda \to 0} 
 \frac{1}{\sqrt \pi \lambda^{3/2}}
 e^{-  \frac{1}{ 4 \lambda} }
\end{eqnarray}

\subsection {Densit\'e de polym\`ere autour de l'interface }

La variable d'\'echelle pour la distance $z$ \`a l'interface
\begin{eqnarray}
Z=  \sqrt{\frac{2 \sigma }{D}} \frac{z}{ 3 T (\ln T)}
\end{eqnarray}
est distribu\'ee avec la loi
\begin{eqnarray}
 R(Z)= 4 \int_0^{\infty} d \lambda P(\lambda) \sqrt{\lambda}
\int_{\frac{Z}{\sqrt{\lambda}}}^{\infty} du e^{- u^2}
 \opsimeq_{Z \to \infty}  \frac{8}{\pi} \sqrt{2 Z} e^{-\pi Z} 
\end{eqnarray}

\subsection { Temp\'erature de d\'elocalisation d'une cha\^{\i}ne finie }

Alors qu'une cha\^{\i}ne infinie reste localis\'ee \`a toute temp\'erature,
une cha\^{\i}ne finie de taille $L$ a une dissym\'etrie d'ordre $\sqrt{L}$
pour sa charge totale et subit donc une d\'elocalisation \`a une temp\'erature finie. Pour caract\'eriser ces effets de taille finie, on peut calculer
la distribution de la temp\'erature de d\'elocalisation $T_{\rm deloc}$ 
sur l'ensemble des cha\^{\i}nes cycliques de taille finie grande $L$ :
la variable d'\'echelle 
\begin{eqnarray}
g=\frac{ 3 }  {\sqrt{\sigma L}} T_{\rm deloc} \ln T_{\rm deloc}
\end{eqnarray}
a pour distribution
\begin{eqnarray}  
D(g)   
&& = \frac{\pi^2}{g^3}\sum_{n=1}^{+\infty} (-1)^{n+1} 
n^2 e^{-   \frac{n^2 \pi^2}{4 g^2}} 
\opsimeq_{g \to 0} 
\frac{\pi^2}{g^3}  e^{-   \frac{ \pi^2}{4 g^2}}\\
&& = \frac{2}{\sqrt{\pi }} 
\sum_{m=-\infty}^{+\infty} 
\left[ 2 (2m+1)^2 g^2 -1 \right] e^{-  (2m+1)^2 g^2}
\opsimeq_{g \to \infty} 
\frac{4}{\sqrt{\pi }} g^2  e^{-   g^2}
\end{eqnarray}

\section{R\'esultats pour le cas dissym\'etrique }

\subsection{Temp\'erature critique }

Dans le cas dissym\'etrique ($q_0 >0$), 
dans la limite $\sigma \gg q_0$,
la temp\'erature critique obtenue par renormalisation
\begin{eqnarray}  
T_c=\frac{4 \sigma }{ 3 q_0}
\end{eqnarray}
est dans le domaine des hautes temp\'eratures, et donc dans le domaine de validit\'e de la m\'ethode.

\subsection{ Singularit\'e essentielle de l' \'energie libre }

La renormalisation conduit \`a la
singularit\'e essentielle suivante pour l' \'energie libre
\begin{eqnarray}  
f(T) -f(T_c) \opsimeq_{T \to T_c^-} 
- 2 q_0 \left(\ln \frac{4 \sigma}{q_0}  \right) 
\exp \left[ - \frac{ \ln \frac{4 \sigma}{q_0} }
{ \left(1-\frac{T}{T_c} \right)} \right]
\end{eqnarray}
ce qui correspond donc \`a une transition d'ordre infinie.

\subsection{ Propri\'et\'es des boucles dans les deux solvants }

La longueur typique $l^{+}_{ blob}(T)$ des boucles dans le solvant pr\'ef\'er\'e
diverge avec une singularit\'e essentielle \`a la transition,
alors que la longueur typique $l^{-}_{ blob}(T)$ des boucles dans le solvant d\'efavorable
diverge alg\'ebriquement
\begin{eqnarray}
&& l^{+}_{ blob}(T) \opsimeq_{T \to T_c^-} 
\frac{\sigma}{q_0^2}  \exp \left[ + \frac{ \ln \frac{4 \sigma}{q_0} }
{ \left( 1-\frac{T}{T_c} \right)} \right] \\
&& l^{-}_{blob}(T) \opsimeq_{T \to T_c^-} 
\frac{\sigma}{q_0^2} \frac{ \ln \frac{4 \sigma}{q_0} }
{\left(1-\frac{T}{T_c} \right)} 
\end{eqnarray}
La variable d'\'echelle $\lambda_+=l_+/ l^{+}_{ blob}(T)$ 
pour les boucles dans le solvant pr\'ef\'er\'e
est distribu\'ee avec la distribution exponentielle $e^{-\lambda_+}$.

\subsection{ Densit\'e de polym\`ere autour de l'interface }

La variable d'\'echelle pour la distance $z$ \`a l'interface dans le solvant $(+)$
\begin{eqnarray}
 Z=  \frac{  z}{ \sqrt{ \frac{ D}{2} } \frac{\sigma}{q_0} 
\exp \left[ + \frac{ \ln \frac{4 \sigma}{q_0} }
{ 2 \left( 1-\frac{T}{T_c} \right)} \right] }
\qquad 
\end{eqnarray}
a pour distribution
\begin{eqnarray}
R^+(Z)= 4 \int_0^{\infty} du e^{- u^2}
\int_{\frac{Z}{u}}^{\infty} dv v^2 e^{- v^2} \opsimeq_{Z \to \infty}
 \sqrt{\pi Z} e^{-2 Z}
\end{eqnarray}

\subsection{ Temp\'erature de d\'elocalisation d'une cha\^{\i}ne finie }

Pour caract\'eriser les effets de taille finie,
on peut calculer la distribution de la temp\'erature de d\'elocalisation
 $T_{\rm deloc}$ 
sur l'ensemble des cha\^{\i}nes cycliques de taille $L$ :
la variable al\'eatoire 
\begin{eqnarray}  
r=\frac{\sigma}{q_0^2 L} 
\left( \frac{4 \sigma}{q_0^2 } \right)^{\frac{T_{\rm deloc}}{T_c-T_{\rm deloc}}}
\end{eqnarray}
est distribu\'ee selon la loi
\begin{eqnarray}  
D^+(r)= \frac{1}{r^2} e^{-\frac{1}{r}} 
\qquad .
\end{eqnarray}
En particulier, la valeur typique pour la temp\'erature 
de d\'elocalisation d'une cha\^{\i}ne de taille $L$ pr\'esente
une correction logarithmique d'ordre $(1/\ln L)$ 
par rapport \`a la temp\'erature critique $T_c$ de la limite thermodynamique
\begin{eqnarray}  
 T_{\rm deloc}^{typ} \sim T_c \left(1-\frac{4 \sigma }{q_0^2 \ln L} \right)
\end{eqnarray}

\section{ Conclusion }

La renormalisation de type Ma-Dasgupta permet d'\'etudier
la localisation d'un h\'et\'eropolym\`ere al\'eatoire autour d'une interface 
\`a haute temp\'erature,
en construisant explicitement la succession des boucles
dans chaque \'echantillon. 
Ces boucles sont des domaines d'Imry-Ma g\'en\'eralis\'es de type  \'energie/entropie. Dans le cas dissym\'etrique, la comp\'etition entre 
les \'energies al\'eatoires gagn\'ees dans les boucles et les entropies
perdues lors des retours \`a l'interface 
 conduit \`a une transition de d\'elocalisation 
caract\'eris\'ee par des singularit\'es essentielles.

\subsection*{  Publication associ\'ee  [P6] }

\chapter
{Potentiel Brownien en pr\'esence d'un confinement quadratique}

\label{chaplandscape}

%  \begin{flushright}
%  \emph{ 
% }\\  ~\\
%  \end{flushright}

\section{Pr\'esentation du mod\`ele}

Le potentiel al\'eatoire unidimensionnel 
\begin{eqnarray}
U_{toy}(x)=\frac{\mu}{2} x^2 + V(x)
\label{deftoy}
\end{eqnarray}
contenant un terme quadratique d\'eterministe
et un terme al\'eatoire Brownien $V(x)$ 
\begin{eqnarray}
\overline{ \left(V(x)-V(y) \right)^2 }= 2  \vert x-y \vert
\label{corre}
\end{eqnarray}
a \'et\'e introduit par Villain {\it et al.}  
comme un `mod\`ele jouet' pour les interfaces en pr\'esence de champ al\'eatoire \cite{villain83}. Par ailleurs, 
dans le cadre de l'\'etude des vari\'et\'es de dimension interne $D$
plong\'ees dans un milieu al\'eatoire de dimension $(N+D)$,
il est consid\'er\'e comme le cas extr\^eme le plus simple $D=0$ et $N=1$ \cite{mezardparisi92}.

Ce mod\`ele simple permet en effet de mieux comprendre
certaines propri\'et\'es de syst\`emes d\'esordonn\'es plus compliqu\'es.

\subsection{ Argument de Imry-Ma }

A cause du confinement quadratique, le minimum absolu du potentiel est fini
et un argument de Imry-Ma, entre l' \'energie \'elastique d'ordre $\mu x^2$
et l' \'energie al\'eatoire d'ordre $\sqrt{x}$, permet de d\'eterminer son ordre de grandeur
\begin{eqnarray} 
x_{min} \sim  \mu^{-2/3} 
\label{scaling1}
\end{eqnarray}
alors que les m\'ethodes usuelles de perturbation
\`a tous les ordres \cite{villain88} ou d'it\'eration \cite{villainsemeria}
n'arrivent pas du tout \`a reproduire ce scaling (\ref{scaling1}).
Dans le cadre de la m\'ethode variationnelle des r\'epliques,
le r\'esultat (\ref{scaling1}) n\'ecessite une brisure de sym\'etrie des r\'epliques \cite{mezardparisi92,engel}.

\subsection{ Sym\'etrie statistique de `tilt' }

Ce mod\`ele pr\'esente aussi une sym\'etrie statistique de `tilt'
( `statistical tilt symmetry' en anglais)
comme les autres mod\`eles en champs al\'eatoires,
qui implique des identit\'es remarquables \cite{identities88} 
pour les moyennes thermiques des cumulants thermiques
de la position, que l'on peut r\'esumer par \cite{identities88}
\begin{eqnarray}
 \overline { \ln < e^{- \lambda x } > } && = T \frac{\lambda^2}{2  \mu}
 \label{genlambda2}
\end{eqnarray}
Cette identit\'e sur la fonction g\'en\'eratrice montre
que le second cumulant est simplement
\begin{eqnarray}
&& \overline { < x^2 > - <x>^2 } = \frac{T}{\mu } \label{second}
\end{eqnarray}
et que toutes les moyennes sur le d\'esordre des cumulants sup\'erieurs
s'annulent! Dans le r\'egime de basse temp\'erature, le r\'esultat
(\ref{second}) implique que les fluctuations thermiques
sont reli\'ees \`a la pr\'esence d'\'etats m\'etastables 
dans des \'echantillons rares \cite{identities88}.
La renormalisation permet en particulier d'\'etudier quantitativement ce ph\'enom\`ene.

\section{ Principe de la renormalisation }

Pour un potentiel al\'eatoire $U(x)$ `Markovien', 
satisfaisant une \'equation de Langevin locale de type
\begin{eqnarray}
\frac{dU(x)}{dx} = F[U(x),x] + \eta(x)
%\label{langevin} 
\end{eqnarray}
o\`u $\eta(x)$ est un bruit blanc Gaussien,
la mesure du paysage renormalis\'e se factorise en blocs
qui satisfont des \'equations de renormalisation ferm\'ees
(Publication [P9]).
Pour le cas de paysages stationnaires, o\`u la force $F$
est ind\'ependante de $x$
\begin{eqnarray}
F[U,x] = F[U] = - \frac{d W[U]}{d U}
\label{defstatio}
\end{eqnarray}
(ce qui g\'en\'eralise le paysage Brownien pur $F=0$
et le paysage Brownien biais\'e $F[U]=F>0$),
et pour le mod\`ele (\ref{deftoy}) qui correspond \`a
une force ind\'ependante de $U$ et lin\'eaire en $x$
\begin{eqnarray}
F[U(x),x] = F[x] = \mu x 
\label{forcetoy}
\end{eqnarray}
on obtient des solutions explicites pour la renormalisation
(Publication [P9]).

En particulier, pour le mod\`ele (\ref{deftoy}) qui nous int\'eresse
dans ce chapitre, la mesure du paysage renormalis\'e
s'exprime en terme des fonctions d'Airy.

\section{ R\'esultats sur la statistique des minima }

\subsection{Position du minimum} 

\label{equilibriumtoy}

A temp\'erature nulle, la particule se trouve au minimum $x_{min}$
du potentiel. L'\'etat final $\Gamma=\infty$ de la renormalisation permet de retrouver que la distribution de $x_{min}$ sur l'ensemble des \'echantillons
est, en accord avec \cite{groeneboom,frachebourgmartin}
\begin{eqnarray}
&& P_{]-\infty,+\infty[}(x) = g(x) g(-x) 
\label{distriminabs}
\end{eqnarray}
et la fonction auxiliaire
\begin{eqnarray}
 g(x) = \int_{-\infty}^{+\infty} \frac{d\lambda}{2 \pi} 
\frac{e^{ - i \lambda x }}{a Ai(b i \lambda)}
\label{defg}
\end{eqnarray}
en utilisant les notations $a=(\mu/2)^{1/3}$ et $b=1/a^2$,

\subsection{ \'Echantillons avec deux minima presque d\'eg\'en\'er\'es}

La probabilit\'e qu'un \'echantillon pr\'esente deux minima
presque d\'eg\'en\'er\'es avec une diff\'erence d' \'energies
 $\Delta E=\epsilon \to 0$
situ\'es en $x_1$ et $x_2$ s'\'ecrit 
\begin{eqnarray}
{\cal D}(\epsilon,x_1,x_2) =  \epsilon g(-x_1) d(x_2-x_1)g(x_2) +O(\epsilon^2)
\label{restwomin}
\end{eqnarray}
en termes de la fonction $g$ (\ref{defg})
et de la fonction
\begin{eqnarray}
d(y) =a  \int_{-\infty}^{+\infty} \frac{d \lambda}{2 \pi}
 e^{i \lambda y }
 \frac{ Ai'(i b \lambda)}{ Ai(i b\lambda)}
\end{eqnarray}

La probabilit\'e d'avoir deux minima s\'epar\'es par une distance $y>0$ 
est donc
\begin{eqnarray}
D(y) && = \int_{-\infty}^{+\infty} dx_1 
\lim_{\epsilon \to 0} \left( \frac{ {\cal D}(\epsilon,x_1,x_1+y)}{\epsilon}
\right)    = b d(y) \int_{-\infty}^{+\infty} \frac{d\lambda}{2 \pi}
\frac{e^{-i \lambda y} }{  Ai^2( i b \lambda)} 
\label{deffDy}
\end{eqnarray}
En particulier, le calcul du second moment donne
\begin{eqnarray}
\int_0^{+\infty} dy y^2 D(y) 
= \frac{1}{\mu}
\label{y2toy}
\end{eqnarray}

\subsection{Contribution au second cumulant \`a basse temp\'erature}

La contribution des \'echantillons avec deux minima d\'eg\'en\'er\'es
au second cumulant thermique de la position (\ref{second})
peut \^etre estim\'e \`a l'ordre $T$ en temp\'erature
comme suit : 
les deux minima ont pour poids de Boltzmann respectifs 
$ p =\frac{ 1 } { 1 + e^{ - \beta \epsilon} }$ and
 $(1-p)=\frac{  e^{ - \beta \epsilon} } { 1 + e^{ - \beta \epsilon} }$.
La variable $(x-<x>)$ est donc $(1-p)(x_1-x_2)$ avec probabilit\'e $p$ et $p (x_2-x_1)$ avec probabilit\'e $1-p$, et donc apr\`es moyenne
on obtient
\begin{eqnarray}
&& \overline{ < (x-<x>)^2 > }  = \overline{ p(1-p) (x_1-x_2)^2 } \\
&&  = \int_{-\infty}^{+\infty} dx_1 \int_{x_1}^{+\infty} dx_2  
\int_{-\infty}^{+\infty} d\epsilon {\cal D}'(\epsilon=0,x_1,x_2) \frac{e^{- \epsilon/T}}{(1 + e^{- \epsilon/T})^2}   (x_2-x_1)^2  \\
&& = T \int_0^{+\infty} dy   D(y)  y^2  
\label{two}
\end{eqnarray}
En utilisant (\ref{y2toy}), on obtient alors exactement le r\'esultat exact (\ref{second}). Ceci montre que les fluctuations thermiques \`a basse temp\'erature viennent enti\`erement des \'etats m\'etastables 
qui existent dans quelques \'echantillons rares.
En particulier, la susceptibilit\'e
\begin{eqnarray}
\chi \equiv \frac{1}{T} \left( <x^2>-<x>^2 \right)
\label{suscepti}
\end{eqnarray}
a une moyenne finie \`a temp\'erature nulle
\begin{eqnarray}
\overline{ \chi } = \frac{1}{\mu}
\end{eqnarray}
mais seuls les \'echantillons avec deux minima d\'eg\'en\'er\'es
contribuent \`a cette valeur moyenne, car les \'echantillons typiques avec un seul minimum ont une susceptibilit\'e qui s'annule
\`a temp\'erature nulle. 

\subsection{Contribution aux moments thermiques \`a basse temp\'erature}

De m\^eme, les moments pairs de la position relative
 $(x-<x>)$ se comportent de la mani\`ere suivante \`a basse temp\'erature
\begin{eqnarray}
 \overline {  < (x-<x>)^{2n} > } && = \frac{T}{n} \int_0^{+\infty} y^{2n} D(y) +O(T^2)
 \label{momentslowT}
\end{eqnarray}
en terme de la fonction $D(y)$ d\'efinie en (\ref{deffDy}).
La comparaison avec l'identit\'e (\ref{genlambda2}) 
montre qu'il y a beaucoup de termes qui se compensent
dans les moyennes sur le d\'esordre des cumulants.

\section{ R\'esultats sur la statistique de la plus grande barri\`ere }

Le temps $t_{eq}$ n\'ecessaire pour atteindre l'\'equilibre
est directement reli\'e \`a la plus grande barri\`ere $\Gamma_{max}=T \ln t_{eq}$ qui existe dans l'\'echantillon.
Plus pr\'ecis\'ement, la probabilit\'e ${\cal P}(t_{eq}<t)$
que le syst\`eme ait d\'ej\`a atteint l'\'equilibre \`a l'instant $t$, 
correspond \`a la probabilit\'e qu'il ne reste plus qu'une vall\'ee
renormalis\'ee \`a l'\'echelle $\Gamma = T \ln t$ : c'est une fonction
de la variable d'\'echelle $\gamma$
\begin{eqnarray}
{\cal P}(t_{eq}<t)  = \Phi \left(\gamma \equiv \left( \frac{\mu}{2} \right)^{1/3} T \ln t \right)
 \label{ptequi}
\end{eqnarray}
qui a pour expression explicite en termes des fonctions d'Airy
\begin{eqnarray}
\Phi \left(\gamma \right)
&&  = \int_{-\infty}^{+\infty} \frac{d\lambda}{2 \pi}
\frac{1}{ Ai^2(i  \lambda) } e^{- 2 \int_{0}^{+\infty}
 df \tilde{\psi}_{\gamma} (f,\lambda)} 
\\
 \tilde{\psi}_{\gamma}(f,\lambda) && = \frac{  Ai(f+\gamma +  i \lambda) }
{\pi Ai(f +  i \lambda) 
\left[ Ai(f +  i \lambda) Bi(f+\gamma +  i \lambda)
 -Bi(f +  i \lambda) 
Ai(f+\gamma +  i \lambda) \right] }
\nonumber
\end{eqnarray}

Ce r\'esultat correspond directement par d\'erivation 
\`a la distribution de probabilit\'e de la variable
r\'eduite $\gamma=  \left( \frac{\mu}{2} \right)^{1/3} \Gamma_{max}$
de la plus grande barri\`ere $\Gamma_{max}$ de l'\'echantillon
\begin{eqnarray}
P_{max}(\gamma) = 
 \frac{d}{d \gamma} \Phi \left(  \gamma \right)
\label{distribarrmax}
\end{eqnarray}

Les comportements asymptotiques de cette distribution sont les suivants
\begin{eqnarray}
P_{max}(\gamma) && \opsimeq_{\gamma \to \infty} \frac{9}{4}   \sqrt{\frac{\pi}{2}}
 \gamma^{5/4} e^{- \frac{3}{2}  \gamma^{3/2}}
\label{distribarrmaxlarge} \\
P_{max}(\gamma) && \opsimeq_{\gamma \to 0} \frac{6 \zeta(3)}{   \gamma^4 }
e^{- 2 \frac{\zeta(3)}{   \gamma^3 } }
\label{distribarrmaxsmall}
\end{eqnarray}
o\`u $\zeta(n)$ d\'esigne la fonction zeta de Riemann.

\section{Conclusion}

Pour le potentiel Brownien avec confinement quadratique,
la solution explicite du paysage renormalis\'e en termes de
fonctions d'Airy permet d'\'etudier la statistique des minima
et des barri\`eres. En particulier, ceci permet de montrer
que les fluctuations thermiques \`a basse temp\'erature sont enti\`erement
gouvern\'es par les \'echantillons rares qui poss\`edent deux minima d\'eg\'en\'er\'es.

\subsection*{ Publication associ\'ee   [P9] }

%%%%%%%%%%%%%%%%%%%%%%%%%%%%%%%%%%%%%%%%%%%%%%%%%%%%%%%%%%%%

\chapter{Le mod\`ele de pi\`eges unidimensionnel  }

\label{chaptrap}

%  \begin{flushright}
%  \emph{ 
% }\\  ~\\
%  \end{flushright}

\section{ Pr\'esentation des mod\`eles de pi\`eges }

Les mod\`eles de pi\`eges proposent un m\'ecanisme tr\`es simple de vieillissement \cite{jpbtraps} : une particule effectue une dynamique
stochastique dans un paysage contenant des pi\`eges
d' \'energies al\'eatoires de distribution exponentielle
\begin{eqnarray}
\rho(E)= \theta(E) \frac{1}{T_g} e^{- \frac{E}{T_g}} 
\label{rhoe} 
\end{eqnarray}
Ce choix de distribution exponentielle est justifi\'e
par les \'etats de basse  \'energie du REM \cite{rem},
la th\'eorie des r\'epliques \cite{replica}, 
et plus g\'en\'eralement par la queue exponentielle
 de la distribution de Gumbel qui repr\'esente une classe d'universalit\'e importante dans la statistique des extr\^emes.
Cette distribution exponentielle des  \'energies (\ref{rhoe}) correspond
pour le temps de pi\'egeage d'Arrh\'enius $\tau = e^{\beta E}$
\`a la loi alg\'ebrique
\begin{eqnarray}
q(\tau)
= \theta(\tau>1)  \frac{\mu}{\tau^{1+\mu}} 
 \label{qtau}
\end{eqnarray}
avec l'exposant 
\begin{eqnarray}
\mu = \frac{T}{T_g}
\label{defmu}
\end{eqnarray}
A basse temp\'erature $T<T_g$, le temps moyen de pi\'egeage $\int d \tau 
\tau q(\tau)$ diverge, ce qui conduit imm\'ediatement \`a des effets de vieillissement.

Les propri\'et\'es de vieillissement ont \'et\'e beaucoup \'etudi\'ees,  
soit dans la version de champ moyen  
\cite{jpbtraps,benarousreview,fielding,junier}, soit dans la version unidimensionnelle \cite{isopi,bertinjp}, qui appara\^{\i}t
naturellement dans divers contextes physiques
 \cite{alexander,jpbreview,bubbledna}, et qui pr\'esente deux \'echelles de temps caract\'eristiques pour le vieillissement. 
L'objet de ce chapitre est de comprendre ce ph\'enom\`ene gr\^ace \`a une renormalisation appropri\'ee.

\section{ Principe de la renormalisation }

Nous avons d\'ej\`a d\'ecrit la renormalisation adapt\'ee au mod\`ele {\it dirig\'e} de pi\`eges, dans le Chapitre \ref{chapsinaibiais}.
Ici, dans le mod\`ele {\it non-dirig\'e }, chaque site peut \^etre visit\'e 
un grand nombre de fois, ce qui conduit \`a un changement essentiel :
un pi\`ege du paysage renormalis\'e va \^etre caract\'eris\'e par deux
temps importants, \`a savoir 

(i) son temps de pi\'egeage $\tau_i$, qui repr\'esente le temps typique
de sortie vers ses voisins imm\'ediats

(ii) son temps d'\'evasion $T_i$, qui repr\'esente le temps n\'ecessaire
pour arriver dans un pi\`ege plus profond. 

\begin{figure}[ht]

\centerline{\includegraphics[height=8cm]{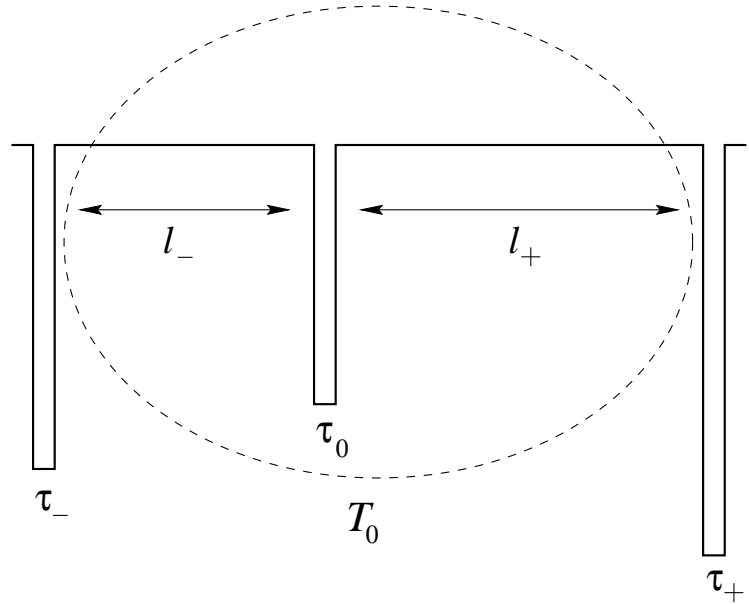}} 
\caption{\it D\'efinition du temps d'\'evasion dans le paysage renormalis\'e :
le pi\`ege de temps de pi\'egeage $\tau_0$ est entour\'e par deux
pi\`eges, le pi\`ege $\tau_+$ \`a la distance $l_+$ et le
pi\`ege $\tau_-$ \`a la distance $l_-$. Le temps d'\'evasion $T_0$ est le temps caract\'eristique n\'ecessaire pour atteindre $\tau_+$
ou $\tau_-$ quand on part de $\tau_0$.   } 
\label{defescape}
\end{figure}

Plus pr\'ecis\'ement, le paysage renormalis\'e \`a l'\'echelle $R$ est d\'efini comme suit : tous les pi\`eges $\tau_i<R$ sont remplac\'es par un paysage `plat', alors tous les pi\`eges $\tau_i>R$ sont inchang\'es.
Dans le paysage renormalis\'e (cf Figure \ref{defescape}), quand la particule quitte le pi\`ege $\tau_0$, elle s'\'echappe vers la droite ou vers la gauche avec probabilit\'e $1/2$. Si elle s'\'echappe vers la gauche, elle va arriver
\`a atteindre le pi\`ege suivant $\tau_-$ avec probabilit\'e $1/l_-$.
Si elle s'\'echappe vers la droite, elle va arriver
\`a atteindre le pi\`ege suivant $\tau_+$ avec probabilit\'e $1/l_+$.
Sinon, elle va \^etre r\'eabsorb\'ee par le pi\`ege $\tau_0$.
Asymptotiquement, le nombre de retours dans le pi\`ege $\tau_0$
avant l'\'evasion va donc \^etre grand, d'ordre $R^{\mu}$. 
On peut montrer que le temps pass\'e \`a l'int\'erieur de $\tau_0$
pendant ces multiples visites est dominant par rapport au temps pass\'e
dans les multiples excursions infructueuses et dans l'excursion
r\'eussie pour s'\'echapper. Finalement, on obtient que
le temps d'\'evasion $t_{esc}$ du pi\`ege $\tau_0$ a une distribution exponentielle
\begin{eqnarray}
P(t_{esc}) 
\opsimeq_{R \to \infty}  \frac{1}{T_0} e^{- \displaystyle \frac{t_{esc}}{T_0} }
\label{pin}
\end{eqnarray}
avec le temps caract\'eristique
\begin{eqnarray}
T_0 = \frac{2}{ \frac{1}{l_+} + \frac{1}{l_-}}  \tau_0
\end{eqnarray}
ce qui explique pourquoi il y a deux \'echelles de temps importantes.
En particulier, pour d\'ecrire la dynamique au temps $t$, 
on souhaite garder les pi\`eges de temps d'\'evasion
$T>t$ dont la particule n'a pas eu le temps
de s'\'echapper, et \'eliminer au contraire les pi\`eges de temps d'\'evasion
$T<t$, ce qui conduit au choix suivant pour l'\'echelle de renormalisation $R(t)$ du paysage en fonction de temps
\begin{eqnarray}
  R(t) \simeq  t^{\frac{1}{1+\mu}}
\end{eqnarray} 

De plus, dans la limite $\mu \to 0$, la dynamique effective suivante devient exacte : \`a l'instant $t$, la particule partie de l'origine se trouve
soit sur le premier pi\`ege renormalis\'e $M_+$ \`a distance $l_+$ sur la droite, soit sur le premier pi\`ege renormalis\'e $M_-$ \`a distance $l_-$ sur la gauche. Le poids de $M_+$ est simplement la probabilit\'e $l_-/(l_+ + l_-)$
d'avoir atteint $M_+$ avant $M_-$
\begin{eqnarray}
 P_{eff}(x,t) \sim  \frac{l_+}{l_+ + l_-} \delta(x+l_-) + \frac{l_-}{l_+ + l_-} \delta(x-l_+)
\label{pefftrap}
\end{eqnarray}

\section{ R\'esultats   }

\subsection{ Distance entre pi\`eges dans le paysage renormalis\'e}

Les distances entre pi\`eges successifs dans le paysage renormalis\'e
sont des variables al\'eatoires ind\'ependantes et la variable
r\'eduite $\lambda=l/\xi(t)$ est distribu\'ee selon la loi exponentielle
 \begin{eqnarray}
P(\lambda)=e^{-\lambda}  
\end{eqnarray}
L'\'echelle de longueur caract\'eristique du paysage renormalis\'e \`a l'instant
$t$ est
 \begin{eqnarray}
 \xi(t)   
 = \xi_0(\mu) \ t^{\frac{\mu}{1+\mu}}  
\end{eqnarray}
avec le pr\'efacteur
 \begin{eqnarray}
 \xi_0(\mu) =1+O(\mu)  
\end{eqnarray}

\subsection{ Forme d'\'echelle du front de diffusion }

Dans la variable d'\'echelle $X=x/\xi(t)$, la moyenne sur les 
\'echantillons du front de diffusion (\ref{pefftrap}) est
\begin{eqnarray}
g_{\mu}(X) = e^{- \vert X \vert}  \int_0^{+\infty} du
e^{-u} \frac{u}{\vert X \vert +u} +O(\mu) 
\end{eqnarray}

\subsection{ Param\`etres de localisation }

Les param\`etres de localisation, qui repr\'esentent les moyennes sur les
\'echantillons des probabilit\'es que $k$ particules ind\'ependantes se 
trouvent au m\^eme site, sont donn\'es par
\begin{eqnarray}
Y_k (\mu) \equiv \lim_{t \to \infty} \overline{  \sum_{n=0}^{+\infty} 
P^k(n,t\vert 0,0) } = 
   \frac{2}{(k+1)} +O(\mu)
\label{resyksummary}
\end{eqnarray}
ce qui est en accord avec les simulations num\'eriques
de Bertin et Bouchaud \cite{bertinjp}
qui ont obtenu $Y_2 \to 2/3$ et $Y_3 \to 1/2$ 
dans la limite $\mu \to 0$.

\subsection{ Fonction g\'en\'eratrice des cumulants thermiques }

La largeur thermique est
\begin{eqnarray}
c_2(\mu)  \equiv \lim_{t \to \infty} \overline{\frac{<n^2>-<n>^2}
{\xi^2(t)} } =   1+O(\mu)
\end{eqnarray}
et plus g\'en\'eralement, les autres cumulants thermiques 
peuvent \^etre d\'eriv\'es de la fonction g\'en\'eratrice  
\begin{eqnarray}
Z_{\mu}(s) \equiv \overline{ \ln < e^{-s \frac{n}{\xi(t)} }>}
 = \int_0^{+\infty} d \lambda e^{- \lambda} \lambda 
\left( \frac{s \lambda}{2} \coth \frac{s \lambda}{2}-1 \right)
+O(\mu)
 \end{eqnarray}

\subsection{ Fonction de corr\'elation \`a deux particules }

La fonction de corr\'elation \`a deux particules prend la forme
\begin{eqnarray}
C(l,t) && \equiv \overline{  \sum_{n=0}^{+\infty} \sum_{m=0}^{+\infty}
P(n,t\vert 0,0) P(m,t\vert 0,0) \delta_{l,\vert n-m \vert} }  \opsimeq_{t \to \infty} Y_2(\mu) \delta_{l,0}
+ \frac{1}{ \xi(t) } {\cal C}_{\mu} \left( \frac{l}{\xi(t)} \right)
\nonumber \label{correform}
 \end{eqnarray}
Le poids de la fonction $\delta$ \`a l'origine correspond
comme il se doit au param\`etre de localisation  $Y_2
=2/3 +O(\mu)$ (\ref{resyksummary}),
alors que la seconde partie est une fonction d'\'echelle de la variable $\lambda=\frac{l}{\xi(t)}$ qui a pour expression
\begin{eqnarray}
 {\cal C}_{\mu} (\lambda)= e^{-\lambda} \frac{\lambda}{3}   +O(\mu) 
\label{correlong} 
\end{eqnarray}

\subsection{ Les deux fonctions de corr\'elations temporelles }

  La probabilit\'e $\Pi(t+t_w,t_w)$ de ne pas sauter pendant l'intervalle de temps $[t_w,t_w+t]$ a une forme d'\'echelle de `sous-vieillissement'
\begin{eqnarray}
\Pi(t+t_w,t_w) = {\tilde \Pi}_{\mu} \left( g= \frac{t}{ R(t_w) } \right)
= {\tilde \Pi}_{\mu} \left( g= [ {\tilde T}_0(\mu) ]^{\frac{1}{1+\mu}}
 \frac{ t }{ t_w^{\frac{1}{1+\mu}}  } \right)
\end{eqnarray}
avec
 \begin{eqnarray}
 {\tilde \Pi}_{\mu}^{(0)} (g)
= \int_0^1 dz \mu z^{\mu-1}
 e^{- z  g }
\label{resumepi0}
\end{eqnarray}
En particulier, on obtient le comportement asymptotique
\begin{eqnarray}
\Pi(t+t_w,t_w) \opsimeq_{\frac{ t }{ t_w^{\frac{1}{1+\mu}}  } \to + \infty} \left( \frac{t}{ t_w^{\frac{1}{1+\mu}}  } \right)^{-\mu}
\left[ \mu +O(\mu^2) \right]
\end{eqnarray}

 La probabilit\'e $C(t+t_w,t_w)$ d'\^etre au temps $(t+t_w)$ dans le pi\`ege  occup\'e \`a l'instant $t_w$ a une forme d'\'echelle de vieillissement
\begin{eqnarray}
C(t+t_w,t_w) = {\tilde C}_{\mu} \left( h= \frac{t}{R^{1+\mu}(t_w)} \right)
= {\tilde C}_{\mu} \left( h= {\tilde T}_0(\mu) \frac{t}{t_w} \right)
\end{eqnarray}
avec 
\begin{eqnarray}
 {\tilde C}_{\mu}^{(0)} (h)
=  {\tilde C}_{\mu} (h)
= \frac{2 \mu}{ ( 2  h )^{ \mu} }
 \int_0^{\sqrt{2  h} }  dz z^{1+2 \mu}
 K_1^2( z )
\end{eqnarray}
En particulier, on obtient le comportement asymptotique
\begin{eqnarray}
C(t+t_w,t_w) \opsimeq_{ \frac{t}{t_w} \to \infty} 
\left( \frac{t}{t_w} \right)^{-\mu} \left[ \mu +O(\mu^2) \right]
\end{eqnarray}

Dans cette approche de renormalisation, la pr\'esence de deux \'echelles de temps diff\'erentes dans les propri\'et\'es de vieillissement 
vient directement de la pr\'esence d'un temps d'\'evasion diff\'erent du temps
de pi\'egeage.
Comme en dimension $d=2$ une marche al\'eatoire est encore r\'ecurrente,
on s'attend l\`a aussi \`a avoir deux \'echelles de temps,
alors que pour $d>2$, il n'y a plus de r\'ecurrence, et la seule \'echelle de temps
pour le vieillissement est l'\'echelle des temps de pi\'egeages.
Il est donc int\'eressant de discuter le cas $d=2$.

\section{ Analyse de scaling pour le mod\`ele de pi\`eges en $d=2$ }

\label{dim2}

\subsection{ Paysage renormalis\'e }

On d\'efinit encore le paysage renormalis\'e \`a l'\'echelle $R$
en ne gardant que les pi\`eges de temps de pi\'egeage $\tau_i>R$.
\'Etant donn\'e un pi\`ege, la distance $l$ au plus proche voisin
a pour distribution
\begin{eqnarray}
  P_R (l) 
\opsimeq_{R \to \infty} \frac{2 \pi l}{R^{\mu} } e^{- 
\displaystyle \frac{ \pi l^2}{R^{\mu}} }
\label{lofrd2}
\end{eqnarray}

En utilisant le fait que la probabilit\'e de s'\'echapper jusqu'\`a une distance $l$ sans \^etre revenu \`a l'origine se comporte en $c/(\ln l)$,
on obtient que le nombre $n$ de retours avant une \'evasion r\'eussie
se comporte en $n \sim \ln l \sim \ln R^{\frac{\mu}{2} }$.
Finalement, le temps d'\'evasion d'un pi\`ege $\tau_0$
du paysage renormalis\'e
est de la forme
 \begin{eqnarray}
T_0 = a \tau_0 \ln R^{\frac{\mu}{2} } 
\label{escaped2}
\end{eqnarray}
o\`u $a$ est une variable al\'eatoire d'ordre 1 qui caract\'erise la g\'eom\'etrie des pi\`eges voisins de $\tau_0$. 
L'\'echelle de renormalisation $R$ doit donc \^etre choisie en
fonction du temps selon
\begin{eqnarray}
t \sim R(t) \ln R^{\frac{\mu}{2} }(t) 
\end{eqnarray}
ce qui donne par inversion \`a l'ordre dominant
\begin{eqnarray}
R(t) \sim \frac{ t}{ \ln t^{ \frac{\mu}{2} } }
\label{rdetd2}
\end{eqnarray}

\subsection{ Deux \'echelles de temps}

La pr\'esence de deux \'echelles de temps distinctes $R(t)$ et $t$ 
va conduire \`a deux comportements de vieillissement :
on s'attend \`a ce que
la corr\'elation $\Pi(t+t_w,t_w)$ soit une fonction de $t/R(t_w)$
avec (\ref{rdetd2}) et \`a ce que
la corr\'elation$C(t+t_w,t_w)$ 
soit une fonction de $t/t_w$, 
ce qui est en accord avec les r\'esultats rigoureux 
\cite{cerny2d,benarousreview}.

\subsection{ Deux \'echelles de longueur}

La grande nouveaut\'e qualitative
par rapport au cas $d=1$ est qu'il existe deux \'echelles de longueur en $d=2$.

\subsubsection{ Distance entre deux pi\`eges renormalis\'es } 

La distance caract\'eristique entre deux pi\`eges voisins
du paysage renormalis\'e est
\begin{eqnarray}
\xi(t) \sim [R(t)]^{\frac{\mu}{2} } \sim 
\left[ \frac{ t}{ \ln t^{\mu } } \right]^{\frac{\mu}{2} }
\label{xid2}
\end{eqnarray}

\subsubsection{ Distance parcourue \`a l'instant $t$ } 

Par ailleurs, l'analyse de scaling \`a partir des sommes de L\'evy
 \cite{argumentlevy} en $d=2$ est la suivante : apr\`es $N$ pas, le nombre de sites distincts visit\'es croit selon $S=N/(\ln N)$ et chaque site 
a \'et\'e visit\'e typiquement $(\ln N)$ fois,
ce qui conduit \`a la correspondance
\begin{eqnarray}
t \sim \sum_{i=1}^{S} \tau_i (\ln N) \sim
(\ln N) \left( \frac{N}{\ln N} \right) ^{\frac{1}{\mu}}
\label{tofN} 
\end{eqnarray}
Comme la distance typique apr\`es $N$ pas est $r \sim \sqrt N $, 
on obtient finalement le comportement en temps 
de la distance parcourue 
\begin{eqnarray}
r(t) \sim t^{\frac{\mu}{2} }  ( \ln t^{\mu} )^{\frac{1-\mu }{2}} 
\label{roftd2} 
\end{eqnarray}

\subsubsection{ Nombre de pi\`eges importants \`a l'instant $t$ } 
 
En $d=2$, la pr\'esence des deux \'echelles de longueur
diff\'erentes $\xi(t)$ (\ref{xid2}) et $r(t)$ (\ref{roftd2}) et $\xi(t)$ (\ref{xid2}) montre que le nombre 
$n(t)$ de pi\`eges renormalis\'es contenus dans le disque de rayon $r(t)$ 
accessible, ne reste pas fini mais croit logarithmiquement en temps 
\begin{eqnarray}
n(t) \sim \frac{ r^2(t)} {\xi^2(t)} \sim  \ln t^{\mu}
\label{numbertraps}
\end{eqnarray}
ce qui est en accord avec le r\'esultat rigoureux
concernant l'absence de localisation en $d=2$ \`a temps infini $t \to \infty$ \cite{benarousreview}.
Cependant, dans le r\'egime $t \to \infty$ et $\mu \to 0$
avec $t^{\mu}$ fix\'e (\ref{numbertraps}), 
on s'attend \`a ce que le front de diffusion
dans un \'echantillon soit une somme de pics $\delta$
dont les poids d\'ependent de la r\'epartition g\'eom\'etrique
des pi\`eges mais pas de leurs  \'energies.
Il serait bien s\^ur int\'eressant d'arriver \`a d\'efinir une dynamique
effective plus pr\'ecise pour d\'ecrire ce r\'egime, \`a partir d'une construction explicite des probabilit\'es d'absorption des diff\'erents pi\`eges
pour une configuration al\'eatoire fix\'ee. 

\section{ Conclusion }

La renormalisation appropri\'ee pour d\'ecrire
la phase de vieillissement $\mu<1$ du mod\`ele de pi\`eges unidimensionnel est 
bas\'ee sur deux \'echelles de temps, 
le temps de pi\'egeage et le temps d'\'evasion,
qui gouvernent deux fonctions de corr\'elations temporelles diff\'erentes.
A l'ordre dominant en $\mu \to 0$, dans un \'echantillon donn\'e,
le front de diffusion est localis\'e seulement sur deux sites,
qui correspondent aux deux pi\`eges renormalis\'es les plus proches de l'origine
\`a l'\'echelle de renormalisation $R(t)$.
Les poids des pi\`eges sont simplement donn\'es par les probabilit\'es respectives
d'atteindre un pi\`ege avant l'autre.
Ceci montre que la dynamique reste toujours hors \'equilibre :
les poids des deux pi\`eges ne sont pas donn\'es par des facteurs de Boltzmann, ils ne d\'ependent m\^eme pas des  \'energies des pi\`eges, mais seulement des distances \`a l'origine.

\subsection*{ Publication associ\'ee  [P12] }

%%%%%%%%%%%%%%%%%%%%%%%%%%%%%%%%%%%%%%%%%%%%%%%%%%%%%%%%%

\chapter{Le mod\`ele de pi\`eges en champ : r\'eponses lin\'eaire et non-lin\'eaire }

\label{chapreponsetrap}

%  \begin{flushright}
%  \emph{ 
% }\\  ~\\
%  \end{flushright}

\section{ Pr\'esentation du mod\`ele }

En pr\'esence d'un champ ext\'erieur $f$,
l'\'equation ma\^{\i}tresse du mod\`ele de pi\`eges s'\'ecrit \cite{jpbreview}
\begin{eqnarray}
\frac{dP_t^{(f)}(n)}{dt} = 
P_t^{(f)}(n+1) \frac{e^{ - \beta \frac{f}{2} }}{2 \tau_{n+1}} + 
P_t^{(f)}(n-1)\frac{e^{+ \beta \frac{f}{2} }}{ 2 \tau_{n-1}} 
- P_t^{(f)}(n) \frac{e^{+ \beta \frac{f}{2} } +e^{ - \beta \frac{f}{2} } }{2 \tau_{n}}
\label{masterequation}
\end{eqnarray}
dans laquelle les taux de transition satisfont la condition de bilan d\'etaill\'e
\begin{eqnarray}
e^{-\beta U(n) } W_{ \{n \to n+1 \}}^{(f)}
= e^{ - \beta U_{n+1} } W_{ \{n+1 \to n \}}^{(f)}
\label{detailedbalance}
\end{eqnarray}
en terme de l' \'energie totale $U_n$ contenant
l' \'energie al\'eatoire $(-E_n)$ du pi\`ege
et l' \'energie potentielle lin\'eaire $(-f n)$ associ\'ee au champ
ext\'erieur $f$
\begin{eqnarray}
 U_n = -E_n -f n   
\end{eqnarray}

La r\'eponse de ce mod\`ele a \'et\'e \'etudi\'ee
par E. Bertin et J.P. Bouchaud \cite{bertinreponse}
avec des arguments de scaling et des simulations num\'eriques.
Leur r\'esultat principal est le suivant :
le Th\'eor\`eme Fluctuation-Dissipation du r\'egime de r\'eponse lin\'eaire
est valable m\^eme dans le secteur de vieillissement,
mais la r\'eponse devient non-lin\'eaire asymptotiquement.
Le but du pr\'esent chapitre est de comprendre ces propri\'et\'es,
en utilisant d'une part une sym\'etrie particuli\`ere
de l'\'equation ma\^{\i}tresse, et d'autre part en g\'en\'eralisant la renormalisation
du chapitre pr\'ec\'edent pour prendre en compte la pr\'esence d'un champ ext\'erieur.

\section{ Une sym\'etrie dynamique qui emp\^eche la violation du th\'eor\`eme Fluctuation-Dissipation }

Dans cette section, les \'equations seront \'ecrites
pour le cas $d=1$ pour simplifier les notations,
mais il est clair que les r\'esultats peuvent \^etre g\'en\'eralis\'es
imm\'ediatement \`a toute dimension finie $d$.

\subsection{ Loi de conservation dans chaque \'echantillon
pour le mod\`ele sans champ }

Le mod\`ele de pi\`eges a une propri\'et\'e tr\`es sp\'eciale \cite{bertinreponse} : en l'absence de champ ext\'erieur $f=0$
la moyenne thermique de la position 
\begin{eqnarray}
 <n>_{f=0}(t)   \equiv  \sum_{n=-\infty}^{+\infty} 
n P_t^{(f=0)}(n)   
\end{eqnarray}
est une quantit\'e conserv\'ee dans tout \'echantillon !
Cela vient du fait qu'en sortant d'un pi\`ege
la particule saute avec des probabilit\'es \'egales
 $(1/2,1/2)$ vers la droite ou vers la gauche.
Cette propri\'et\'e de conservation est d'autant plus remarquable
que le d\'esordre brise la sym\'etrie $n \to -n$
dans chaque \'echantillon. 

\subsection{ Un th\'eor\`eme de fluctuation non-lin\'eaire pour $t_w=0$ }

Dans un \'echantillon donn\'e, si on compare les fronts de diffusion
en pr\'esence de $(+f)$ et de $(-f)$, on obtient la relation tr\`es simple
\begin{eqnarray}
\frac{ P_t^{(+f)}(n) }{ P_t^{(-f)}(n)} = e^{ \beta f n }
\label{ratio+-}
\end{eqnarray}
ce qui permet d'exprimer la diff\'erence des moyennes thermiques de la position
\begin{eqnarray}
<n>_{ +f}(t) - <n>_{ - f}(t) && \equiv
\sum_{n=-\infty}^{+\infty} n \left[ P_t^{(+f)}(n)  - P_t^{(-f)}(n) \right] 
\nonumber \\
&& = \sum_{n=-\infty}^{+\infty} n (1-e^{- \beta f n} ) P_t^{(+f)}(n) 
\label{diffmeann}
\end{eqnarray}

\subsubsection{ Relation de Fluctuation-Dissipation dans le r\'egime lin\'eaire}

La premi\`ere cons\'equence (\ref{diffmeann}) est que la relation d'Einstein
\begin{eqnarray}
 <n>_{f}(t)  \opsimeq_{f \to 0}  \frac{ \beta f }{2}  < n^2>_{f=0} (t)
\label{einsteint}
\end{eqnarray}
est valable pour $t$ fix\'e dans la limite $f \to 0$.

\subsubsection{ Fin du r\'egime de r\'eponse lin\'eaire }

La deuxi\`eme cons\'equence de (\ref{diffmeann}) est que pour $f$ fix\'e,
le r\'egime de r\'eponse lin\'eaire n'est valable
que s'il est l\'egitime de lin\'eariser le facteur $(1-e^{- \beta f n} )$
ce qui donne comme crit\`ere
\begin{eqnarray}
\beta f \xi_{(f=0)}(t) \ll 1
\end{eqnarray}
o\`u $\xi_{(f=0)}(t)$ est le d\'eplacement typique sans champ
\`a l'instant $t$.
Le r\'egime de r\'eponse lin\'eaire est donc limit\'e par un temps
caract\'eristique associ\'e \`a la force $f$
\begin{eqnarray}
t \ll t_{\mu}( f) \equiv \left( \frac{1}{\beta f} \right)^{ \frac{1+\mu}{\mu} }
\label{tmuf}
\end{eqnarray}

\subsubsection{ Propri\'et\'e du front de diffusion }

Une troisi\`eme cons\'equence de (\ref{diffmeann}) concerne la dissym\'etrie
de la moyenne sur les \'echantillons
du front de diffusion (Publication [P13])
\begin{eqnarray}
\overline{ P_t^{(f)}(-n) }=  e^{ - \beta f n } 
\overline {P_t^{(f)}(n) }
\label{averageddifffront}
\end{eqnarray}

\subsection{ Un th\'eor\`eme de fluctuation non-lin\'eaire pour $t_w$ quelconque}

Les r\'esultats pr\'ec\'edents peuvent \^etre g\'en\'eralis\'es
au cas o\`u le syst\`eme \'evolue d'abord sans champ ext\'erieur
 $f=0$ pendant l'intervalle $[0,t_w]$,
avant d'\^etre soumis au champ $f$ pour $t \geq t_w$. 
On introduit le front de diffusion \`a deux temps  
$P^{(f;t_w)}(n,t;n_w,t_w \vert 0,0)$ qui repr\'esente
la probabilit\'e jointe d'\^etre en $n_w$ \`a $t_w$ 
et en $n$ \`a $t$ (on suppose $t >t_w$).

La g\'en\'eralisation de (\ref{ratio+-})
\begin{eqnarray}
\frac{ P^{(+f;t_w)}(n,t;n_w,t_w \vert 0,0) }
{ P^{(-f;t_w)}(n,t;n_w,t_w \vert 0,0) } 
= e^{ \beta f (n-n_w) }
\label{ratio+-t_w}
\end{eqnarray} 
permet d'obtenir les trois cons\'equences suivantes.

\subsubsection{ R\'egime lin\'eaire pour $(t,t_w)$ fix\'es }

Pour des temps $(t,t_w)$ fix\'es, la relation d'Einstein est valide
sous la forme 
\begin{eqnarray}
 <(n-n_w)>_{f,t_w}(t)  \opsimeq_{f \to 0}  
\frac{ \beta f }{2} < (n-n_w)^2>_{f=0} (t)
\label{einsteinttw}
\end{eqnarray}

\subsubsection{ Fin du r\'egime de r\'eponse lin\'eaire }

Pour un temps $t_w$ fix\'e et un champ ext\'erieur $f$ fix\'e,
la validit\'e du r\'egime lin\'eaire est limit\'ee aux temps $t$ tels que
\begin{eqnarray}
\beta f \xi_{f=0}(t,t_w) \ll  1
\end{eqnarray}
o\`u $\xi_{f=0}(t,t_w)$ repr\'esente le d\'eplacement caract\'eristique entre $t_w$ et $t$ en l'absence de champ, et donc finalement
\begin{eqnarray}
t-t_w \ll  t_{\mu}( f) 
\end{eqnarray}
ce qui g\'en\'eralise (\ref{tmuf}). 

\subsubsection{ Front de diffusion \`a deux temps }

L'asym\'etrie induite par $f$ sur 
du front diffusion moyen est simplement
\begin{eqnarray}
\overline{ P^{(f)}(-n,t;-n_w,t_w \vert 0,0) }=  e^{ - \beta f (n-n_w) } 
\overline {P^{(f)}(n,t;n_w,t_w \vert 0,0) }
\label{averageddifffrontttw}
\end{eqnarray}

\subsection{ Discussion }

Si on consid\`ere une \'equation ma\^{\i}tresse g\'en\'erale,
la condition pour d\'eriver un th\'eor\`eme de fluctuation non-lin\'eaire analogue \`a celui qui existe pour le mod\`ele de pi\`eges
est la suivante (Publication [P13]) : pour toute configuration ${\cal C}$,
le taux de transition global pour en sortir en pr\'esence de $(+f)$
doit \^etre \'egal au taux de transition global pour en sortir en pr\'esence de $(-f)$
  \begin{eqnarray}
w_{out}^{(+f)}({\cal C}) = w_{out}^{(-f)}({\cal C}) 
\ \ \ {\rm \ pour \ toute \ configuration \ } \ \ {\cal C}
\label{symcondition}
\end{eqnarray}
A l'ordre lin\'eaire en $f$, cette condition implique
que l'observable $b$ lin\'eairement coupl\'e au champ $f$
(qui g\'en\'eralise la position dans le mod\`ele de pi\`eges)
satisfait une loi de conservation :
sa moyenne thermique doit \^etre conserv\'ee en l'absence de champ
pour une condition initiale arbitraire.

Existe-t-il d'autres mod\`eles 
int\'eressants qui poss\`edent cette sym\'etrie?
La fa\c{c}on la plus simple de satisfaire la condition  
(\ref{symcondition})
serait de pouvoir associer \`a toute transition ${\cal C} \to {\cal C'}$
une autre transition ${\cal C} \to {\cal C''}$
qui a le m\^eme taux de transition en champ nul
$W^{(0)} ( {\cal C} \to {\cal C}'' )
=W^{(0)} ( {\cal C} \to {\cal C}' )$ et telle que la variation
de $b$ soit exactement l'oppos\'ee 
$b({\cal C}'') -b({\cal C} )
= -( b ( {\cal C}') - b({\cal C} ))$.
Cependant, c'est une propri\'et\'e tr\`es forte
car elle doit \^etre vraie pour toute configuration initiale ${\cal C}$.

\section{  \'Etude de la r\'eponse par renormalisation  }

Dans le Chapitre pr\'ec\'edent sur le mod\`ele de pi\`eges sans biais,
nous avons d\'ecrit une proc\'edure de renormalisation \`a
partir de la notion du temps d'\'evasion et des probabilit\'es respectives
de s'\'echapper vers le pi\`ege renormalis\'e de droite ou vers le pi\`ege renormalis\'e de gauche. En g\'en\'eralisant cette approche \`a la pr\'esence
d'un champ ext\'erieur $f$, on peut obtenir
beaucoup de r\'esultats exacts sur la dynamique.

\subsection{ R\'esultats explicites pour $t_w=0$  }

La renormalisation permet de d\'efinir une \'echelle de longueur
$\xi(t,f)$ qui repr\'esente la distance moyenne entre deux pi\`eges
dans le paysage renormalis\'e associ\'e \`a l'instant $t$.
Cette \'echelle interpole entre les deux comportements
\begin{eqnarray}
&& \xi(t,f) \opsimeq_{t \ll t_{\mu}(f)} t^{\frac{\mu}{1+\mu}}
\left[ 1+O(\mu)  \right]  \\
&& \xi(t,f) \opsimeq_{t \gg t_{\mu}(f)} (\beta f t)^{\mu} \left[  1+O(\mu) \right]
\label{xifbig}
\end{eqnarray}

\subsubsection{ Front de diffusion  }

La moyenne sur les \'echantillons du front de diffusion
prend la forme d'\'echelle
\begin{eqnarray}
\overline{ P_t(n,f) }
 = \frac{1}{\xi(t,f)} 
g_{\mu} \left( X= \frac{n}{\xi(t,f)} , F= \beta f\xi(t,f)  \right)  
\end{eqnarray}
et la fonction d'\'echelle $g_{\mu}$ a pour expression, dans la limite 
 $\mu \to 0$, 
\begin{eqnarray}
g_0(X,F) 
&& = e^{- \vert X \vert }\left[ \theta(X>0)   + \theta(X<0) 
e^{- F \vert X \vert } \right]
 \int_0^{+\infty} d \lambda
  e^{ - \lambda} 
\frac{1-e^{-F \lambda} }
{ 1- e^{-F ( \vert X \vert+ \lambda) }} \\
&& = e^{- \vert X \vert }\left[ \theta(X>0)   + \theta(X<0) 
e^{- F \vert X \vert } \right]
\sum_{n=0}^{+\infty} \frac{F}{ (1+F n) (1+F+F n)}
e^{-n F \vert X \vert } 
\nonumber
\end{eqnarray}

\subsubsection{ Position moyenne  }

La moyenne sur les \'echantillons de la moyenne thermique de la position 
est de la forme
\begin{eqnarray}
\overline{ <x(t,f)> } \equiv  \sum_{x=-\infty}^{+\infty}
 x  \overline{ P_t(x,f) }
= \xi(t,f) {\cal X}_{\mu} (F= \beta f\xi(t,f) ) 
\end{eqnarray}
et la fonction d'\'echelle $ {\cal X}_{\mu}$ a pour expression, dans la limite 
 $\mu \to 0$, 
\begin{eqnarray}
{\cal X}_{0}(F)    = \int_{0}^{+\infty} d\lambda \lambda
e^{-  \lambda  } \frac{ \frac{ F \lambda}{2} \coth \frac{ F \lambda}{2} -1}{ F} = 1- \frac{1}{F}-\frac{1}{F^3} \psi'' \left( 1+\frac{1}{F} \right)  
\label{calx0f}
\end{eqnarray}

\subsubsection{ Largeur thermique  }

La moyenne sur les \'echantillons de la largeur thermique  
est de la forme
\begin{eqnarray}
\overline{ <\Delta x^2(t,f)> } 
 = \xi^2(t,f) \Delta_{\mu}(F=\beta f \xi(t,f)) 
\label{defthermalwidtf} 
\end{eqnarray}
et la fonction d'\'echelle $ {\Delta}_{0}$ est en fait directement reli\'ee
\`a la fonction $ {\cal X}_{0}$ 
\begin{eqnarray}
\Delta_0(F) && = \left[ \frac{1}{F} + \frac{d}{d F}  \right]
{\cal X}_0(F)
\label{simpledeltafrommean}
\end{eqnarray}

\subsection{ R\'esultats explicites pour $t_w>0$  }

\subsubsection { Dynamique effective en fonction de $(f,t,t_w)$  }

L'\'etat atteint au temps $t_w$ en l'absence de champ
doit \^etre consid\'er\'e comme une condition initiale
pour la dynamique en pr\'esence du champ $f>0$
en fonction du nouveau temps $(t-t_w)$.
Comme \`a l'instant $t_w$, la particule est typiquement dans un pi\`ege $\tau>R(t_w,f=0)$, la dynamique effective ne recommence que lorsque
la proc\'edure de d\'ecimation redevient active   
pour $R(t-t_w,f)>R(t_w,f=0)$.
Il est donc utile d'introduire le param\`etre
\begin{eqnarray} 
\alpha(t,t_w,f) \equiv \frac{ \xi(t-t_w,f) }{ \xi(t_w,f=0)}
= \frac{ R^{\mu}(t-t_w,f) }{  R^{\mu}(t_w,f=0)}
\label{defalpha}
\end{eqnarray}
qui mesure les \'echelles de longueur des deux paysages renormalis\'es
\`a $t_w$ et \`a $t$.
Dans le secteur $\alpha<1$, la r\'eponse est domin\'e par des \'ev\`enements rares, alors que dans le secteur $\alpha(t,t_w,f)>1$, la r\'eponse
est domin\'e par la dynamique effective.
Le domaine en $(t-t_w)$ correspondant au secteur $\alpha>1$
d\'epend des valeurs relatives de $t_w$
et du temps $t_{\mu}(f)$.

(i) Pour $t_w \ll t_{\mu}(f)$, 
le secteur $\alpha>1$ correspond au domaine temporel
$(t-t_w)>t_w$, et le param\`etre $\alpha$ se comporte selon
\begin{eqnarray} 
\alpha(t,t_w,f) && =  
\left( \frac{ t-t_w }{ t_w} \right)^{\frac{\mu}{1+\mu}}
\ \ \rm{for} \ \  t_w < t-t_w \ll   t_{\mu}(f)
\label{ashort} 
\end{eqnarray}
et
\begin{eqnarray} 
\alpha(t,t_w,f) && =  
\left( \frac{  \frac{\beta f}{2}   (t-t_w) }
{ t_w^{\frac{1}{1+\mu}}} \right)^{\mu}
\ \ \rm{for} \ \   t-t_w \gg   t_{\mu}(f)
\label{along}
\end{eqnarray}

(ii) Pour $t_w \gg t_{\mu}(f)$, le secteur $\alpha>1$ correspond
au domaine
$(t-t_w)>\frac{2}{\beta f} t_w^{\frac{1}{1+\mu}}$, et 
le param\`etre $\alpha$ se comporte comme (\ref{along})
partout.

En \'etudiant la statistique du paysage renormalis\'e \`a deux \'echelles de renormalisation successives, on peut alors obtenir des r\'esultats explicites 
valables dans tout le secteur $\alpha>1$ dans la limite $\mu \to 0$

\subsubsection { Loi du d\'eplacement entre $t_w$ et $t$  }

En termes des variables d'\'echelle $Y=\frac{ x(t)-x(t_w) }{\xi(t-t_w,f)}$, 
$F=\beta f \xi(t-t_w,f)$ et $\alpha= \xi(t-t_w,f)/\xi(t_w,f=0)$,
la moyenne sur les \'echantillons de la distribution de $Y$ est
de la forme 
\begin{eqnarray}  
P \left(  Y ; F, \alpha \right)   = 
 \frac{1}{ \alpha}  \delta(Y)
+ \left[ \theta(Y \geq 0 ) +
e^{-F \vert Y \vert }  \theta(Y \leq 0 ) \right] 
G_{ns} \left( \vert Y \vert, F, \alpha \right)
\label{reslawy}
\end{eqnarray}
La partie singuli\`ere en $\delta$ repr\'esente les particules
qui n'arrivent pas \`a s'\'echapper vers un autre pi\`ege renormalis\'e
entre $t_w$ et $t$. La partie r\'eguli\`ere
 est d\'efinie en termes de la fonction
\begin{eqnarray}  
 G_{ns} \left(  Y , F, \alpha \right)
&& =  \frac{e^{-Y}}{2}    \int_0^{+\infty} du \frac{1-e^{-F u } }
{ 1-e^{-F \left( Y+ u\right) }}    
\left[ ( 2 -  e^{-  (\alpha-1) Y  } )  e^{-  u }
   -  e^{- \alpha  u }    \right] 
%\\ && =  e^{-Y} \sum_{n=0}^{+\infty} e^{ -Fn Y} 
%\left[ \frac{H}{(1+F n) (1+F+F n) } (2-e^{- (\alpha-1) Y} ) - \frac{H}{(a+F n) %(a+F+F n) } \right]
\end{eqnarray}

Plus g\'en\'eralement, on peut calculer ainsi toutes les fonctions d'\'echelle souhait\'ees dans la limite $\mu \to 0$, en particulier
pour la position moyenne et la largeur thermique (Publication [P14]).

\section{ Conclusion}

C'est la sym\'etrie particuli\`ere de l'\'equation ma\^{\i}tresse
du mod\`ele de pi\`eges qui emp\^eche la violation du Th\'eor\`eme
de Fluctuation-Dissipation m\^eme dans le r\'egime de vieillissement.
Par ailleurs, la renormalisation bas\'ee
 sur les temps d'\'evasion, qui devient exacte dans la limite $\mu \to 0$,
permet d'\'etudier en d\'etail la r\'eponse non-lin\'eaire 
dans les diff\'erents r\'egimes qui existent en fonction des deux temps $(t_w,t)$ et du champ ext\'erieur $f$.

\subsection*{ Publications associ\'ees}  

Sur le th\'eor\`eme de Fluctuation-Dissipation non-lin\'eaire : [P13]

Sur les calculs explicites par renormalisation : [P14]

%%%%%%%%%%%%%%%%%%%%%%%%%%%%%%%%%%%%%%%%%%%%%%%%%%%%%%%%%%%%

\chapter
{Cha\^{\i}ne de spin quantique $S=1$ 
antiferromagn\'etique al\'eatoire}

\label{chapquantique1}

\section{La cha\^{\i}ne antiferromagn\'etique $S=1$ sans d\'esordre }

\subsection{ Diff\'erences entre spin demi-entier et spin entier}

Alors que la cha\^{\i}ne antiferromagn\'etique $S= 1/2 $ pr\'esente des corr\'elations en loi de puissance
et des excitations sans gap, la cha\^{\i}ne antiferromagn\'etique $S = 1$ est caract\'eris\'ee par des corr\'elations \`a d\'ecroissance exponentielle et par un gap pour les excitations.
Une fa\c{c}on simple de comprendre ces diff\'erences qui existent plus g\'en\'eralement entre spins demi-entiers et spins entiers
est la fonction d'onde ``Valence-Bond-Solid" (VBS), qui permet de 
mettre en \'evidence un ordre \`a longue port\'ee pour un param\`etre d'ordre
topologique qui est non-local en termes de spins.

\subsection{ La fonction d'onde VBS }

 \begin{figure}[b]
\centerline{\includegraphics[height=3cm]{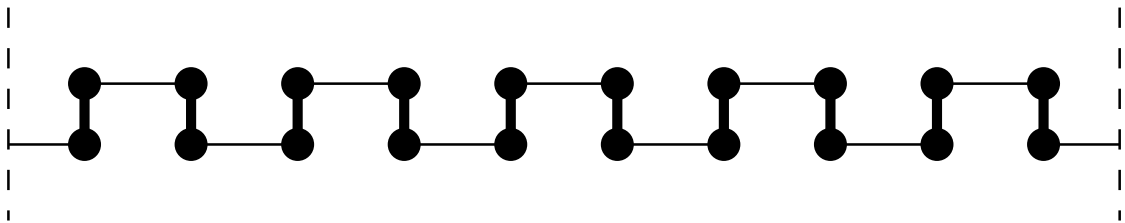}} 
\caption{\it Fonction d'onde VBS :
chaque spin $S=1$ est repr\'esent\'e par la sym\'etrisation de 2 spins 1/2 (les boules noires reli\'ees par un trait vertical gras);
les lignes horizontales qui relient deux spins 1/2 repr\'esentent les singulets entre spins 1/2 constitutifs.   } 
\label{figdefvbs}
\end{figure}

Si on repr\'esente chaque spin $S=1$ comme la sym\'etrisation de deux spins \'el\'ementaires $S=1/2$, la fonction d'onde VBS \cite{AKLT} est l'\'etat 
obtenu en format un singulet sur chaque lien entre deux spins $1/2$ constitutifs (Figure \ref{figdefvbs}). 
Cette fonction d'onde est l'\'etat fondamental exact de l'Hamiltonien \cite{AKLT}
\begin{eqnarray} 
H_{AKLT} = \sum_i P_2( \vec S_i + \vec S_{i+1} ) 
=\sum_i \left[ \frac{1}{2} \vec S_i .\vec S_{i+1} 
+ \frac{1}{6} (\vec S_i .\vec S_{i+1})^2 + \frac{1}{3} \right]
\end{eqnarray} 
o\`u $P_2$ est le projecteur sur le sous-espace $s=2$.

\subsection{  Le param\`etre d'ordre topologique }

 Le param\`etre d'ordre \cite{SOP} qui permet de
caract\'eriser l'ordre topologique de la fonction d'onde VBS
est non local dans les spins
\begin{eqnarray} 
t_{ij}= - \langle \psi_o
\big \vert S_i^z \exp\left[i \pi \sum_{i<k<j} S_k^z \right]
S_{j}^z\big\vert \psi_0\rangle 
\label{tij}
\end{eqnarray} 
Pour l'\'etat pur VBS sur une cha\^{\i}ne cyclique de $N$ spins, 
on obtient $t_{ij}=4/9+O(3^{-N})$.
Pour le fondamental de la cha\^{\i}ne antiferromagn\'etique pure,
ce param\`etre d'ordre tend aussi vers une valeur finie,
ce qui indique une parent\'e avec la fonction d'onde VBS.

\section{Construction des r\`egles de renormalisation }

\subsection{Renormalisation d'un lien antiferromagn\'etique }

L' Hamiltonien d'un lien antiferromagn\'etique entre deux spins $S=1$ 
\begin{eqnarray} 
h_0= J_1 \vec S_1 . \vec S_2={J_1 \over
2} \left[\left( \vec S_1+ \vec S_{2} \right)^2 -\vec S_1^2-\vec
S_{2}^2 \right] ={J_1 \over 2} \left[\left( \vec S_1+ \vec S_{2}
\right)^2 - 4\right] 
\end{eqnarray}
pr\'esente trois niveaux d' \'energie index\'es par la valeur
$s=0,1,2$ du spin total
\begin{eqnarray} 
e_{s}={J_1 \over 2} \left[s(s+1)-4 \right] 
\end{eqnarray}
il y a donc un singulet $e_0=-2J_1$,
un triplet $e_1=-J_1$ et un quintuplet $e_2=J_1$.

Si on g\'en\'eralise na\"{\i}vement la r\`egle de Ma-Dasgupta du cas $S=1/2$
en projetant sur le singulet, le nouveau couplage effectif est
\begin{eqnarray} 
J_0'={4 \over 3} {{J_0 J_2} \over {J_1}} 
\end{eqnarray}
Le coefficient ${4 \over 3}$ \'etant plus grand que $1$, cette r\`egle n'est pas automatiquement consistante, et la proc\'edure ne peut \^etre justifi\'ee que si on part d'un d\'esordre assez large.

Nous avons donc propos\'e de g\'en\'eraliser la proc\'edure de Ma-Dasgupta 
avec la r\`egle suivante : au lieu de projeter sur le niveau le plus bas
de $h_0$, il faut en fait \'eliminer le niveau le plus haut.
Ainsi, pour l'hamiltonien $h_0$, on \'elimine le quintuplet, mais on garde le singulet et le triplet,
en rempla\c{c}ant les deux spins $S=1$ 
par deux spins $S=1/2$.
Cette d\'ecimation partielle \'elargit l'espace initial, mais il est possible de d\'efinir
une renormalisation ferm\'ee avec 4 types de liens.

\subsection{Proc\'edure de renormalisation avec 4 types de liens }

La proc\'edure de renormalisation que nous avons propos\'ee est d\'efinie
pour l' ensemble \'elargi des cha\^{\i}nes constitu\'ees de spins de taille
$S=1/2$ ou $S=1$, dans lesquelles les couplages $\{J_i\}$ sont soit ferromagn\'etiques (F) soit antiferromagn\'etiques (AF), avec la contrainte suivante :
pour tout segment $\{i,j\}$, l' aimantation classique doit satisfaire $\vert m_{i,j} \vert \leq 1$. 
Cette condition pour deux voisins $j=i+1$ montre qu'il y a 4 types de liens possibles : 

1) Lien F entre deux spins S=1/2,

2) Lien AF entre deux spins S=1/2,

3) Lien AF entre un spin S=1 et un spin S=1/2, 

4) Lien AF entre deux spins S=1. 

Les quatre r\`egles de renormalisation correspondantes sont les suivantes :

\bigskip
\bigskip
\hbox to 450pt {(1)\hfill
\vbox{\hsize=10pt \centerline{$s_0$}\par \centerline{$\bullet$}} 
\kern -7pt \raise 2pt
\vtop{\hsize=45pt \centerline{\hrulefill} \par \centerline{$J_0$}} 
\kern -7pt
\vbox{\hsize=10pt \centerline{$s_1={1\over 2}$}\par 
\centerline{$\bullet$}} \kern -7pt \raise 2pt
\vtop{\hsize=45pt \centerline{\hrulefill} \par \centerline{$J_1<0$}} 
\kern -7pt
\vbox{\hsize=10pt \centerline{$s_2={1\over 2}$}\par 
\centerline{$\bullet$}} \kern -7pt \raise 2pt
\vtop{\hsize=45pt \centerline{\hrulefill} \par \centerline{$J_2$}} 
\kern -7pt
\vbox{\hsize=10pt \centerline{$s_3$}\par \centerline{$\bullet$}} 
\hfill$\longrightarrow$\hfill
\vbox{\hsize=10pt \centerline{$s_0$}\par \centerline{$\bullet$}} 
\kern -7pt \raise 2pt
\vtop{\hsize=67pt \centerline{\hrulefill} \par 
\centerline{$J^\prime_0={J_0\over 2}$}} \kern -7pt
\vbox{\hsize=10pt \centerline{$s^\prime_1=1$}\par 
\centerline{$\bullet$}} \kern -7pt \raise 2pt
\vtop{\hsize=67pt \centerline{\hrulefill} \par 
\centerline{$J^\prime_1={J_2\over 2}$}} \kern -7pt
\vbox{\hsize=10pt \centerline{$s_3$}\par \centerline{$\bullet$}} 
\hfill}
\bigskip
\bigskip
\hbox to 450pt {(2)\hfill
\vbox{\hsize=10pt \centerline{$s_0$}\par \centerline{$\bullet$}} 
\kern -7pt \raise 2pt
\vtop{\hsize=45pt \centerline{\hrulefill} \par \centerline{$J_0$}} 
\kern -7pt
\vbox{\hsize=10pt \centerline{$s_1={1\over 2}$}\par 
\centerline{$\bullet$}} \kern -7pt \raise 2pt
\vtop{\hsize=45pt \centerline{\hrulefill} \par \centerline{$J_1>0$}} 
\kern -7pt
\vbox{\hsize=10pt \centerline{$s_2={1\over 2}$}\par 
\centerline{$\bullet$}} \kern -7pt \raise 2pt
\vtop{\hsize=45pt \centerline{\hrulefill} \par \centerline{$J_2$}} 
\kern -7pt
\vbox{\hsize=10pt \centerline{$s_3$}\par \centerline{$\bullet$}} 
\hfill$\longrightarrow$\hfill
\vbox{\hsize=10pt \centerline{$s_0$}\par \centerline{$\bullet$}} 
\kern -7pt \raise 2pt
\vtop{\hsize=135pt \centerline{\hrulefill} \par 
\centerline{$J^\prime_0={J_0\,J_2\over 2\,J_1}$}} \kern -7pt
\vbox{\hsize=10pt \centerline{$s_3$}\par \centerline{$\bullet$}} 
\hfill}
\bigskip
\bigskip
\hbox to 450pt {(3)\hfill
\vbox{\hsize=10pt \centerline{$s_0$}\par \centerline{$\bullet$}} 
\kern -7pt \raise 2pt
\vtop{\hsize=45pt \centerline{\hrulefill} \par \centerline{$J_0$}} 
\kern -7pt
\vbox{\hsize=10pt \centerline{$s_1=1$}\par \centerline{$\bullet$}} 
\kern -7pt \raise 2pt
\vtop{\hsize=45pt \centerline{\hrulefill} \par \centerline{$J_1>0$}} 
\kern -7pt
\vbox{\hsize=10pt \centerline{$s_2={1\over 2}$}\par 
\centerline{$\bullet$}} \kern -7pt \raise 2pt
\vtop{\hsize=45pt \centerline{\hrulefill} \par \centerline{$J_2$}} 
\kern -7pt
\vbox{\hsize=10pt \centerline{$s_3$}\par \centerline{$\bullet$}} 
\hfill$\longrightarrow$\hfill
\vbox{\hsize=10pt \centerline{$s_0$}\par \centerline{$\bullet$}} 
\kern -7pt \raise 2pt
\vtop{\hsize=45pt \centerline{\hrulefill} \par 
\centerline{$J^\prime_0={4\,J_0\over 3}$}} \kern -7pt
\vbox{\hsize=10pt \centerline{$s^\prime_1={1\over 2}$}\par 
\centerline{$\bullet$}} \kern -7pt \raise 2pt
\vtop{\hsize=90pt \centerline{\hrulefill} \par 
\centerline{$J^\prime_1=-{J_2\over 3}$}} \kern -7pt
\vbox{\hsize=10pt \centerline{$s_3$}\par \centerline{$\bullet$}} 
\hfill}
\bigskip
\bigskip

\hbox to 450pt {(4)\hfill
\vbox{\hsize=10pt \centerline{$s_0$}\par \centerline{$\bullet$}} 
\kern -7pt \raise 2pt
\vtop{\hsize=45pt \centerline{\hrulefill} \par \centerline{$J_0$}} 
\kern -7pt
\vbox{\hsize=10pt \centerline{$s_1=1$}\par \centerline{$\bullet$}} 
\kern -7pt \raise 2pt
\vtop{\hsize=45pt \centerline{\hrulefill} \par \centerline{$J_1>0$}} 
\kern -7pt
\vbox{\hsize=10pt \centerline{$s_2=1$}\par \centerline{$\bullet$}} 
\kern -7pt \raise 2pt
\vtop{\hsize=45pt \centerline{\hrulefill} \par \centerline{$J_2$}} 
\kern -7pt
\vbox{\hsize=10pt \centerline{$s_3$}\par \centerline{$\bullet$}} 
\hfill$\longrightarrow$\hfill
\vbox{\hsize=10pt \centerline{$s_0$}\par \centerline{$\bullet$}} 
\kern -7pt \raise 2pt
\vtop{\hsize=45pt \centerline{\hrulefill} \par 
\centerline{$J^\prime_0=J_0$}} \kern -7pt
\vbox{\hsize=10pt \centerline{$s^\prime_1={1\over 2}$}\par 
\centerline{$\bullet$}} \kern -7pt \raise 2pt
\vtop{\hsize=45pt \centerline{\hrulefill} \par 
\centerline{$J^\prime_1=J_1$}} \kern -7pt\vbox{\hsize=10pt 
\centerline{$s^\prime_2={1\over 2}$}\par \centerline{$\bullet$}} 
\kern -7pt \raise 2pt
\vtop{\hsize=45pt \centerline{\hrulefill} \par 
\centerline{$J^\prime_2=J_2$}} \kern -7pt
\vbox{\hsize=10pt \centerline{$s_3$}\par \centerline{$\bullet$}} 
\hfill}
\bigskip
\bigskip

\subsection{Interpr\'etation de la renormalisation en termes d'amas VBS}

\begin{figure}[ht]
\centerline{\includegraphics[height=3cm]{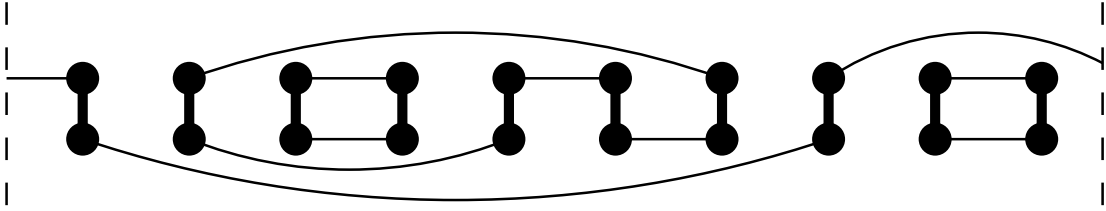}} 
\caption{\it L'\'etat fondamental obtenu par renormalisation, en termes d'amas  VBS : chaque spin $S=1$ est repr\'esent\'e par la sym\'etrisation de 2 spins 1/2 (les boules noires reli\'ees par un trait vertical gras);
les lignes qui relient deux spins 1/2 repr\'esentent des singulets.
 Les spins 1 initiaux sont ainsi regroup\'es en amas : ici, par exemple,
il y a deux amas de taille 2, un amas de taille 4, et un amas plus grand qui d\'epasse les limites de la figure.    } 
\label{figamasvbs}
\end{figure}

Si on repr\'esente chaque spin $S=1$ initial comme
la sym\'etrisation de deux spins $S=1/2$, les r\`egles 2, 3, et 4 pour les liens AF s' interpr\`etent
comme la formation d' un singulet entre deux spins $S=1/2$ constitutifs. 
La r\`egle 1 pour un lien F entre deux spins $S=1/2$ correspond a leur sym\'etrisation. 
A la fin de la proc\'edure de renormalisation, lorsqu il n'y a plus de spin libre,
la cha\^{\i}ne est d\'ecompos\'ee en un ensemble d'amas disjoints
qui ont une structure VBS
 ( Figure \ref{figamasvbs}).

Pour la cha\^{\i}ne d\'esordonn\'ee, on a donc un param\`etre
 d'ordre (\ref{tij}) 
$t_{ij}=4/9$ si les deux sites appartiennent au m\^eme amas
et $t_{ij}=0$ sinon.
Pour une cha\^{\i}ne de taille N, la moyenne spatiale
$\sum_{i,j} t_{ij} /N^2$ du param\`etre $t_{ij}$ est proportionnelle \`a
la probabilit\'e T que deux spins appartiennent au m\^eme amas
\begin{eqnarray} 
T \equiv \sum_c {n_c^2\over N^2}={9\over 4} {1\over
N^2} \sum_{i,j} t_{ij} +O(1/N) 
\end{eqnarray} 
o\`u $n_c$ est le nombre de spins dans un amas c,  
qui n'est pas directement li\'ee \`a son extension spatiale.
Ce param\`etre d'ordre T ne peut \^etre non-nul dans la limite thermodynamique
que s'il existe un amas VBS qui contient une fraction finie des spins de la cha\^{\i}ne.

\section{ \'Etude num\'erique de la renormalisation  }

Nous avons \'etudi\'e num\'eriquement la proc\'edure de renormalisation
avec 4 types de liens en partant de cha\^{\i}nes cycliques de N spins
(par exemple $N=2^{22} \sim 4.10^6$ ), avec des couplages initiaux
$J_i$  uniform\'ement distribu\'es dans l'intervalle $[1,1+d]$. 
Le param\`etre $d$ repr\'esente donc la largeur du d\'esordre initial. 
Pour chaque taille, nous avons effectu\'e des moyennes
sur un certain nombre  d'\'echantillons (typiquement 100) .

Nous avons \'etudie le flot des quantit\'es suivantes en fonction de l' \'echelle $\Gamma$ :

(i) le nombre $N(\Gamma)$ de spins effectifs $S=1/2$ et $S=1$ encore pr\'esents 
 \`a l' \'echelle $\Gamma$ 

(ii) la proportion $\{ N_{(S=1) (\Gamma)}/ N
(\Gamma)\}$ de spins $S=1$ parmi les spins effectifs pr\'esents 

(iii) les proportions $\rho_i(\Gamma)=\{N_i(\Gamma)/N(\Gamma)\}$
des liens de type $i=1,2,3,4$  

(iv)  les distributions de probabilit\'e $P_i(J,\Omega)$ des couplages $J$ 
pour les quatre types de liens $i=1,2,3,4$.

\begin{figure}[ht]
\centerline{\includegraphics[height=5cm]{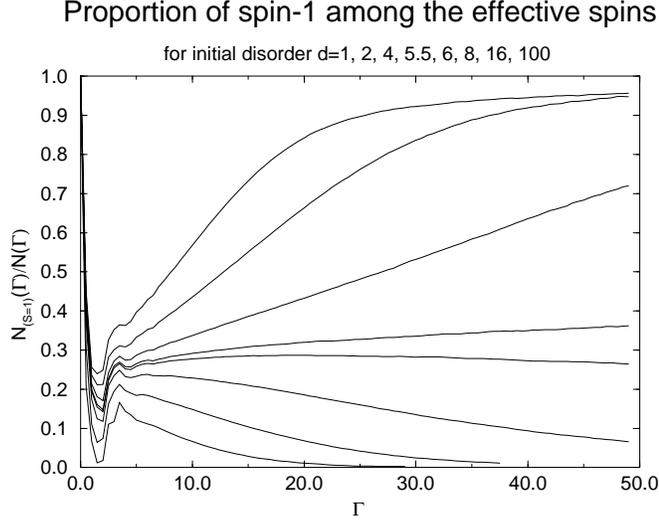}} 
\caption{\it  Proportion de spins $S=1$ parmi les spins effectifs \`a l'\'echelle $\Gamma$ pour diff\'erentes valeurs du d\'esordre initial $d=1,2,3,4,5.5,6,8,16,100$ : cette proportion tend vers 
$0$ dans la phase de faible d\'esordre, et vers $1$ dans la phase de fort d\'esordre. Au point critique $d_c \simeq 5.75(5)$ la proportion
reste stationnaire autour de la valeur interm\'ediaire $0.315(5)$.  } 
\label{figs1global}
\end{figure}

Les r\'esultats de la renormalisation pr\'esentent un changement qualitatif pour une certaine valeur critique $d_c \simeq 5.75(5)$ du d\'esordre initial.
Ainsi, la Figure \ref{figs1global} repr\'esente le flot de la proportion $ { {
N_{(S=1)} (\Gamma)} \over { N (\Gamma)} } $ de spins $S=1$ 
pour diff\'erentes valeurs du d\'esordre initial :
il y a deux valeurs attractives, \`a savoir $0$ \`a faible d\'esordre
et $1$ \`a fort d\'esordre.

\newpage

 \thispagestyle{empty}

\subsection{ R\'esultats dans la phase de fort d\'esordre }

\begin{figure}[p]
\centerline{\includegraphics[height=6cm]{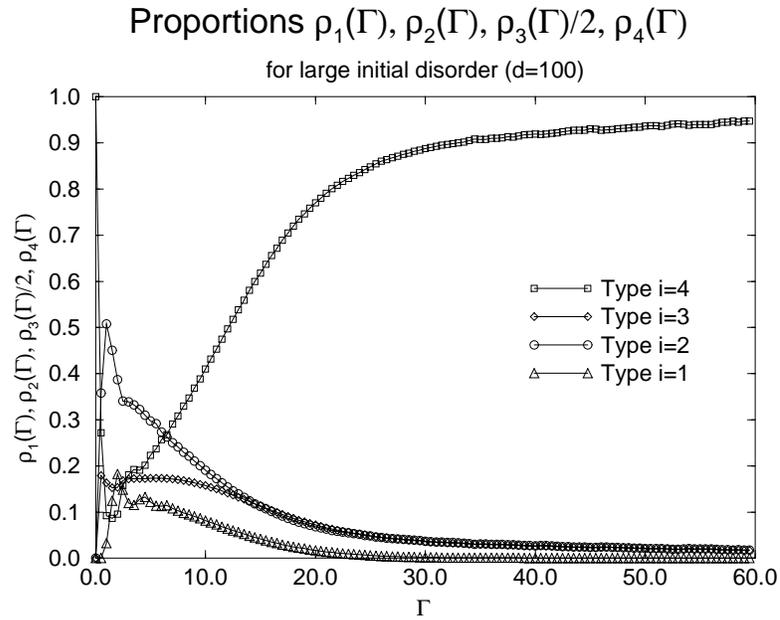}} 
\caption{\it Les proportions $\rho_i(\Gamma)$ des 4 types de liens $i=1,2,3,4$ of en fonction de l'\'echelle $\Gamma$ pour un d\'esordre initial fort $d=100$.   } \label{figrho100}
\end{figure}

Dans la phase de fort d\'esordre $d > d_c$, le nombre $N(\Gamma)$ de spins effectifs
 d\'ecro\^{\i}t comme dans la phase appel\'ee ``Random Singlet" de la cha\^{\i}ne $S=1/2$ :
\begin{eqnarray} 
N(\Gamma) \oppropto_{\Gamma \to \infty} { 1 \over
\Gamma^2} .
\label{ngafort} 
\end{eqnarray}
et les proportions $\rho_i(\Gamma)$ des quatre types de liens
 convergent vers un r\'egime asymptotique caract\'eris\'e par les proportions simples
 (Fig \ref{figrho100}) 
\begin{eqnarray} 
\rho_1(\Gamma)\sim 0 \qquad \ \rho_2(\Gamma)
\sim \epsilon(\Gamma) \qquad \ \rho_3(\Gamma) \sim 2 \epsilon(\Gamma)
\qquad \ \rho_4(\Gamma) \sim 1-3\epsilon(\Gamma) 
\label{rhofort} 
\end{eqnarray}
o\`u $\epsilon(\Gamma)$ tend lentement vers $0$ en $\Gamma \to \infty$.
Il y a donc presque partout des liens de type 4, avec quelques d\'efauts
de type (lien de type 3,lien de type 2,lien de type 3)
qui viennent de la d\'ecimation partielle momentan\'ee des liens de type $4$. 
A fort d\'esordre, la renormalisation converge donc vers la
phase ``Random Singlet" de Ma-Dasgupta.

\subsection{ R\'esultats dans la phase de faible d\'esordre }

\begin{figure}[p]
\centerline{\includegraphics[height=6cm]{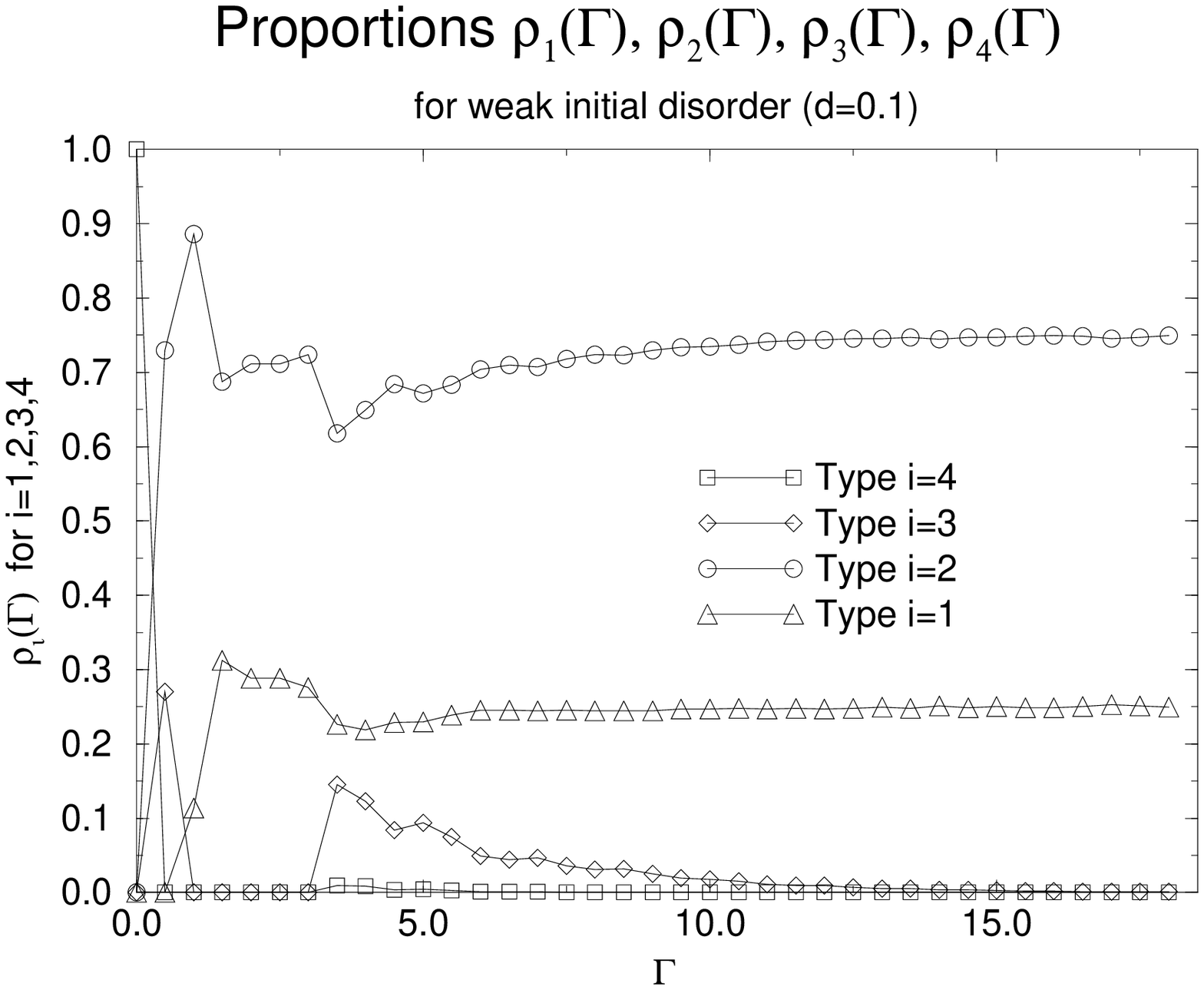}} 
\caption{\it Les proportions $\rho_i(\Gamma)$ des 4 types de liens $i=1,2,3,4$ of en fonction de l'\'echelle $\Gamma$ pour un d\'esordre initial faible $d=0.1$   } 
\label{figrho01}
\end{figure}

Dans la phase de faible d\'esordre, le nombre $N(\Gamma)$ de
de spins effectifs d\'ecro\^{\i}t exponentiellement  
\begin{eqnarray}
N(\Gamma) \oppropto_{\Gamma \to \infty} e^{-\alpha(d) \Gamma}
\label{ngafaible} 
\end{eqnarray}
avec un coefficient $\alpha(d)$ qui est une fonction d\'ecroissante de $d$ 
qui s' annule \`a la transition $d \to d_c^-$. 
Les
proportions $\rho_i(\Gamma)$ des quatre types de liens convergent vers le r\'egime asymptotique
 (Figure \ref{figrho01}) 
\begin{eqnarray}
\rho_1(\Gamma)\simeq 0.25 \qquad \ \rho_2(\Gamma) \simeq 0.75 \qquad
\ \rho_3(\Gamma) \simeq 0 \qquad \ \rho_4(\Gamma) \simeq 0
\label{rhofaible} 
\end{eqnarray}

\subsection{ R\'esultats num\'eriques au point critique }

\subsection{ \'Etude des types de liens}

\begin{figure}[ht]
\centerline{\includegraphics[height=8cm]{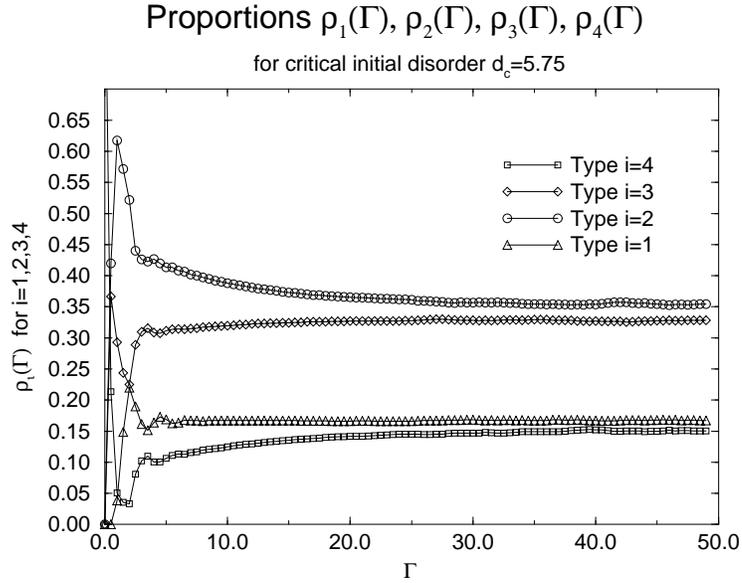}} 
\caption{\it Les proportions $\rho_i(\Gamma)$ des 4 types de liens $i=1,2,3,4$ of en fonction de l'\'echelle $\Gamma$ pour un d\'esordre initial critique $d_c=5.75$ .   } 
\label{figrhocrit}
\end{figure}

Au point critique, le nombre de spins effectifs d\'ecro\^{\i}t alg\'ebriquement
\begin{eqnarray} 
N(\Gamma) \oppropto_{\Gamma \to \infty} { 1 \over
\Gamma^3} .
\label{ngacriti} 
\end{eqnarray}
et les proportions $\rho_i(\Gamma)$ des quatre types de lien convergent vers le r\'egime asymptotique
(Figure \ref{figrhocrit}) 
\begin{eqnarray} 
\rho_1(\Gamma)\sim 0.17 \, ,\qquad \
\rho_2(\Gamma) \sim 0.35 \, ,\qquad \ 
\rho_3(\Gamma) \sim 0.33 \, ,\qquad \
\rho_4(\Gamma) \sim 0.15 \, .
\label{rhocnume} 
\end{eqnarray}

\subsection{ \'Etude num\'erique de la transition de percolation des amas VBS}

Du point de vue des amas VBS, la transition de phase quantique correspond \`a une transition de percolation : dans la phase de fort d\'esordre, il n'y a que des amas finis,
alors que dans la phase de faible d\'esordre, il appara\^{\i}t un amas macroscopique
qui contient une fraction finie des spins.

\subsubsection{ Param\`etre d'ordre}

\begin{figure}[ht]
\centerline{\includegraphics[height=5cm]{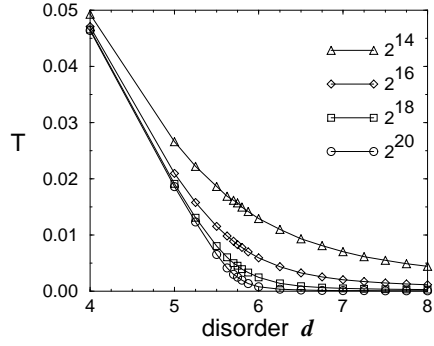}} 
\caption{\it  Param\`etre d'ordre topologique T en fonction du d\'esordre initial d pour des cha\^{\i}nes de tailles $N=2^{14}-2^{20}$. 
Le point critique est situ\'e en $d_c = 5.76(2)$  } 
\label{figordret}
\end{figure}

Si on note $\beta$
l'exposant qui gouverne l' annulation de la fraction 
$n_1 /N$ des spins dans l'amas macroscopique, le param\`etre d'ordre 
se comporte en $T \sim (d_c -d)^{2\beta}$ pour $d<d_c$. 
 L'\'etude de finite size scaling de la Figure (\ref{figordret})
conduit \`a l'estimation
\begin{eqnarray}
2\beta = 1.0(1). 
\label{nbeta} 
\end{eqnarray}

\subsubsection{ Susceptibilit\'e}

\begin{figure}[ht]
\centerline{\includegraphics[height=5cm]{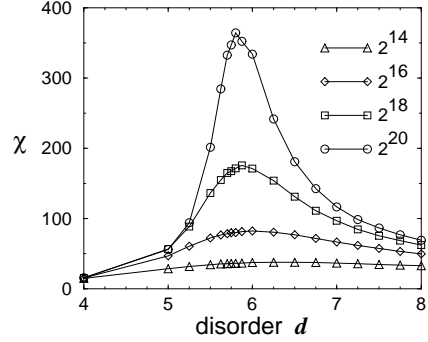}} 
\caption{\it Susceptibilit\'e $\chi$ en fonction du d\'esordre initial d pour des cha\^{\i}nes de tailles $N=2^{14}-2^{20}$.   } 
\label{figchi}
\end{figure}

La Figure (\ref{figchi})  repr\'esente la taille moyenne des amas finis, qui joue le  r\^ole d'une  susceptibilit\'e 
\begin{eqnarray} 
\chi \equiv 
\sum_{c >1} {n_c^2 \over N} . 
\end{eqnarray} 
 (c=1 est l'amas le plus grand). Si on note $\gamma$ l'exposant critique
qui gouverne sa divergence $\chi \sim |d_c -d
|^{-\gamma}$, on obtient l'estimation
\begin{eqnarray}
 \gamma =1.2(1). 
\label{numegamma}
\end{eqnarray}

\subsubsection{ Distribution des amas au point critique }

\begin{figure}[ht]
\centerline{\includegraphics[height=5cm]{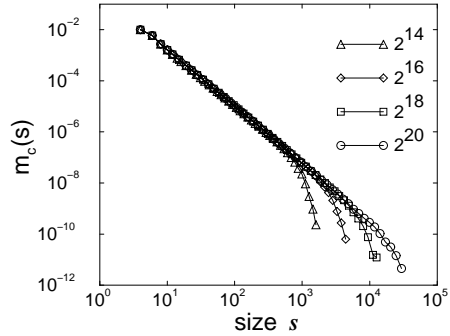}} 
\caption{\it Distribution $m_c (s)$ de la taille $s$ des amas au point critique :
la mesure de la pente conduit \`a l'estimation $\tau =2.2(1)$   } 
\label{figdistricriti}
\end{figure}

Au point critique, la distribution $m_c(s)$ de la taille $s$
des amas a une d\'ecroissance alg\'ebrique (Figure \ref{figdistricriti})
\begin{eqnarray} 
m_c(s) \sim \
{ 1 \over {s^{\tau}}} \label{scaling} 
\end{eqnarray} 
avec un exposant 
\begin{eqnarray} 
\tau= 2.2(1).
\label{numetau} 
\end{eqnarray}

\section{ Calcul des exposants critiques exacts }

Il existe un mod\`ele effectif plus simple de cha\^{\i}ne de spin
\cite{HY} qui se trouve \^etre dans la m\^eme classe d'universalit\'e (Publications [P1]). 
En introduisant une variable auxiliaire $\mu$
 pour \'etudier la taille des amas VBS dans ce mod\`ele plus simple, nous avons
calcul\'e 
les exposants exacts de la transition
\begin{eqnarray} 
&& 2\beta  = { {4(3-\sqrt 5)} \over {\sqrt{13} -1} } = 
1.17278...,
\\
&& \gamma={ {2(2\sqrt 5-3)} \over
{\sqrt{13} -1}}=1.13000...,
\\
&& \tau =1+{3 \over \sqrt{5}} =2.34164...,
\end{eqnarray} 
qui sont en tr\`es bon accord avec nos estimations num\'eriques (\ref{nbeta}), (\ref{numegamma}), (\ref{numetau}).

\section{ Conclusion}

La g\'en\'eralisation de la m\'ethode de Ma-Dasgupta
\`a la cha\^{\i}ne de spin $S=1$ antiferromagn\'etique al\'eatoire
conduit \`a un mod\`ele \`a 4 types de liens que nous avons \'etudi\'e
num\'eriquement. La transition de phase observ\'ee
en fonction du d\'esordre initial s'interpr\`ete comme
une transition de percolation pour les amas VBS de l'\'etat fondamental.
En utilisant un mod\`ele effectif dans
la m\^eme classe d'universalit\'e, nous avons calcul\'e 
les exposants critiques exacts de cette transition de phase quantique.

\section*{ Publications associ\'ees }

Sur la transition de percolation des amas VBS : [P1]

Sur les degr\'es de libert\'e effectifs de basse \'energie dans les diff\'erentes phases : [P2]

%%%%%%%%%%%%%%%%%%%%%%%%%%%%%%%%%%%%%%%%%%%%%%%%%%%%%%%%%%%%%

\chapter
{Cha\^{\i}ne d'Ising avec couplages et champs transverses al\'eatoires}

\label{chapquantique2}

%  \begin{flushright}
%  \emph{ 
%  }\\  ~\\
%  \end{flushright}

\section{ Pr\'esentation du mod\`ele }

\subsection{ Transition de phase quantique }

La cha\^{\i}ne d'Ising quantique
\begin{eqnarray}
H=-\sum_i J_i S_i^z S_{i+1}^z - \sum_i h_i S_i^x
\label{defrtfic}
\end{eqnarray}
avec couplages $J_i>0$ et champs transverses $h_i>0$ al\'eatoires
(les signes peuvent \^etre fix\'es par une transformation de jauge),
est le mod\`ele d\'esordonn\'e le plus simple qui pr\'esente une transition de phase quantique \`a temp\'erature nulle :
le point critique situ\'e en
\begin{eqnarray}
\overline{\ln h } = \overline{\ln J }
\label{criticality}
\end{eqnarray}
s\'epare une phase ferromagn\'etique ($ \overline{\ln h} < \overline{\ln J} $)
d'une phase d\'esordonn\'ee ($ \overline{\ln h} > \overline{\ln J} $).
Ce mod\`ele peut \^etre \'etudi\'e en d\'etail
par la renormalisation de type Ma-Dasgupta \cite{danielrtfic}
comme nous l'avons d\'ej\`a expliqu\'e dans l'Introduction (Page \pageref{introdaniel}). Dans ce Chapitre, nous allons nous int\'eresser
aux propri\'et\'es de taille finie au voisinage de la transition,
dans deux ensembles possibles pour le d\'esordre.

\subsection{ Ensembles microcanonique et canonique pour un syst\`eme d\'esordonn\'e fini }

Dans l'\'etude des syst\`emes d\'esordonn\'es, on consid\`ere
g\'en\'eralement que les variables al\'eatoires de d\'esordre
dans un \'echantillon sont des variables ind\'ependantes :
cet ensemble d'\'echantillons sera appel\'e ici
`canonique' comme dans les r\'ef\'erences \cite{igloi98,dharyoung}.
(ce qui correspond \`a l'ensemble appel\'e `grand canonique'
dans d'autres r\'ef\'erences \cite{paz1,aharonyharris,domany,paz2}). 
Cependant, il est apparu qu'il pouvait \^etre int\'eressant \cite{paz1}
de consid\'erer aussi un ensemble d'\'echantillons
appel\'e `microcanonique' dans les r\'ef\'erences
\cite{igloi98,dharyoung} ( qui correspond \`a l'ensemble appel\'e `canonique'
dans les r\'ef\'erences \cite{paz1,aharonyharris,domany,paz2}). 
Dans cet ensemble microcanonique, il existe une contrainte globale 
sur les variables al\'eatoires de d\'esordre dans un \'echantillon de $N$ sites.
Le premier exemple important concerne un syst\`eme pur avec une fraction
 $p \in [0,1]$ d'impuret\'es \cite{paz1}  :
dans l'ensemble canonique, le nombre total d'impuret\'es
pr\'esente des fluctuations d'ordre $\sqrt{N}$ 
autour de sa valeur moyenne $p N$, alors que dans l'ensemble microcanonique,
 le nombre total d'impuret\'es est exactement $p N$ dans tous les \'echantillons,
et le d\'esordre restant ne concerne que les positions des impuret\'es.
  Il est clair que l'ensemble microcanonique
est beaucoup moins d\'esordonn\'e que l'ensemble canonique,
et en effet il conduit parfois \`a des r\'eductions spectaculaires 
de bruit dans les \'etudes num\'eriques \cite{paz1,paz2}.

Pour la Cha\^{\i}ne d'Ising avec couplages et champs transverses al\'eatoires,
l'ensemble microcanonique au point critique (\ref{criticality})
est d\'efini par la contrainte globale \cite{igloi98,dharyoung}
\begin{eqnarray}
\sum_{i=1}^L ( \ln J_i - \ln h_i) =0
\label{microcriticality}
\end{eqnarray}
alors que dans l'ensemble canonique, 
la quantit\'e du membre de gauche pr\'esente des fluctuations
d'ordre $\sqrt L$ autour de sa valeur moyenne nulle.
Les \'echantillons de l'ensemble microcanonique sont donc `plus proches'
du point critique dans un certain sens, et permettent d'obtenir
des r\'esultats num\'eriques avec moins de fluctuations.

Mais il y a un d\'ebat sur l'int\'er\^et physique
de cet ensemble microcanonique.
D'un c\^ot\'e, certains auteurs \cite{paz1,paz2} pensent
que l'ensemble microcanonique doit \^etre pr\'ef\'er\'e \`a l'ensemble canonique
qui introduirait un bruit suppl\'ementaire susceptible de masquer les propri\'et\'es `intrins\`eques' du syst\`eme.
D'un autre c\^ot\'e, les fluctuations d'ordre
$\sqrt l$ d'une somme de $l$ variables al\'eatoires 
sont pr\'ecis\'ement une propri\'et\'e essentielle des syst\`emes d\'esordonn\'es, qui appara\^{\i}t par exemple dans les divers arguments physiques
discut\'es dans l'Introduction (Page \pageref{argumentslocaux}).
De plus, si on divise un syst\`eme de taille $L$ 
en deux sous-syst\`emes de taille $L/2$,
chaque moiti\'e pr\'esente des fluctuations d'ordre $\sqrt L$ 
dans les deux ensembles :
dans l'ensemble canonique, ces deux moiti\'es sont ind\'ependantes, 
alors que dans l'ensemble microcanonique, les deux moiti\'es sont compl\`etement corr\'el\'ees et ont des fluctuations oppos\'ees.
De ce point de vue, l'ensemble microcanonique peut donc appara\^itre comme artificiel.

Bien s\^ur, on s'attend \`a ce que les deux ensembles 
soient \'equivalents dans la limite thermodynamique, mais comme
dans la th\'eorie des lois d'\'echelle en taille finie, on mesure les exposants
thermodynamiques sur les propri\'et\'es de taille finie,
la discussion sur les deux ensembles conduit en fait au probl\`eme
de la th\'eorie du `finite-size scaling' pour les syst\`emes d\'esordonn\'es
\cite{paz1,domany,paz2}.

\subsection{ Th\'eorie du `finite-size scaling' pour les syst\`emes d\'esordonn\'es ? }

En pr\'esence de d\'esordre, la question centrale est la suivante \cite{paz1,domany,paz2} :
si on obtient par simulations num\'eriques des valeurs pour une observable
comme la susceptibilit\'e $\chi(i,L)$, pour plusieurs \'echantillons $(i)$
pour diverses tailles $L$, quelle est la meilleure proc\'edure pour analyser ces donn\'ees?
La proc\'edure usuelle consiste \`a moyenner l'observable
sur les \'echantillons de taille fix\'ee $\chi_m(L)=<\chi(i,L)>_i$ 
et \`a analyser la d\'ependance en $L$ de la valeur moyenne $\chi_m(L)$
comme dans la th\'eorie du `finite-size scaling' d'un syst\`eme pur.
Cependant, les r\'ef\'erences \cite{paz1,domany,paz2} qui ont \'etudi\'e
 la question sont arriv\'ees \`a la conclusion que la proc\'edure
correcte consiste \`a introduire une pseudo-temp\'erature critique
 $T_c(i,L)$ qui d\'epend de l'\'echantillon.
En effet, ces \'etudes ont montr\'e que les fluctuations de $\chi(i,L)$
entre les \'echantillons venaient en grande partie des fluctuations de $T_c(i,L)$, et que l'emploi de la variable r\'eduite de la variable $(T-T_c(i,L))$ dans l'analyse de `finite-size scaling' 
permettait d'obtenir de bien meilleurs r\'esultats
(cf par exemple les Fig.2 et Fig.3 dans la r\'ef\'erence \cite{paz2}).
De ce point de vue, l'avantage de l'ensemble microcanonique par rapport
\`a l'ensemble canonique est justifi\'e
si la contrainte microcanonique d\'etermine en fait $T_c(i,L)$
ou du moins r\'eduit consid\'erablement ses fluctuations entre \'echantillons,
comme dans le cas des distributions binaires \cite{paz1,paz2}.
Pour les transitions de phase quantiques, la pseudo-temp\'erature critique
$T_c(i,L)$ doit \^etre remplac\'e par un pseudo point critique
comme dans la cha\^{\i}ne d'Ising avec couplages et champs transverses al\'eatoires,
avec la contrainte (\ref{microcriticality}).

 \subsection{ Observables int\'eressantes en taille finie  }

Les \'etudes num\'eriques sur les propri\'et\'es
de taille finie dans les deux ensembles \cite{igloi98,dharyoung}
se sont int\'eress\'ees aux observables suivantes : 

$\bullet$ les aimantations de surface d\'efinies comme l'aimantation d'un bord
lorsque le spin \`a l'autre extr\'emit\'e est fix\'e
\begin{eqnarray}
m^s_1 && \equiv < \sigma_1^z> \vert_{\sigma_L^z=1} \\
m^s_L && \equiv < \sigma_L^z > \vert_{\sigma_1^z=1} 
\label{defmsms}
\end{eqnarray}
et leur corr\'elation $( m^s_1 m^s_L)$.

$\bullet$ la corr\'elation spin-spin entre les deux bords
\begin{eqnarray}
C(L) \equiv < \sigma_1^z \sigma_L^z >
\label{defcl}
\end{eqnarray}

$\bullet$ le `gap', qui repr\'esente la diff\'erence d' \'energies entre 
les deux premiers niveaux
\begin{eqnarray}
\Delta(L) \equiv E_1-E_0
\label{defgapl}
\end{eqnarray}

Le but de ce chapitre est de comparer analytiquement
les comportements de ces diverses observables dans les deux ensembles, au point critique et dans son voisinage.

\section{ R\'esultats sur l'aimantation de surface   }

\subsection{ Aimantation de surface et variable de Kesten  }

L'aimantation de surface 
(\ref{defmsms}) dans un \'echantillon de taille $(L+1)$
a en fait une expression exacte en fonction des couplages
al\'eatoires \cite{igloi98,dharyoung}
\begin{eqnarray}
m_1^S= \left[ 1 + Z_L \right]^{-1/2}
\label{msdef}
\end{eqnarray}
o\`u
\begin{eqnarray}
Z_L \equiv \sum_{i=1}^L \prod_{j=1}^i \left( \frac{h_i}{J_i} \right)^2
\label{kestenzl}
\end{eqnarray}
a une structure sp\'ecifique de somme de produits de nombres al\'eatoires
connue sous le nom de variable al\'eatoire de Kesten \cite{kestenetal}.
Ce type de variables appara\^{\i}t aussi dans les versions discr\`etes du mod\`ele de Sinai \cite{solomon,sinai,derridapomeau},
et dans divers probl\`emes formul\'es en termes de 
produit de matrices de transfert
al\'eatoires $2 \times 2$ \cite{derridahilhorst,calan}.
La version continue d'une variable de Kesten
est la fonctionnelle exponentielle \cite{jpbreview,flux}
\begin{eqnarray}
Z_L =  \int_0^L dx e^{- U(x)}
\label{zlinte}
\end{eqnarray}
du mouvement Brownien $U(x)=  \int_0^{x} dy F(y)$
o\`u le processus $\{F(x)\}$ correspond \`a la version continue
des variables $( -2 \ln (h_i/J_i))$.

\subsection{ Distributions asymptotiques de l'aimantation de surface  }

\label{saddlems}

\subsubsection{ M\'ethode du col dans chaque \'echantillon  }

Il est naturel d'\'evaluer l'int\'egrale (\ref{zlinte}) pour $L$ grand
par une m\'ethode du col  
\begin{eqnarray}
Z_L \opsimeq_{L \to \infty} e^{ E_L } 
\label{saddlezl}
\end{eqnarray}
o\`u $(-E_L)<0$ est le minimum du processus $U(x)$ 
sur l'intervalle $[0,L]$.
Au point critique, le scaling $E_L \sim \sqrt{L}$
montre qu'il existe une distribution limite pour la variable d'\'echelle
 $(\ln Z_L)/\sqrt L \sim \ln m^s_L/\sqrt L \sim E_L/\sqrt{L}$.

\subsection{ Lois d'\'echelle dans les deux ensembles  }

Plus g\'en\'eralement, au voisinage du point critique
on obtient les formes d'\'echelle suivantes 
pour la distribution de probabilit\'e $(-\ln m^s_1)$
\begin{eqnarray}
P_L( -\ln m^s_1) = \frac{2}{ \sqrt{  \sigma L}} Q 
\left( w  = \frac{ -2 \ln m^s_1 }{ \sqrt{  \sigma L}}
 ; \gamma  = \mu {\sqrt { \sigma L }} \right)
\end{eqnarray}
o\`u les fonctions d'\'echelle sont respectivement donn\'ees
dans les deux ensembles par 
\begin{eqnarray}
Q_{cano} 
( w  ; \gamma  )
&& =  \theta(w>0)
\left[   \frac{ 1 }{ \sqrt {\pi  } } 
e^{- \frac{(w+\gamma)^2}{4 } }  
+ \gamma e^{- \gamma w } 
\int_{\frac{w-\gamma}{2}}^{+\infty} \frac{dz}{\sqrt{\pi}} e^{-z^2} \right] \\
Q_{micro} 
\left( w  ; \gamma  \right)
&& = \theta(w >0) \theta(w> - \gamma )
  \left( 2 w+\gamma  \right) 
e^{- \gamma w - w^2 } 
\label{reswithrescaling}
\end{eqnarray}
Ces r\'esultats ont aussi \'et\'e obtenus dans la r\'ef\'erence \cite{dharyoung}
par une m\'ethode diff\'erente.
L'avantage de la formulation par la m\'ethode du col (\ref{saddlezl})
est d'une part de clarifier la validit\'e des formules (\ref{reswithrescaling})
et d'autre part,
de pouvoir \^etre g\'en\'eralis\'ee \`a d'autres observables,
comme la corr\'elation entre les deux aimantations de surface.

\subsubsection{ Distribution asymptotique de la corr\'elation $(m^s_1 m^s_L)$  }

De m\^eme, si on \'evalue par une m\'ethode du col dans chaque \'echantillon
la corr\'elation  
 $(m^s_1 m^s_L)$ entre les deux aimantations de surface, on obtient
\begin{eqnarray}
\ln ( m^s_1 m^s_L ) 
\opsimeq_{L \to \infty} - \frac{ U(0)-U_{min} + U_{max}-U(L)  }{2} 
\label{lnmsmswithd}
\end{eqnarray}
Le calcul de la distribution asymptotique de $(m^s_1 m^s_L)$
se ram\`ene donc au calcul de la loi jointe de l'amplitude
 $A(L)=U_{max}-U_{min}$ et du point final $U(L)$ d'une trajectoire Brownienne partant de $U(0)=0$.
Les r\'esultats pour les deux ensembles (canonique et microcanonique) sont donn\'es dans la Publication [P15].

\subsection{ Moyenne de l'aimantation de surface  }

\subsubsection{ \'Ev\`enements rares  }

On vient de voir qu'au point critique, l'aimantation de surface
a pour comportement typique 
\begin{eqnarray}
m^s_1 = e^{ - \frac{\sqrt {  \sigma L }}{2} w }
\end{eqnarray}
o\`u $w$ est une variable al\'eatoire d'ordre 1.
La moyenne sur les \'echantillons de surface va donc \^etre domin\'ee 
par les \'echantillons rares qui pr\'esentent une aimantation
anormalement grande d'ordre 1, et sa d\'ecroissance avec la taille $L$
va \^etre gouvern\'ee par la d\'ependance en $L$ de la mesure de ces \'echantillons rares.
Ainsi le comportement dans l'ensemble canonique \cite{igloi98,dharyoung} 
\begin{eqnarray}
[m^s_1]_{cano} \oppropto_{L \to \infty} \frac{1}{\sqrt L}
\end{eqnarray}
refl\`ete directement la probabilit\'e qu'une marche al\'eatoire ne repasse pas par l'origine dans l'intervalle $[0,L]$.
En revanche, le comportement diff\'erent dans l'ensemble microcanonique
\cite{igloi98,dharyoung} 
\begin{eqnarray}
[m^s_1]_{micro} \oppropto_{L \to \infty} \frac{1}{ L}
\end{eqnarray}
peut \^etre interpr\'et\'e comme le rapport entre
  
(i) la probabilit\'e $1/L^{3/2}$ pour la probabilit\'e de premier retour 

(ii) la probabilit\'e $1/\sqrt{L}$ d'\^etre \`a l'origine en $L$.

\subsection{ R\'esultats exacts par un calcul d'int\'egrale de chemin  }

Gr\^ace \`a une m\'ethode d'int\'egrale de chemin avec contraintes
(Publication [P15]),
on peut calculer les distributions exactes de l'aimantation de surface
en taille finie
(c'est \`a dire sans faire la m\'ethode du col (\ref{saddlezl})) et obtenir ainsi exactement les aimantations moyennes dans les deux ensembles.

Dans l'ensemble canonique, on trouve
\begin{eqnarray}
[m^s_1 ]_{cano} (\gamma = \mu {\sqrt { \sigma L }}) \opsimeq_{L \to \infty}    
\frac{ e^{\frac{1 }{ 2 \sigma }}
  K_0 \left( \frac{1}{2 \sigma} \right)  }
{ { \sqrt { \pi \sigma L} }  } 
\left[  e^{- \frac{\gamma^2}{4} } +  \gamma  
\int_{-\frac{  \gamma }{2} }^{+\infty} dv e^{-v^2}   \right]
+ O\left(\frac{1}{L} \right)
\label{meanmscanogamma}
\end{eqnarray}
alors que dans l'ensemble microcanonique, on trouve que les deux variables d'\'echelle sont
diff\'erentes des deux c\^ot\'es de la transition,
avec les fonctions d'\'echelle suivantes
\begin{eqnarray}
&& [m^s_1 ]_{micro} (\gamma = \mu {\sqrt { \sigma L }}>0) \opsimeq_{L \to \infty} 
 \frac{ \gamma e^{\frac{1 }{ 2 \sigma }}
  K_0 \left( \frac{1}{2 \sigma} \right)  }{ \sqrt{ \sigma L} } \\
&& [m^s_1 ]_{micro} (\rho= \mu \sigma L<0) \opsimeq_{L \to \infty}    
 \frac{ 4} { {\sqrt \pi} \sigma L  } \int_0^{+\infty} dp p^{-1/2} e^{-p} 
K_{0} \left( 2 \sqrt{\frac{p}{\sigma}} \right)
K_{0} \left( 2 \sqrt{\frac{p}{\sigma}} e^{- \frac{\rho}{2} } \right)  
+ O \left( \frac{1}{L^2 } \right)
\nonumber 
\label{meanmsmicrorho}
\end{eqnarray}

Le r\'egime critique du c\^ot\'e d\'esordonn\'e est donc gouvern\'e par la variable d'\'echelle
$ \rho= \mu \sigma L <0 $ dans l'ensemble microcanonique, 
alors qu'il est gouvern\'e par la variable d'\'echelle
$ \gamma=\mu \sqrt{\sigma L}<0$ dans l'ensemble canonique !
C'est parce qu'il existe en fait deux exposants de corr\'elation.

\subsubsection{ Discussion des deux exposants de corr\'elation}

Il appara\^{\i}t deux exposants de corr\'elation distincts $\nu=2$
et $\tilde{\nu}=1$ dans l'\'etude de la cha\^{\i}ne
dans la limite thermodynamique \cite{danielrtfic} :
en particulier, la fonction de corr\'elation moyenne
est gouvern\'ee par l'exposant $\nu=2$,
alors que la corr\'elation typique est gouvern\'ee
par l'exposant ${\tilde \nu}=1$ \cite{danielrtfic}.
Plus g\'en\'eralement, ces deux \'echelles de longueur
appara\^{\i}ssent aussi dans des mod\`eles reli\'es,
en particulier dans l'\'etude des fonctions propres
de l'op\'erateur de Fokker-Planck du mod\`ele de Sinai  
(cf la discussion Page \pageref{twolengtheigen}) et
dans les mod\`eles unidimensionnels de transport pour les fermions
\cite{balents_fisher,eigenhuse}.

La pr\'esence de ces deux exposants 
 peut \^etre 
expliqu\'e comme suit \cite{danielrtfic}.
La premi\`ere \'echelle de longueur correspond
\`a la longueur ${ \tilde \xi}$
sur laquelle la valeur moyenne  $\overline{U(L)-U(0)}=F_0 L=\mu \sigma L$ 
devient d'ordre 1, ce qui donne
\begin{eqnarray}
{ \tilde \xi } \sim \frac{ 1}{\sigma \mu^{\tilde \nu} } \ \ \ {\rm avec} \ \ {\tilde \nu}=1
\label{deftildexi}
\end{eqnarray}
La seconde \'echelle de longueur correspond
\`a la longueur $\xi$
sur laquelle la plupart des \'echantillons
sont effectivement caract\'eris\'es par
une diff\'erence de potentiel 
$(U(L)-U(0))$ qui est du m\^eme signe que la valeur moyenne :
 l'\'echelle $\sqrt{ \sigma L}$ des fluctuations 
doit \^etre du m\^eme ordre que la valeur moyenne, ce qui donne
\begin{eqnarray}
 \xi \sim \frac{ 1}{\sigma \mu^{\nu}} \ \ \ {\rm avec} \ \  \nu=2
\label{defxi}
\end{eqnarray}

Cette analyse montre qu'il n'est pas surprenant de trouver
qu'une m\^eme observable puisse \^etre gouvern\'ee par deux exposants
diff\'erents dans les deux ensembles : en effet, 
si on cherche la longueur sur laquelle la plupart des \'echantillons
`conna\^{\i}ssent' vraiment le signe de $F_0$,  
on voit que c'est la longueur $\xi$ (\ref{defxi}) 
dans l'ensemble canonique,
mais que c'est la longueur $ \tilde \xi$ (\ref{deftildexi})
dans l'ensemble microcanonique. En conclusion, la d\'efinition
m\^eme des deux longueurs $\xi$ et $ \tilde \xi$
montre que la contrainte microcanonique est susceptible de jouer un r\^ole
dans les comportements d'\'echelle en taille finie.

\section{ Gap et corr\'elation entre spins de bord dans les deux ensembles }

\label{gapcorre}

Contrairement aux aimantations de surface discut\'ees jusqu'\`a pr\'esent,
le gap $\Delta(L)$ (\ref{defgapl}) et la corr\'elation $C(L)$ (\ref{defcl})
entre les deux spins de bord ne peuvent pas \^etre exprim\'es
simplement en termes de tous les couplages al\'eatoires de la cha\^{\i}ne.
Cependant, ils peuvent \^etre \'etudi\'es \cite{rgfinitesize} gr\^ace \`a la formulation en taille finie de la renormalisation de type Ma-Dasgupta, ce qui a permis de faire des pr\'edictions d\'etaill\'ees pour l'ensemble
canonique \cite{rgfinitesize}, qui sont en excellent accord avec les donn\'ees num\'eriques \cite{rgfinitesize,dharyoung}. 
Le but de cette section est de comparer ces r\'esultats de l'ensemble
canonique avec les r\'esultats pour l'ensemble microcanonique calcul\'es dans la Publication [P15].

\subsection{D\'etermination du gap et de la corr\'elation par renormalisation }

L'analyse par renormalisation \cite{danielrtfic,rgfinitesize}
permet de d\'eterminer le gap $\Delta(L)$ (\ref{defgapl}) et la corr\'elation $C(L)$ (\ref{defcl}) dans chaque \'echantillon d\'ecrit par
le potentiel Brownien $U(x)$ for $0 \leq x \leq L$ :
le gap $\Delta(L)$ est asymptotiquement d\'etermin\'e
par la derni\`ere barri\`ere renormalis\'ee ascendante $G$ selon 
\begin{eqnarray}
\ln \Delta (L) = -G 
\end{eqnarray}
alors que la corr\'elation $C(L)$ est asymptotiquement donn\'ee
par la variable $\Lambda \equiv G-U(L)+U(0)$ selon 
\begin{eqnarray}
\ln C (L) = - \Lambda
\end{eqnarray}
En particulier, la renormalisation pr\'edit une relation tr\`es simple
dans chaque \'echantillon entre les deux observables
\begin{eqnarray}
G-\Lambda= U(L)-U(0)
\end{eqnarray}
Cette pr\'ediction a \'et\'e confirm\'ee num\'eriquement au point critique,
avec la convergence de la loi de la variable $(G-\Lambda)/\sqrt{L}$
vers une distribution Gaussienne dans l'ensemble canonique,
et vers une distribution $\delta$ dans l'ensemble microcanonique
\cite{dharyoung}.

\subsection{ R\'esultats au point critique  }

\subsubsection{ Gap dans les deux ensembles  }

Au point critique, les distributions de probabilit\'e de la variable
$G=-\ln \Delta (L)$
ont respectivement pour transform\'ees de Laplace 
par rapport \`a la longueur $L$ dans les deux ensembles
\begin{eqnarray}
&& \int_0^{+\infty} dL e^{-p L} 
\left[  P_L^{\mu=0}(G) \right]_{cano} 
=  \frac{ \sinh \sqrt{p} G  }{\sqrt{p}   \cosh^2 \sqrt{p} G  } \\
&& \int_0^{+\infty} dL e^{-p L} 
 \left[ \frac{ P_L^{(\mu=0) }(G) }{ \sqrt{ 4 \pi  L }} \right]_{micro} 
=   \frac{\sqrt{p} G}{\sinh \sqrt{p} G} e^{ -  \sqrt{p} G \coth \sqrt{p} G } 
\label{resgcriti}
\end{eqnarray}

En particulier, les comportements en singularit\'es essentielles \`a l'origine
obtenus apr\`es l'inversion des transform\'ees de Laplace, conduisent
\`a des comportements en exponentielles \'etir\'ees pour les valeurs moyennes
\begin{eqnarray}
&& \left[ \Delta(L) \right]_{cano}^{scaling} \oppropto_{L \to \infty} 
L^{1/6} e^{- \frac{3}{2} \left( \frac{\pi^2}{2} L \right)^{1/3} }  \\
&& \left[ \Delta_L \right]_{micro}^{scaling}  
= \left[ C_L \right]_{micro}^{scaling}
 \oppropto_{L \to \infty} L^{-1/6} e^{ - \frac{3}{2} ( 2 \pi^2 L)^{1/3}
+ 2 ( 2 \pi^2 L)^{1/6}   }
\label{resgmeancriti}
\end{eqnarray}

\subsubsection{ Corr\'elation dans les deux ensembles  }

Dans l'ensemble canonique, la variable $\Lambda=-\ln C(L)$
a une distribution tr\`es simple
\begin{eqnarray}
P_{cano}^{\mu=0}(\Lambda)   
= \frac{\Lambda}{2 L} e^{- \frac{\Lambda^2}{4 L } }
\end{eqnarray}
qui conduit \`a une d\'ecroissance alg\'ebrique pour la valeur moyenne 
\begin{eqnarray}
\left[  C(L) \right]_{cano}^{scaling}   
\oppropto_{L \to \infty} \frac{1}{L}
\end{eqnarray}
alors que dans l'ensemble microcanonique, les variables $G=\Lambda$
sont identiques : la distribution (\ref{resgcriti}) est tr\`es diff\'erente,
et la valeur moyenne d\'ecro\^{\i}t selon l'exponentielle \'etir\'ee 
(\ref{resgmeancriti}).
  
\subsection{ Comportements dans la r\'egion critique }

De m\^eme, on peut comparer les r\'esultats explicites sur le gap et
la corr\'elation au voisinage du point critique,
du c\^ot\'e ordonn\'e et du c\^ot\'e d\'esordonn\'e (Publication [P15]).
On trouve que les distributions de probabilit\'e sont gouvern\'ees 
par l'exposant critique $\nu=2$ dans les deux ensembles, 
mais que l'exposant $\tilde \nu=1$ gouverne les observables
suivantes dans l'ensemble microcanonique, au lieu de l'exposant $\nu=2$ de l'ensemble  canonique :

$\bullet$ la moyenne de la corr\'elation dans la phase d\'esordonn\'ee

$ \bullet$   la moyenne du gap dans la phase ordonn\'ee.

De nouveau, ces diff\'erences entre les deux ensembles
viennent de la pr\'esence de deux \'echelles de longueur 
(\ref{deftildexi}) et (\ref{defxi}).

\section{ Conclusion}

Cette \'etude explicite des propri\'et\'es de taille finie 
de la transition de phase quantique, 
permet de clarifier l'origine des diff\'erences entre les deux ensembles,
canonique et microcanonique, pour le d\'esordre.
Les variables d'\'echelle sont les m\^emes dans les deux ensembles,
mais les lois de probabilit\'e r\'eduites sont diff\'erentes,
notamment dans leurs comportements asymptotiques.
En cons\'equence, les moyennes des observables qui sont gouvern\'ees
par des \'ev\`enements rares peuvent avoir des comportements
critiques tr\`es diff\'erents.

Par ailleurs, l'\'etude de l'aimantation de surface qui admet une
expression ferm\'ee en termes des variables de d\'esordre, \'eclaire la signification de la m\'ethode de type Ma-Dasgupta.
En effet, dans le cas particulier de l'aimantation de surface
\'ecrite sous la forme de variable de Kesten, 
la m\'ethode de Ma-Dasgupta correspond exactement \`a la m\'ethode du col dans chaque \'echantillon. Plus g\'en\'eralement, on peut donc voir la m\'ethode
de Ma-Dasgupta pour les autres observables
comme un acc\`es direct `au col' dans chaque \'echantillon,
alors qu'on ne conna\^{\i}t pas d'expressions ferm\'ees sur lesquelles
on pourrait effectuer une m\'ethode du col usuelle.

\subsection*{  Publication associ\'ee  [P15] }

%%%%%%%%%%%%%%%%%%%%%%%%%%%%%%%%%%%%%%%%%%%%%%%%%%%%%%%%%%%%%%%%%

\chapter*{Conclusion}
\thispagestyle{plain}
\addcontentsline{toc}{chapter}{Conclusion}
\markboth{Conclusion}{Conclusion}

\label{conclusion}

\section*{ L'int\'er\^et des approches de type Ma-Dasgupta }

\addcontentsline{toc}{section}{L'int\'er\^et des approches de type Ma-Dasgupta }

Dans ce m\'emoire, nous avons d\'ecrit en d\'etail comment diff\'erents types
de syst\`emes d\'esordonn\'es pouvaient \^etre \'etudi\'es
par des proc\'edures de renormalisation de type Ma-Dasgupta.
L'int\'er\^et de ces m\'ethodes vient des propri\'et\'es
sp\'ecifiques suivantes.  \\

$ \bullet $ { \bf Leur domaine de validit\'e : }

Elles permettent d'\'etudier les syst\`emes d\'esordonn\'es 
dans lesquels les h\'et\'erog\'en\'eit\'es spatiales du d\'esordre 
dominent \`a grande \'echelle par rapport aux fluctuations thermiques ou quantiques. Pour les syst\`emes gouvern\'es par un point fixe de d\'esordre infini, elles donnent des r\'esultats exacts.
Pour les syst\`emes gouvern\'es par un point fixe
de d\'esordre fini suffisamment fort,
elles constituent l'ordre dominant d'un d\'eveloppement syst\'ematique
par rapport \`a l'inverse du d\'esordre.  \\

$ \bullet $ { \bf Leur structure pour les r\`egles de renormalisation : } 

Contrairement aux m\'ethodes usuelles de renormalisation qui traitent
l'espace de mani\`ere homog\`ene, les proc\'edures de type Ma-Dasgupta
sont inhomog\`enes dans l'espace pour mieux s'adapter aux r\'ealisations
locales du d\'esordre : c'est une variable extr\^eme de d\'esordre
qui est d\'ecim\'ee de mani\`ere it\'erative et qui engendre un flot
de renormalisation pour une distribution de probabilit\'e.
En dimension $d=1$, cette structure particuli\`ere permet d'obtenir
des solutions explicites pour de nombreux mod\`eles.
En revanche, en dimension sup\'erieure $d>1$, 
l'application des r\`egles de renormalisation
n'a pu \^etre faite, du moins jusqu'\`a pr\'esent, que de mani\`ere num\'erique
(cf Annexe). \\

$ \bullet $ { \bf Leur sens physique : }

Elles ont un sens physique tr\`es clair, car les r\`egles de renormalisation
op\`erent dans l'espace r\'eel et sont bas\'ees sur des arguments physiques simples.
Ainsi, dans les mod\`eles classiques dynamiques, la renormalisation est bas\'ee sur le temps n\'ecessaire pour franchir une barri\`ere de potentiel ou pour s'\'echapper d'un pi\`ege.
 Dans les mod\`eles d'\'equilibre thermodynamique,
comme les cha\^{\i}nes de spins classiques ou l'h\'et\'eropolym\`ere \`a une interface, la renormalisation peut se voir comme un prolongement
des arguments de type Imry-Ma sur l'\'energie al\'eatoire que le syst\`eme
est susceptible de gagner dans un domaine. 
Enfin, dans les mod\`eles quantiques, la renormalisation permet de construire des amas de spins fortement corr\'el\'es. 

$ \bullet $ { \bf Leur r\'esultats : }

Pour tous les mod\`eles o\`u ces m\'ethodes peuvent \^etre appliqu\'ees,
elles donnent une description tr\`es compl\`ete des ph\'enom\`enes.
En effet, comme la proc\'edure de renormalisation est effectu\'ee \'echantillon par \'echantillon, il est possible d'\'etudier par une m\^eme analyse
de tr\`es nombreuses observables. Souvent, elles donnent m\^eme acc\`es aux distributions de probabilit\'e des observables par rapport \`a l'ensemble des \'echantillons, ce qui permet de bien comprendre l'influence des \'ev\`enements rares sur certaines quantit\'es physiques.

$ \bullet $ { \bf Leur efficacit\'e : }

Si elles fournissent des r\'esultats qu'on ne peut pas obtenir par d'autres m\'ethodes, c'est finalement parce qu'elles reposent sur deux piliers
tr\`es diff\'erents, qui ne cohabitent g\'en\'eralement pas \`a l'int\'erieur 
d'une m\^eme approche, \`a savoir des arguments physiques heuristiques 
d'une part et des calculs exacts de renormalisation d'autre part.
L'id\'ee qualitative fructueuse des approches de type Ma-Dasgupta 
est d'int\'egrer ces deux aspects de mani\`ere claire et coh\'erente :
(i) elles utilisent des arguments physiques simples pour identifier
les degr\'es de libert\'e importants \`a grande \'echelle,
et pour d\'efinir une renormalisation directement sur ces degr\'es de libert\'e
(ii) elles \'etudient ensuite exactement le flot de cette renormalisation.

\section*{ Probabilit\'es et renormalisation  }

\addcontentsline{toc}{section}{ Probabilit\'es et renormalisation }

Pour conclure par une r\'eflexion plus g\'en\'erale,
la renormalisation 
est un langage naturel pour l'\'etude des probabilit\'es.
En effet, de m\^eme que certains font de la prose sans le savoir, 
la th\'eorie des probabilit\'es a fait de la `renormalisation'
bien avant que les physiciens n'introduisent ce mot,
avec le th\'eor\`eme de la Limite Centrale, les notions de lois stables 
et de bassins d'attraction.

Dans le cadre des syst\`emes d\'esordonn\'es, les questions 
probabilistes sous-jacentes rel\`event de la statistique des extr\^emes 
\cite{derridayk,bouchaudm-extreme}  : 
l'\'equilibre \`a basse temp\'erature est gouvern\'e par la statistique
des \'etats de basse \'energie \cite{rem,derridayk,bouchaudm-extreme,carpentier2,deanmaj}, alors que la dynamique \`a grand
temps est gouvern\'ee par la statistique des grandes barri\`eres
\cite{feigelman,drosselbarrier,balents,vinokuretal}.

De ce point de vue, les approches de type Ma-Dasgupta
donnent un nouvel \'eclairage en proposant de mani\`ere g\'en\'erale
de renormaliser des distributions de probabilit\'e par d\'ecimation
locale d'une variable al\'eatoire extr\^eme.
En particulier, pour le cas des potentiels al\'eatoires unidimensionnels,
elles proposent de calculer
par renormalisation la distribution de probabilit\'e jointe
des positions des extrema et des barri\`eres qui les s\'eparent.
Parmi les solutions explicites que nous avons d\'ecrites, 
la classe d'universalit\'e la plus importante
est la classe d'universalit\'e des extrema Browniens, qui
permet de comprendre de mani\`ere unifi\'ee
l'\'equilibre ou la dynamique
de diff\'erents types de syst\`emes d\'esordonn\'es.

%%%%%%%%%%%%%%%%%%%%%%%%%%%%%%%%%%%%%%%%%%%%%%%%%%

\appendix

\chapter*{ Annexe :
Autres mod\`eles \'etudi\'es dans la litt\'erature }

\addcontentsline{toc}{chapter}{ Annexe : Autres mod\`eles \'etudi\'es dans la litt\'erature}

\thispagestyle{plain}
\markboth{Annexe}{Annexe}

\label{appendiceq}

Comme nous l'avons d\'ej\`a mentionn\'e dans l'introduction
(Page \pageref{introdaniel}), les renormalisations de type Ma-Dasgupta ont \'et\'e tr\`es utilis\'ees
pour les syst\`emes quantiques depuis les travaux de Daniel Fisher
en 1994-1995 \cite{danielrtfic,danielantiferro}.
Le but de cette Annexe est de mentionner tr\`es bri\`evement 
les diverses directions qui ont \'et\'e suivies dans la litt\'erature.
(Des discussions g\'en\'erales sur  
la physique des syst\`emes quantiques gouvern\'es par des points
fixes de d\'esordre infini se trouvent dans les r\'ef\'erences
 \cite{danielreview,husephysrep}). \\

\section*{ Diverses cha\^{\i}nes de spins quantiques }

La renormalisation de Ma-Dasgupta pour la cha\^{\i}ne
antiferromagn\'etique $S=1/2$ isotrope a \'et\'e g\'en\'eralis\'ee
dans les directions suivantes :

$ \bullet$ en pr\'esence d'anisotropie \cite{danielantiferro}

$ \bullet$ en pr\'esence de dim\'erisation \cite{hyman-dimer}

$ \bullet$ pour des spins plus \'elev\'es, en particulier $S=1$ (cf Chapitre \ref{chapquantique1}) et $S=3/2$ \cite{spin3/2}. \\

La r\'ef\'erence \cite{damlegenechain} contient une discussion du diagramme de phases des cha\^{\i}nes antiferromagn\'etiques en fonction
des divers param\`etres anisotropie/spin/d\'esordre.
La renormalisation a \'et\'e aussi utilis\'ee
pour caract\'eriser les propri\'et\'es de transport de ces cha\^{\i}nes \cite{rsrgdyna}. \\

Parmi les autres types de cha\^{\i}nes de spins \'etudi\'ees, on peut citer

$ \bullet$ la cha\^{\i}ne avec couplages et champ al\'eatoires \cite{danielrtfic} (cf Chapitre \ref{chapquantique2}).

$ \bullet$ la cha\^{\i}ne de Potts et la cha\^{\i}ne `clock' \`a $q$ \'etats 
en champs al\'eatoires \cite{senthil}

$ \bullet$ le mod\`ele Ashkin-Teller
 \cite{carlon-clock-at}

$ \bullet$ les \'echelles de spins \cite{melinladders}.

Mentionnons enfin pour terminer une \'etude r\'ecente sur un mod\`ele de bosons
\cite{bosonrefael}.
Cette liste non-exhaustive montre d\'ej\`a 
les nombreuses possibilit\'es d'\'etude
pour les mod\`eles quantiques unidimensionnels.

\newpage

\section*{Mod\`eles quantiques en dimension sup\'erieure $d >1$ }

Pour les mod\`eles quantiques antiferromagn\'etiques, la proc\'edure 
de renormalisation de Ma-Dasgupta
a \'et\'e \'etudi\'ee num\'eriquement depuis longtemps 
en dimension sup\'erieure \cite{bhatt}.
L'\'etude r\'ecente \cite{numedlin} fait le point sur
les points fixes obtenus en dimensions $d=2$ et $d=3$ pour le mod\`ele antiferromagn\'etique
sur r\'eseau hypercubique, ainsi que pour diff\'erentes variantes 
qui pr\'esentent
de la dim\'erisation ou de la frustration.

Pour le mod\`ele d'Ising avec couplages et champs magn\'etiques
al\'eatoires (discut\'e en dimension $d=1$ dans le Chapitre \ref{chapquantique2}), la renormalisation de type Ma-Dasgupta
 a \'et\'e \'etudi\'e num\'eriquement en dimensions $d=2$ et $d=3$
\cite{motrunich,danielreview}. Le r\'esultat essentiel est que
la transition de phase quantique
\`a temp\'erature nulle est encore gouvern\'ee par des points fixes de d\'esordre infini en $d=2$ et $d=3$.

 Y a-t-il un espoir de calculer un jour exactement les exposants
critiques en $d=2$ ?
D'un c\^ot\'e, la proc\'edure de renormalisation de type Ma-Dasgupta
d\'etruit compl\`etement le r\'eseau r\'egulier initial 
qui devient un r\'eseau al\'eatoire, et
introduit de plus des corr\'elations dans les variables de d\'esordre,
ce qui complique beaucoup l'analyse.
D'un autre c\^ot\'e, la renormalisation de type Ma-Dasgupta
transforme le probl\`eme de la transition de phase quantique 
en un probl\`eme de percolation de certains amas bidimensionnels \cite{motrunich}.
Le calcul exact des exposants critiques en $d=2$ sera donc 
peut-\^etre possible un jour ?

\section*{ Mod\`eles classiques \'etudi\'es dans la litt\'erature}

Dans le cadre de la physique statistique classique,
en dehors des mod\`eles discut\'es dans le pr\'esent m\'emoire,
les autres approches r\'ecentes 
de type Ma-Dasgupta concernent les probl\`emes suivants : 

$ \bullet$ la percolation 2D avec d\'esordre constant par colonne \cite{igloiperco}

$ \bullet$ le mod\`ele de Potts 2D dans la limite $q \to \infty$ \cite{igloipottsq}

$ \bullet$ la transition de phase vers un \'etat absorbant en pr\'esence de d\'esordre \cite{igloicontact}

$ \bullet$ un mod\`ele d'exclusion en pr\'esence de d\'esordre \cite{igloipartial}.

%%%%%%%%%%%%%%%%%%%%%%%%%%%%%%%%%%%%%%%%%%%%%%%%%%%%%%%%%%%%%%

%%%%%%%%%%%%%%%%%%%%%%%%%%%%%%%%%%%%%%%%%%%%%%%%%%%%%%%%%%%

\chapter*{ Liste des Publications associ\'ees }
\thispagestyle{plain}
\addcontentsline{toc}{chapter}{Liste des Publications associ\'ees}
\markboth{Liste des Publications associ\'ees}{Liste des Publications associ\'ees}

\label{listpubli}

Ce m\'emoire est bas\'e sur les publications suivantes, dans l'ordre chronologique : \\

\begin{tabbing}

 [P1] \ \= 
 C. Monthus, O. Golinelli and Th. Jolicoeur,  Phys. Rev. Lett. 79 (1997) 3254
 \\ \>
``Percolation transition in the random antiferromagnetic spin-1 chain".
\\ \>  \\

 [P2] \>
 C. Monthus, O. Golinelli and Th. Jolicoeur,  Phys. Rev. B 58  (1998) 805
\\  \> ``Phases of random antiferromagnetic spin-1 chains". \\  \> \\

 [P3] \>
 D. Fisher, P. Le Doussal and C. Monthus, Phys. Rev. Lett. 80  (1998) 3539
\\ \> 
``Random Walks, Reaction-Diffusion, and Nonequilibrium Dynamics \\ \>
of Spin Chains in One-dimensional Random Environments". \\  \> \\

 [P4] \>
 D. S. Fisher, P. Le Doussal and C. Monthus,   Phys. Rev. E 59  (1999) 4795
\\  \>
``Random Walkers in One-dimensional Random Environments : \\ \>
Exact Renormalization Group Analysis",\\  \> \\

 [P5] \>
 P. Le Doussal and C. Monthus,  Phys. Rev. E 60 (1999) 1212\\ \> 
``Reaction Diffusion models in one dimension with disorder". \\  \> \\

 [P6] \>
 C. Monthus,  Eur. Phys. J. B 13 (2000) 111 \\ \> 
``On the localization of random heteropolymers at the interface \\ \> 
 between two selective solvents".\\ \> \\ 

 [P7] \>
 D. S. Fisher, P. Le Doussal and C. Monthus,  Phys. Rev. E 64 (2001) 66107\\  \>
``Nonequilibrium Dynamics of Random Field Ising Spin 
Chains : \\ \> exact results via real space RG". \\  \> \\

 [P8] \>
C. Monthus and  P. Le Doussal,   Phys. Rev. E 65 (2002) 66129\\  \>
``Localization of thermal packets and metastable states in the Sinai model."
\\  \> \\

 [P9] \>
P. Le Doussal and C. Monthus,  Physica A  317  (2003) 140\\ \> 
``Exact solutions for the statistics of extrema of some random 1D landscapes, \\
\> Application to the equilibrium and the dynamics of the toy model."
\\  \> \\

 [P10] \>
 C. Monthus and P. Le Doussal, Physica A 334 (2004) 78 \\ \> 
``Energy dynamics in the Sinai model."
\\  \> \\

 [P11] \>
 C. Monthus,   Phys. Rev. E  67 (2003) 046109 \\ \> 
``Localization properties of the anomalous diffusion phase  \\ \>
  in the directed trap model and in the Sinai diffusion with bias." \\
\> \\ 

 [P12] \>
 C. Monthus,  Phys. Rev. E  68 (2003) 036114 \\  \>
``Anomalous diffusion, Localization, Aging and Sub-aging effects 
\\ \> in trap models at very low temperature". \\
\> \\

 [P13] \>
 C. Monthus,  J. Phys. A  36 (2003) 11605\\  \>
``On a non-linear Fluctuation Theorem for the aging dynamics \\  \>
of disordered trap models ". \\
\> \\ 

 [P14] \>
 C. Monthus, Phys. Rev. E 69, 026103 (2004)  \\ \> 
``Non-linear Response of the trap model in the aging regime: \\
\>  Exact results in the strong disorder limit". \\
\>  \\

 [P15] \>
 C. Monthus, Phys. Rev. B 69, 054431 (2004)  \\  \>
``Finite-size scaling properties of random transverse-field Ising chains :
\\ \>
comparison between canonical and microcanonical ensembles for the disorder
". \\
\> 

\end{tabbing}

% Il faut mettre le résumé sur la quatrième de couverture. 
% Suivant que votre liste de figures comporte un nombre pair 
% ou impair de pages il faudra mettre un nombre pair ou impair
% de commandes \newpage.
% \newpage\thispagestyle{empty}\addtocounter{page}{-1}
% ~
% \newpage\thispagestyle{empty}\addtocounter{page}{-1}
% ~
\newpage\thispagestyle{empty}\addtocounter{page}{-1}

\section*{Résumé}

Les proc\'edures de renormalisation dans l'espace r\'eel
de type Ma-Dasgupta permettent d'\'etudier des syst\`emes d\'esordonn\'es
gouvern\'es par des points fixes de fort d\'esordre.
Apr\`es une pr\'esentation g\'en\'erale des id\'ees physiques importantes
et des m\'ethodes de calcul, ce m\'emoire d\'ecrit les r\'esultats explicites exacts que ces proc\'edures de renormalisation
 permettent d'obtenir dans diff\'erents mod\`eles unidimensionnels,
classiques ou quantiques, dynamiques ou statiques.
La majeure partie du m\'emoire est consacr\'ee
\`a des mod\`eles de physique statistique, avec notamment \\
(i) la dynamique hors \'equilibre d'une particule dans un potentiel Brownien 
ou dans un paysage de pi\`eges al\'eatoires, \\
(ii)
la dynamique de croissance de domaines et l'\'equilibre thermodynamique
 des cha\^{\i}nes de spins d\'esordonn\'ees classiques, \\
(iii) la transition de d\'elocalisation d'un polym\`ere al\'eatoire \`a une interface. \\
La derni\`ere partie du m\'emoire concerne
deux cha\^{\i}nes de spins quantiques d\'esordonn\'ees qui pr\'esentent une transition de phase \`a temp\'erature nulle
 en fonction du d\'esordre, \`a savoir \\
(a) la cha\^{\i}ne de spin $S=1$ antiferromagn\'etique al\'eatoire \\
(b) la
cha\^{\i}ne d'Ising avec couplages et champs transverses al\'eatoires.

\section*{Abstract}

The Ma-Dasgupta real-space renormalization methods allow to study
disordered systems which are governed by strong disorder fixed points.
After a general introduction to the qualitative ideas
and to the quantitative renormalization rules, we describe the explicit
exact results that can be obtained in various one-dimensional models,
either classical or quantum, either for dynamics or statics.
The main part of this dissertation
is devoted to statistical physics models, with special attention to \\
(i) the off-equilibrium dynamics of a particle
diffusing in a Brownian potential or in a trap landscape, \\ 
(ii) the coarsening dynamics and the equilibrium of classical disordered spin chains, \\
(iii) the delocalization transition
 of a random polymer at an interface. \\
The last part of the dissertation deals with two disordered quantum spin chains
which exhibit a zero-temperature phase transition as the disorder varies,
namely \\
(a) the random antiferromagnetic $S=1$ spin chain, \\
(b) the random transverse field Ising chain.

\end{document}